\RequirePackage{fix-cm}
\documentclass[twocolumn]{svjour3}          
\smartqed  
\usepackage{epsfig}
\usepackage{amssymb}
\usepackage{graphicx}
\usepackage{subfig}
\usepackage{epstopdf}
\usepackage{amssymb} 
\usepackage{dcolumn}
\usepackage{wrapfig}
\usepackage{picinpar}
\usepackage{enumitem}
\usepackage{url}
\usepackage{bm}
\usepackage{amsmath}
\usepackage{relsize}
\usepackage{multirow}
\usepackage{color}
\journalname{Computational Mechanics}
\def\beq{\begin{equation}}
\def\eeq{\end{equation}}

\begin{document}

\title{Performance of preconditioned iterative linear solvers for cardiovascular simulations in rigid and deformable vessels}
\author{Jongmin Seo        \and
        Daniele~E.~Schiavazzi \and
        Alison~L.~Marsden 
}
\institute{J. Seo\at
           Department of Pediatrics and Institute for Computational and Mathematical Engineering(ICME), Stanford University, Stanford, CA, USA\\
           \email{jongminseo@stanford.edu}          
           \and
           D.~E.~Schiavazzi \at
           Department of Applied and Computational Mathematics and Statistics, University of Notre Dame, IN, USA\\
            \email{dschiavazzi@nd.edu}
           \and
           A.~L.~Marsden\at
           Department of Pediatrics, Bioengineering and ICME, Stanford University, Stanford, CA, USA\\
           \email{amarsden@stanford.edu}
}  

\date{Received: September 15th 2018 / Accepted: January 21st 2019}

\maketitle

\begin{abstract}
Computing the solution of linear systems of equations is invariably the most time consuming task in the numerical solutions of PDEs in many fields of computational science. In this study, we focus on the numerical simulation of cardiovascular hemodynamics with rigid and deformable walls, discretized in space and time through the variational multiscale finite element method. We focus on three approaches: the problem agnostic generalized minimum residual (GMRES) and stabilized bi-conjugate gradient (BICGS) methods, and a recently proposed, problem specific, bi-partitioned (BIPN) method.
We also perform a comparative analysis of several preconditioners, including diagonal, block-diagonal, incomplete factorization, multigrid, and resistance based methods. Solver performance and matrix characteristics (diagonal dominance, symmetry, sparsity, bandwidth and spectral properties) 
are first examined for an idealized cylindrical geometry with physiologic boundary conditions and then successively tested on several patient-specific anatomies representative of realistic cardiovascular simulation problems. Incomplete factorization preconditioners provide the best performance and results in terms of both strong and weak scalability. The BIPN method was found to outperform other methods in patient-specific models with rigid walls. In models with deformable walls, BIPN was outperformed by BICG with diagonal and Incomplete LU preconditioners.   
\end{abstract}

\section{Introduction}
\label{intro}
\noindent Cardiovascular simulations are increasingly used in clinical decision making, surgical planning, and medical device design.
In this context, numerous modeling approaches have been proposed ranging from lumped parameter descriptions of the circulatory system to fully three-dimensional patient-specific representations. 
Patient-specific models are generated through a pipeline progressing from segmentation of medical image data, to branch lofting, boolean union, application of physiologic boundary conditions tuned to match patient data, and hemodynamics simulation. 
In diseased vessels, e.g., characterized by localized stenosis or aneurysms, computational fluid dynamics (CFD) has been widely used to diagnose important clinical indicators, such as pressure drop or flow reduction~\cite{Taylor2013}.
Measures of shear stress on the vessel lumen have also been correlated with the risk of endothelial damage and thrombus formation~\cite{Segupta2012,Nesbitt2009}. These quantities are determined by discretization in space and time of the incompressible Navier Stokes equations. 
Multiscale models have been developed to simulate the local flow field in thee-dimensional patient-specific anatomies, while accounting for the presence of the peripheral circulation through closed-loop circuit models providing time dependent boundary conditions~\cite{Lagana2002,Corsini2011,Marsden2015,Esmaily2012}.
In addition, several approaches for fluid-structure interaction (FSI) have been suggested to account for vessel wall deformability~\cite{Figueroa2006,Long2013,Ramachandra2017}.
Recently, hemodynamic models have been used in the solution of complex problems in optimization~\cite{Marsden2008,Marsden2014} and uncertainty quantification~\cite{Sankaran2016,Schiavazzi2016,Schiavazzi2017a,Schiavazzi2017,Tran2017}.

Efforts to improve realism in numerical simulations, however, often lead to an increase in the computational cost. Implementation of the coupled-momentum method (CMM) for FSI increases the simulation run time by roughly twice compared to rigid wall assumptions~\cite{Figueroa2006}, and the cost can be substantially higher for Arbitrary Lagrangian and Eulerian (ALE) FSI. 
On top of this, optimization and uncertainty quantification studies often require the solution of a large number of simulations to obtain converged solutions. These requirements all point to a pressing need to reduce the computational cost to enable future integration of these tools in the clinical setting.

Preconditioned iterative approaches are widely used to solve linear systems, $\mathbf{Ay}=\mathbf{b}$, resulting from discretizations using variational multiscale finite element methods. However, few studies in the literature have examined in detail how linear solver performance depends on the properties of the coefficient matrix, to provide a concrete grounding for the choice and development of more efficient solvers. In addition, even fewer studies have carried out this analysis in the context of computational hemodynamics, i.e., the specific geometries, boundary conditions, mesh and material properties used to create numerical approximations of blood flow in rigid and deformable vessels. In this study, we investigate reductions in computational cost achievable through solving the discretized linear system efficiently, as this cost is well known to dominate the execution time. The objective of this study is to perform a systematic comparative analysis of existing linear solver strategies, linking their performance with the distributed coefficient matrix characteristics of cardiovascular modeling. 

Krylov subspace based iterative solvers are typically preferred for the solution of large linear systems from CFD, due to their superior scalability and memory requirements compared to direct methods~\cite{Benzi2002}.
Popular Krylov subspace iterative solvers include the conjugate gradient method (CG) for symmetric positive definite (SPD) coefficient matrices, and the generalized minimum residual method (GMRES) or the bi-conjugate gradient stabilized method (BICGS) in the non-symmetric case. Alternatively, a recently proposed bi-partitioned linear solver (BIPN)~\cite{Esmaily2015} leverages the block-structure in the coefficient matrix, separating contributions from the Navier-Stokes momentum and continuity equations. In BIPN, the coefficient matrix $\mathbf{A}_f$ arising from finite element spatial discretizations and the time discretization for the Navier-Stokes equations consists of four blocks, 
\beq
\centering
\mathbf{A}_f = 
\begin{bmatrix}
   \mathbf{K} & \mathbf{G}\\
   \mathbf{D} & \mathbf{L} 
\end{bmatrix}, 
\eeq
in which $\mathbf{K}$ and $ \mathbf{G}$ stem from the momentum equation, $ \mathbf{D}$ and $ \mathbf{L}$ stem from the continuity equation and stabilization. BIPN solves the matrix block $\mathbf{K}$ using GMRES, while the rest is transformed to a Schur complement form as $\mathbf{L}-\mathbf{D}\mathbf{K}^{-1}\mathbf{G}$, approximated by a SPD matrix, $\mathbf{L}_*+\mathbf{G}_*^T\mathbf{G}_*$, in which the star subscript indicates the symmetric Jacobi scaling with diagonals of $\mathbf{K}$ and $\mathbf{L}$. The Schur complement form is solved with CG, and the solution time for CG takes more than 90$\%$ of the total computing time in benchmark testing\cite{Esmaily2015}. 

It is also well known that preconditioning plays a key role in accelerating the convergence of Krylov subspace methods~\cite{Trefethen,Saad2003} by transforming the original linear system $\mathbf{A}\mathbf{y}=\mathbf{b}$ to  $\mathbf{M}^{-1}\mathbf{A}\mathbf{y}=\mathbf{M}^{-1}\mathbf{b}$ (left preconditioning),  $\mathbf{A}\mathbf{M}^{-1}\mathbf{z}=\mathbf{b}, \mathbf{y} = \mathbf{M}^{-1}\mathbf{z}$ (right preconditioning), or {\color{black}$\mathbf{M}_1^{-1}\mathbf{A}\mathbf{M}_2^{-1}\mathbf{y}=\mathbf{M}_{1}^{-1}\mathbf{b}$,  $\mathbf{x} = \mathbf{M}_2^{-1}\mathbf{y}$ (left and right preconditioning)}. {\color{black} In many cases, $\mathbf{M}$ is constructed in a way that $\mathbf{M}^{-1}$ approximates $\mathbf{A}^{-1}$.} In general, an ideal preconditioner should be relatively cheap to apply and effective to reduce the overall solution time. 
In its simplest form, left, right or left-right Jacobi (diagonal) preconditioners are effective in shrinking the eigenvalue spectrum of diagonally dominant matrices.
Preconditioners based instead on incomplete factorization (ILU) provide an approximate decomposition in the form $\mathbf{M}=\bar{\mathbf{L}}\bar{\mathbf{U}}$, where the sparsity pattern of $\mathbf{A}$ is preserved in the factors.
ILUT preconditioners are slightly more general approaches allowing for adjustable inclusion of fill-ins, but require the user to specify an additional threshold parameter.
We note that the efficiency of an ILU preconditioner results from a trade off between fewer Krylov iterations needed for convergence and the cost of incomplete factorization~\cite{Saad2003}.
Application-specific preconditioners have also been proposed in cardiovascular hemodynamics to improve performance when the model outlets are coupled through a resistance boundary condition, an RCR circuit, or for more general multi-domain configurations. In what follows, we refer to an in-house implementation of this class of preconditioners as \emph{resistance-based} preconditioner (RPC)~\cite{Esmaily2013a,Esmaily2015}. Additional preconditioning techniques for cardiovascular simulations with FSI are suggested in \cite{Manguoglu2010,Deparis2016}. Finally, algebraic multigrid preconditioners have also received significant recent interest~\cite{Santos2004}.

Despite the availability of open-source implementations of iterative solvers and preconditioners, few studies have systematically compared the performance of these solvers for cardiovascuar models with rigid and deformable vessels.
In addition, a thorough understanding of the factors affecting the performance of iterative linear solvers (e.g., diagonal dominance, condition number, sparsity pattern, symmetry, and positive-definiteness) is an important prerequisite for optimal choice of solver and for the development of new algorithms with improved performance.

In the current study, we first compare the performance of various iterative linear solvers and preconditioners for an idealized flow through a cylindrical vessel with a resistance outflow boundary condition. We then test our findings using three representative patient-specific cardiovascular models.
The Trilinos software library~\cite{Trilinos}, developed at Sandia National Laboratory is coupled with the SimVascular svFSI open source finite element code to provide linear solvers such as GMRES and BICGS, as well as a variety of preconditioners: diagonal (Diag), block-diagonal (BlockD), incomplete LU (ILU), thresholded incomplete LU (ILUT), incomplete Cholesky (IC), and algebraic multigrid (ML). We use Kylov linear solvers and Diag, BlockD, ILU, ILUT preconditioners from the AztecOO package, and IC is implemented via the IFPACK package. The block-diagonal preconditioner scales the block matrix via the Trilinos Epetra Vbr class. The incomplete factorization methods use Additive Schwarz domain decomposition for the parallelization. The Trilinos ML package for algebraic multigrid is implemented on the AztecOO package. For detailed information on parallelization and preconditioning options, we refer readers to the Trilinos project \cite{Trilinos}.   
BIPN and RPC are integrated and implemented directly in our flow solver with source code available through the SimVascular open source project (www.simvascular.org) \cite{Updergrove2016}.  

This paper is organized as follows. 
In section~\ref{sec:formulation}, we review the formulation of the coefficient matrices resulting from finite element weak forms in fluid and solid mechanics, before discussing the performance of various linear solvers and preconditioners on a simple pipe benchmark in section~\ref{sec:pipe}.
In section~\ref{sec:Scaling} we report the results of strong and weak scaling for BIPN with RPC and BICG with ILU, while in section~\ref{sec:CharMat} we examine the properties of the coefficient matrix.
The effect of preconditioning on these properties is reported in section~\ref{sec:PC}. 
In section~\ref{sec:PS} we compare performance of linear solvers in patient-specific models.
We draw conclusions and discuss future work in section~\ref{sec:Conclusions}.

\section{Linear systems in cardiovascular simulation}\label{sec:formulation}
%
\noindent We begin by introducing the space-time discretization of the equations governing fluid and solid mechanics following an Arbitrary-Lagrangian-Eulerian (ALE) description of the interaction between  fluid and structure~\cite{Huges1981,Bazilevs2009,Esmaily2015,Vedula2017}.  These equations are discretized with a variational multiscale finite element method, and are provided in the svFSI solver of the SimVascular open source project~\cite{Updergrove2016}.

\subsection{Linear system for fluid mechanics}\label{sec:LSf}
%
\noindent Consider a domain $\Omega_f \in\mathbb{R}^{3}$, occupied by a Newtonian fluid whose evolution in space and time is modeled through the incompressible Navier-Stokes equations in ALE form, 
\begin{equation}
\begin{split}
\rho\,\frac{\partial \mathbf{u}}{\partial t}|_{\hat{ \mathbf{x}}}+\rho\,{\mathbf{v}}\cdot\nabla\,{\mathbf{u}}  &= \rho\,\mathbf{f}+ \nabla \cdot {\mathbf{\sigma}_f}\\
\nabla  \cdot {\mathbf{u}} &= 0
\end{split}
\quad\text{in}\,\,\Omega_f,
\end{equation}
where $\rho, \mathbf{u}=\mathbf{u}(\mathbf{x},t)$, and $\mathbf{f}$ are fluid density, velocity vector, and body force, respectively. 
The fluid stress tensor is $\mathbf{\sigma}_f=-p\,{\mathbf{I}} + \mu\,(\nabla{\mathbf{u}}+\nabla{\mathbf{u}}^T) = -p\,{\mathbf{I}} + \mu\,\nabla^{s}{\mathbf{u}}$, $\mu$ is the kinematic viscosity, $p=p(\mathbf{x},t)$ the pressure, $\mathbf{v}=\mathbf{u-\hat{u}}$ is the fluid velocity relative to the velocity of the domain boundary $\mathbf{\hat{u}}$. Variables are interpolated in space at time $t^n$ as
\begin{equation}
\begin{split}
\mathbf{w}(\mathbf{x})=\sum_{a\in {I}_a} {N}^a(\mathbf{x})\,\mathbf{w}^a, \\
\mathbf{q}(\mathbf{x})=\sum_{a\in {I}_a} {N}^a(\mathbf{x})\,\mathbf{q}^a, 
\end{split}
\end{equation}
\begin{equation}
\begin{split}
\mathbf{u}(\mathbf{x},{t=t^n)}=\mathbf{u}^n(\mathbf{x}) = \sum_{a\in {I}_a} {N}^a(\mathbf{x})\,\mathbf{u}^{a,n}, \\
{p}(\mathbf{x},{t=t^n)}={p}^n(\mathbf{x}) = \sum_{a\in {I}_a} {N}^a(\mathbf{x})\,p^{a,n}, 
\end{split}
\end{equation}
in which $\mathbf{I}_a, N^a, \mathbf{w}^a, q^a, \mathbf{u}^a$, and $p^a$ are the nodal connectivity set, interpolation functions at node $a$, test function weights, velocity and pressure at node $a$, respectively. In this study, we employ P1-P1 type (linear and continuous) spatial approximations of the fluid velocity and pressure. 
We consider a stabilized finite element discretization based on the variational multiscale method~\cite{Bazilevs2008,Esmaily2015}, leading to the weak form of the Navier-Stokes momentum and continuity residuals
\begin{multline}\label{eq:Rm}
\mathbf{R}_m^a(\mathbf{\dot{u}},\mathbf{u},\mathbf{v},p)=\\
\sum_{e\in \mathbf{I}_e}\,\int_{\Omega^e}\,\rho\, N^a
\left(\mathbf{\dot{u}}-\mathbf{f}+(\mathbf{v}+\mathbf{u}_p)\cdot\nabla\mathbf{u}\right)\,d\Omega \\
+\sum_{e\in \mathbf{I}_e}\int_{\Omega^e} (\nabla N^a)^T  
(-p\mathbf{I}+\mu \nabla^s\mathbf{u} +\rho{\tau}_B \mathbf{u}_p \otimes (\mathbf{u}_p \cdot \nabla \mathbf{u})\\
-\rho\, \mathbf{u}\otimes\mathbf{u}_p+\rho \tau_C \nabla \cdot \mathbf{u})\, d\Omega-\int_{\Gamma_h} N^a\,\mathbf{h}\,d\Gamma,
\end{multline}

\begin{equation}\label{eq:Rc}
{R}_c^a(\mathbf{\dot{u}},\mathbf{u},p)=\int_{\Omega}\left(N^a \nabla \cdot \mathbf{u} -(\nabla N^a)^T\mathbf{u}_p\right)\,d\Omega, 
\end{equation}
in which $\mathbf{R}_m^a$ and $R_c^a$ are momentum and continuity residuals at node $a$, and $\mathbf{h}$ is the surface traction on the Neumann boundary $\Gamma^h$. The stabilization parameters are defined as
\beq
\begin{split}
\mathbf{u}_p  =-\tau_M\left( \dot{\mathbf{u}}+\mathbf{v}\cdot \nabla \mathbf{u}+\frac{1}{\rho}\nabla p -\frac{\mu}{\rho} \nabla^2 \mathbf{u} -\mathbf{f}\right),\\
\tau_M = \left( \frac{4}{\Delta t^2} +\mathbf{v}\cdot \mathbf{G} \mathbf{v}+C_I\left(\frac{\mu}{\rho}\right)^2\mathbf{G}:\mathbf{G}\right)^{-1/2},\\
{\tau}_{B}  =(\mathbf{u}_p\cdot \mathbf{G}\,\mathbf{u}_p)^{-1/2},\,\,
\tau_{C}=(\tau_M\,\boldsymbol{g}\cdot\boldsymbol{g})^{-1},\\
G_{ij}=\sum_{k=1}^{3}\frac{\partial \xi_k}{\partial x_i}\frac{\partial \xi_k}{\partial x_j},\\
\boldsymbol{g}\cdot\boldsymbol{g}=\sum_{i=1}^{3}g_i\, g_i,\,\,g_i=\sum_{k=1}^3 \frac{\partial \xi_k}{\partial x_i},
\end{split}
\label{eq:stab}
\eeq
where $C_I$ is a constant set to 3, $\Delta t$ is the time step size, and $\xi$ represents natural coordinates.
Integration in time is performed using the unconditionally stable, second order accurate generalized-$\alpha$ method~\cite{jansen2000generalized}, consisting of four steps: predictor, initiator, Newton-Raphson, and corrector. 
Initial values for accelerations, velocities and pressures at time $t^{n+1}$ are set in the prediction step as
\begin{equation}
\dot{\mathbf{u}}^{a,n+1} =\frac{\gamma -1}{\gamma}\,\dot{\mathbf{u}}^{a,n},\,\,\mathbf{u}^{a,n+1} =\mathbf{u}^{a,n},\,\,p^{a,n+1} =p^{a,n},
\end{equation}
where $\gamma=0.5+\alpha_m-\alpha_f$, $\alpha_m=1/(1+\rho_\infty)$, and $\alpha_f=(3-\rho_\infty)/(2+2\rho_\infty)$ are the generalized-$\alpha$ method coefficients, while $\rho_\infty$ is the spectral radius set to $\rho_\infty=0.2$ in this study.
In the initiator step, accelerations and velocities are computed at an intermediate stage $n+\alpha_m$ and $n+\alpha_f$, 
\beq
\label{eq:interm1}
\begin{split}
\dot{\mathbf{u}}^{a,n+\alpha_m}=(1-\alpha_m)\,\dot{\mathbf{u}}^{a,n}+\alpha_m\,\dot{\mathbf{u}}^{a,n+1},\\
\mathbf{u}^{a,n+\alpha_f}=(1-\alpha_f)\,\mathbf{u}^{a,n}+\alpha_f\,\mathbf{u}^{a,n+1}.
\end{split}
\eeq
A Newton-Raphson iteration is performed based on Equations~\eqref{eq:Rm} and~\eqref{eq:Rc}, using $\dot{\mathbf{u}}^{n+\alpha_m}$, ${\mathbf{u}}^{n+\alpha_f}$, ${p}^{n+1}$ from \eqref{eq:interm1} by solving a linear system of the form
\beq
\begin{split}
\mathbf{K}\,\Delta\mathbf{u}+\mathbf{G}\,\Delta\mathbf{p} & =-\mathbf{R}_m\left(\dot{\mathbf{u}}^{n+\alpha_m}, \mathbf{u}^{n+\alpha_f}, p^{n+1}\right), \\
\mathbf{D}\,\Delta\mathbf{u}+\mathbf{L}\,\Delta\mathbf{p} & =-\mathbf{R}_c\left(\dot{\mathbf{u}}^{n+\alpha_m}, \mathbf{u}^{n+\alpha_f}, p^{n+1}\right), 
\end{split}
\label{eq:LS}
\eeq
where the blocks $\mathbf{K}$, $\mathbf{G}$, $\mathbf{D}$, and $\mathbf{L}$ partition the tangent coefficient matrix with blocks for nodes $a$ and $b$ equal to
\beq\label{eq:LHS}
\begin{split}
\mathbf{K}^{ab} \approx \frac{\partial \mathbf{R}_m^a}{\partial \Delta \mathbf{u}^b},\,\,\mathbf{G}^{ab}\approx \frac{\partial \mathbf{R}_m^a}{\partial \Delta p^b},\\
\mathbf{D}^{ab} \approx \frac{\partial {R}_c^a}{\partial \Delta  \mathbf{u}^b},\,\,\mathbf{L}^{ab}\approx \frac{\partial {R}_c^a}{\partial \Delta p^b}.
\end{split}
\eeq
We re-write this linear system in matrix form as
\beq
\mathbf{A}_f\,\mathbf{y}=-\mathbf{R}_f,
\label{eq:tangent_f}
\eeq
where
\beq
\mathbf{A}_f = 
\begin{bmatrix}
   \mathbf{K} & \mathbf{G}\\
   \mathbf{D} & \mathbf{L}
\end{bmatrix},\,\mathbf{y} = 
\begin{bmatrix}
   \Delta\mathbf{u}\\
  \Delta\mathbf{p}
\end{bmatrix},\,
\mathbf{R}_f = 
\begin{bmatrix}
   \mathbf{R}_m\\
   \mathbf{R}_c
\end{bmatrix}, 
\label{eq:blockmatrix}
\eeq
with blocks $\mathbf{K}$, $\mathbf{G}$, $\mathbf{D}$, $\mathbf{L}$ of size ($3 N_{nd}\times3 N_{nd}$), $(N_{nd}\times3 N_{nd}$), ($3 N_{nd}\times N_{nd}$), ($N_{nd}\times N_{nd}$), respectively. 
Here $N_{nd}$ is the total number of nodes, while $\Delta\mathbf{u}\in\mathbb{R}^{3\,N_{nd}}$ and $\Delta\mathbf{p}\in\mathbb{R}^{N_{nd}}$ contain nodal velocities and pressure increments. We note that the major focus of our study is on solving the linear system in equation (\ref{eq:tangent_f}). 
Once the momentum and continuity residual norms drop below a given tolerance, the unknowns at the next time step are determined through the corrections 
\beq
\begin{split}
\dot{\mathbf{u}}^{a,n+1}\leftarrow \dot{\mathbf{u}}^{a,n+1}+\Delta\mathbf{u}^a,\\
\mathbf{u}^{a,n+1}\leftarrow \bm{u}^{a,n+1}+\gamma\,\Delta t\,\Delta\mathbf{u}^a,\\
p^{a,n+1}\leftarrow p^{a,n+1}+\alpha_f\,\gamma\,\Delta t \,\Delta{p}^a,\\
\end{split}
\label{eq:corrector}
\eeq
$\forall\,a\in \mathbf{I}_a$. 
Finally, detailed expressions for each block of the coefficient matrix from Eq. (\ref{eq:Rm}), Eq. (\ref{eq:Rc}), Eq. (\ref{eq:LHS}), and Eq. (\ref{eq:corrector}) are 
\begin{multline}
\mathbf{K}^{ab}=\sum_{e\in \mathbf{I}_e} \int_{\Omega^e} \Big[ \rho \alpha_m N^a N^b \mathbf{I}^{ab} + \rho \tilde{\alpha}_fN^a(\mathbf{v}+\mathbf{u}_p) \cdot \nabla N^b \mathbf{I}^{ab}\\
+ \mu \tilde{\alpha}_f( \nabla N^a \cdot  \nabla N^b \mathbf{I}^{ab}+ \nabla N^b \otimes  \nabla N^a)\\
+\rho \tilde{\alpha}_f {\tau}_B\mathbf{u}_p \cdot  \nabla N^a \mathbf{u}_p\cdot  \nabla N^b \mathbf{I}^{ab} \\
+ \rho \tau_M \mathbf{u} \cdot  \nabla N^a (\alpha_m N^b + \tilde{\alpha}_f\,\mathbf{u} \cdot  \nabla N^b )\mathbf{I}^{ab} \\
+\rho \tilde{\alpha}_f\tau_C  \nabla N^a \otimes  \nabla N^b\Big] d\Omega, 
\label{eq:Kblock}
\end{multline}

\beq
\mathbf{G}^{ab}=\sum_{e\in \mathbf{I}_e} \int_{\Omega^e}\Big[ -\tilde{\alpha}_f \nabla N^a N^b +  \tilde{\alpha}_f \tau_M \mathbf{u} \cdot\nabla N^a \nabla N^b\Big] d\Omega, 
\label{eq:Gblock}
\eeq

\begin{multline}
\mathbf{D}^{ab}=\sum_{e\in\mathbf{I}_e} \int_{\Omega^e} \Big[ \tilde{\alpha}_f  N^a \nabla N^b +\tilde{\alpha}_f \tau_M  \bm{u} \cdot  \nabla N^a\nabla N^b\\
+  \tau_M  \nabla N^a \alpha_m N^b \Big] d\Omega, 
\label{eq:Dblock}
\end{multline}

\beq
\mathbf{L}^{ab}=\sum_{e\in \mathbf{I}_e} \int_{\Omega^e}\Big[ \frac{\tilde{\alpha}_f \tau_M }{\rho} \nabla N^a\cdot \nabla N^b \Big] d\Omega,  
\label{eq:Lblock}
\eeq
in which $\tilde{\alpha}_f=\gamma\, \Delta t\,\alpha_f$ and $\mathbf{I}_e$ is the list of elements containing nodes $a$ and $b$.

First we observe that the matrix $\mathbf{K}$ is diagonally dominant. Except for entries related to the stabilization terms, which are typically small, the most significant off-diagonal contribution is provided by the viscous term, which is also typically smaller than the acceleration and advection terms in cardiovascular flows. Second, the small magnitude of the stabilization terms suggests that $\mathbf{G}$ is similar to $-\mathbf{D}^T$. We also observe that the matrices $\mathbf{K}$ and $\mathbf{A}_f$ are non-symmetric, while $\mathbf{L}$ is symmetric and singular since it has an identical structure to the matrices arising from the discretization of generalized Laplace operators. We also note that $\mathbf{L}$ is characterized by small entries compared to the other blocks, since it only consists of stabilization terms.

\subsection{Linear system for solid mechanics}\label{sec:LSs}
%
\noindent In the solid domain, we start by introducing measures of deformation induced by a displacement field $\mathbf{d} = \mathbf{x} - \mathbf{X}$, i.e., the difference between the current and material configurations $\mathbf{x}\in\mathbb{R}^{3}$ and  $\mathbf{X}\in\mathbb{R}^{3}$, respectively
\begin{equation}
\begin{split}
\mathbf{F}=\nabla \mathbf{d}+\mathbf{I},\quad\mathbf{C}=\mathbf{F}^{T}\mathbf{F},\quad\mathbf{E}=\frac{1}{2} (\mathbf{C}-\mathbf{I}), 
\end{split}
\end{equation}
where $\mathbf{F}$, $\mathbf{C}$, $\mathbf{E}$, represent the deformation gradient, the Cauchy-Green deformation tensor and the Green strain tensor. The Jacobian is also defined as $J=\text{det}(\mathbf{F})$.
We relate the second Piola-Kirchhoff stress tensor $\mathbf{S}$ with the Green strain tensor $\mathbf{E}$ through the Saint Venant-Kirchhoff hyperelastic constitutive model 
\begin{equation}
\mathbf{S}=\lambda\,\text{tr}(\mathbf{E})\,\mathbf{I} +2\,\mu\,\mathbf{E},\\
\end{equation}
where $\lambda=\frac{\nu E_s}{(1+\nu)(1-2\nu)},\,\mu=\frac{E_s}{2(1+\nu)}$, $E_s$ and $\nu$ represent the Young's modulus and Poisson's ratio, respectively. 
The equilibrium equation is 
\begin{equation}
\begin{split}
\rho_s\,{\frac{\partial \mathbf{u}}{\partial t}}=\rho_s\,\mathbf{f} + \nabla\cdot \sigma_s
\end{split}
\quad\text{in}\,\,\Omega_s,
\end{equation}
where $\rho_s$ and $\sigma_s$ denote the density and solid stress tensor, respectively. This leads to the weak form
\beq\label{equ:solidWeakForm}
\int_{\Omega_s^0}\,\left[\rho_s^0\,\mathbf{w}{(\mathbf{\dot{u}}-\mathbf{f})} +\nabla \mathbf{w} : \mathbf{P}\right]\, d\Omega=\mathbf{0},
\eeq
where $\mathbf{P}=\mathbf{F}\mathbf{S}$ is the first Piola-Kirchhoff stress tensor, $\mathbf{w}$ is a virtual displacement and $\Omega^{0}_s$ is the solid domain in the \emph{reference} configuration.
Discretization of \eqref{equ:solidWeakForm} leads to the residual
\beq\label{eq:solidR}
\mathbf{R}^a_m(\dot{\mathbf{u}}, \mathbf{d})=\int_{\Omega_s^0}\,\left[\rho_s^0N^a{(\mathbf{\dot{u}-f})} + \mathbf{F\,S}\nabla N^a\right]\,d\Omega.
\eeq
Using the generalized-$\alpha$ method, the displacements at time $t^{n+1}$ are predicted as
\beq
\mathbf{d}^{a,n+1}=\mathbf{d}^{a,n}+\mathbf{u}^{a,n+1}\Delta t + \frac{0.5\gamma-\beta}{\gamma-1}\mathbf{\dot{u}}^{a,n+1}\Delta t^2,
\eeq
in which $\beta=\frac{1}{4}(1+\alpha_f-\alpha_m)^2$. In the initiator step, the intermediate displacements are provided by
\beq
{\mathbf{d}}^{a,n+\alpha_f} = (1-\alpha_f)\,{\mathbf{d}}^{a,n} + \alpha_f \,{\mathbf{d}}^{a,n+1}. 
\eeq
Solving \eqref{eq:solidR} with the Newton-Raphson method, we obtain the linear system
\beq
\mathbf{K}_s\,\Delta\mathbf{d}=-\mathbf{R}_m\,(\mathbf{\dot{u}}^{n+\alpha_m}, \mathbf{d}^{n+\alpha_f}),
\label{eq:tangent_s}
\eeq
where $\mathbf{K}_s^{ab}\approx \frac{\partial \mathbf{R}_m^a}{\partial \Delta d^b}$,
with tangent stiffness matrix  
\begin{multline}
\label{eq:solid}
\mathbf{K}_s^{ab}=
\int_{\Omega_s^0}\Big[ \rho_s^0 \alpha_m N^a N^b \mathbf{I} +
\hat{\alpha}_f \big(\mathbf{S}\nabla N^a\big) \cdot \nabla N^b \mathbf{I} \\
+ \lambda\,\hat{\alpha}_f (\mathbf{F}\nabla N^a) \otimes (\mathbf{F}\nabla N^b) 
+ \mu\,\hat{\alpha}_f (\mathbf{F}\nabla N^b) \otimes (\mathbf{F} \nabla N^a)\\
+\mu\,\hat{\alpha}_f \mathbf{F} \mathbf{F}^T\nabla N^a \cdot \nabla N^b\Big]d\Omega, 
\end{multline}
and $\hat{\alpha}_f=\alpha_f\,\beta\,\Delta t^2$. 
At this point, $\mathbf{d}$ is corrected using 
\begin{equation}
\mathbf{d}^{a,n+1}\leftarrow \mathbf{d}^{a,n+1}+\beta\,\Delta t^2 \Delta\mathbf{d}^a,\,\,\forall a \in \mathbf{I}_a.
\end{equation}
Again, we note that the focus of this work is on solving the linear system in equation (\ref{eq:tangent_s}) together with solving equation (\ref{eq:tangent_f}) when including FSI. 
Finally, we observe that most terms in \eqref{eq:solid} (i.e., the third, fourth, and fifth terms) contribute to the off-diagonals of $\mathbf{K}_s$ and that the matrix $\mathbf{K}_s$ is symmetric. 

\section{Linear solver performance on pipe flow benchmark}\label{sec:pipe}
\begin{figure}[htbp]
\vspace{-6pt}
\centering
\includegraphics[width=0.4\textwidth, keepaspectratio]{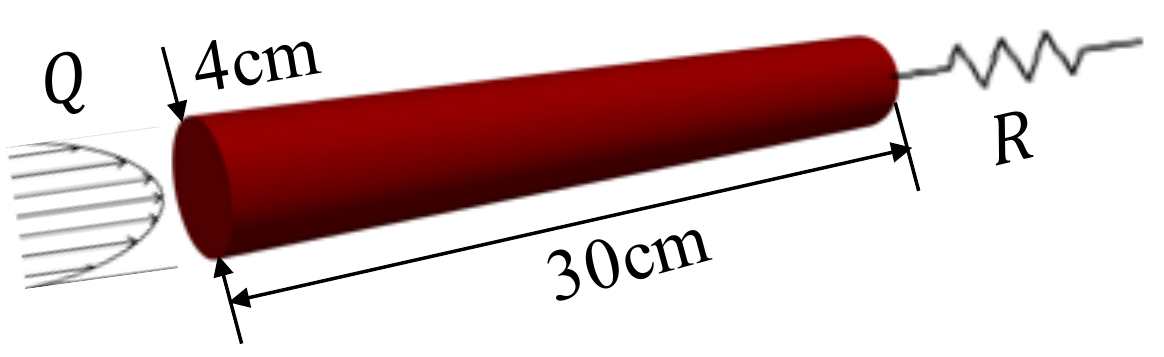}
\includegraphics[width=0.35\textwidth, keepaspectratio]{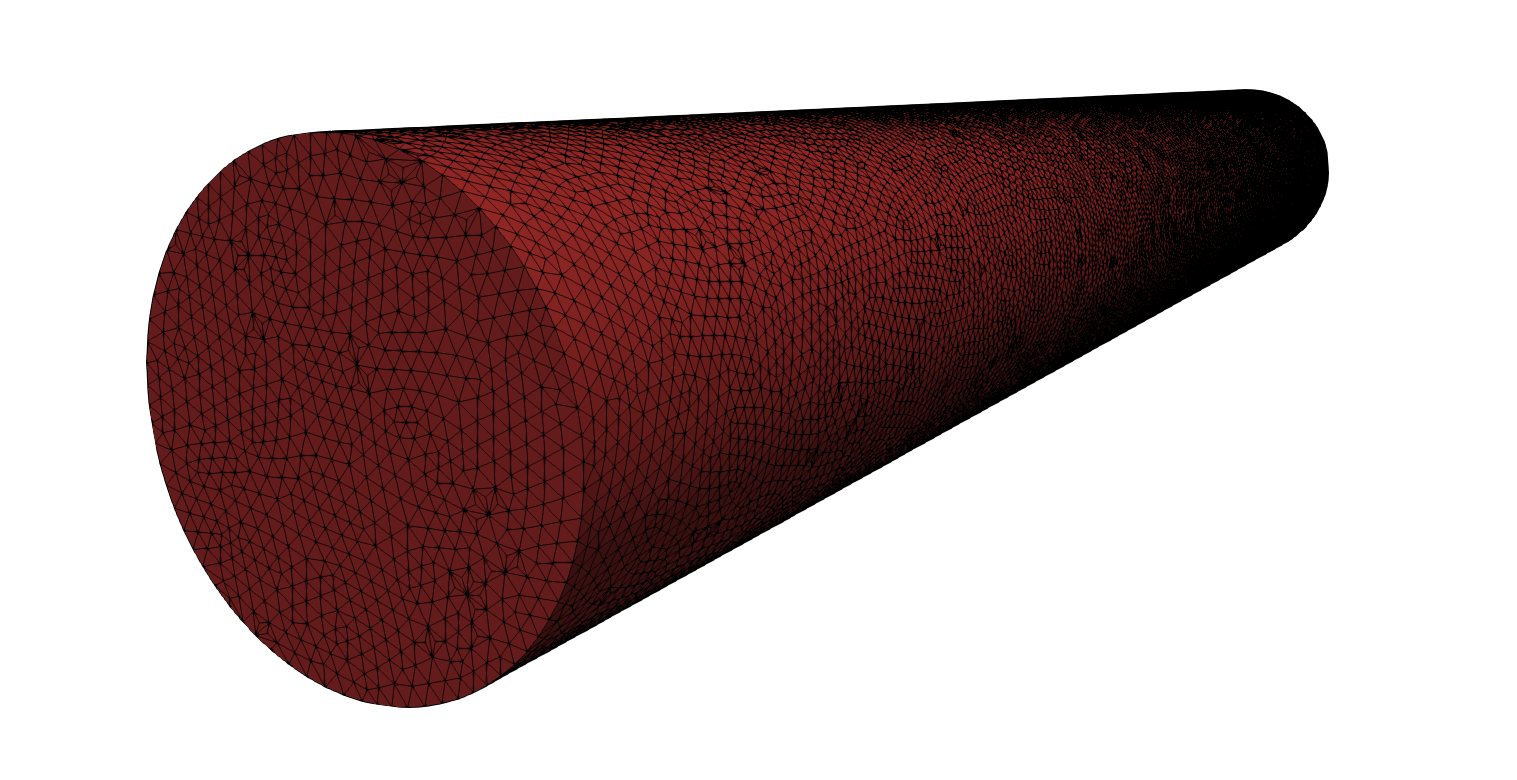}
\includegraphics[width=0.4\textwidth, keepaspectratio]{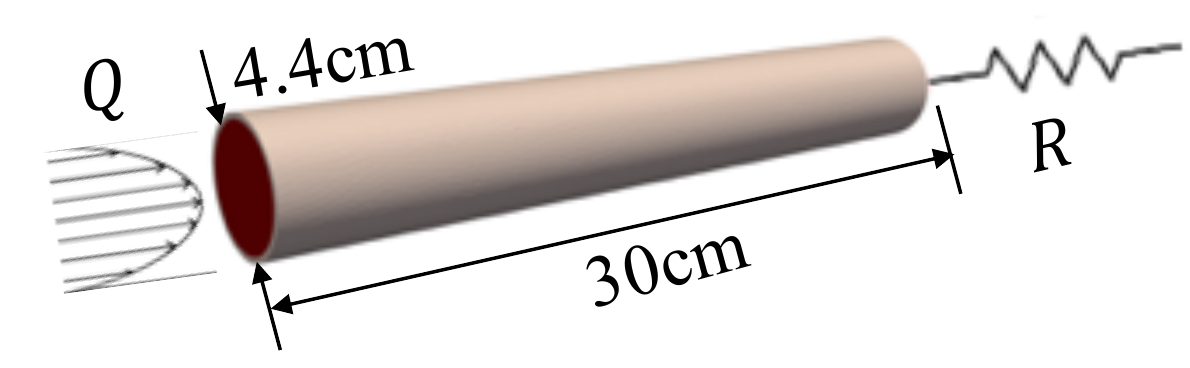}
\includegraphics[width=0.35\textwidth, keepaspectratio]{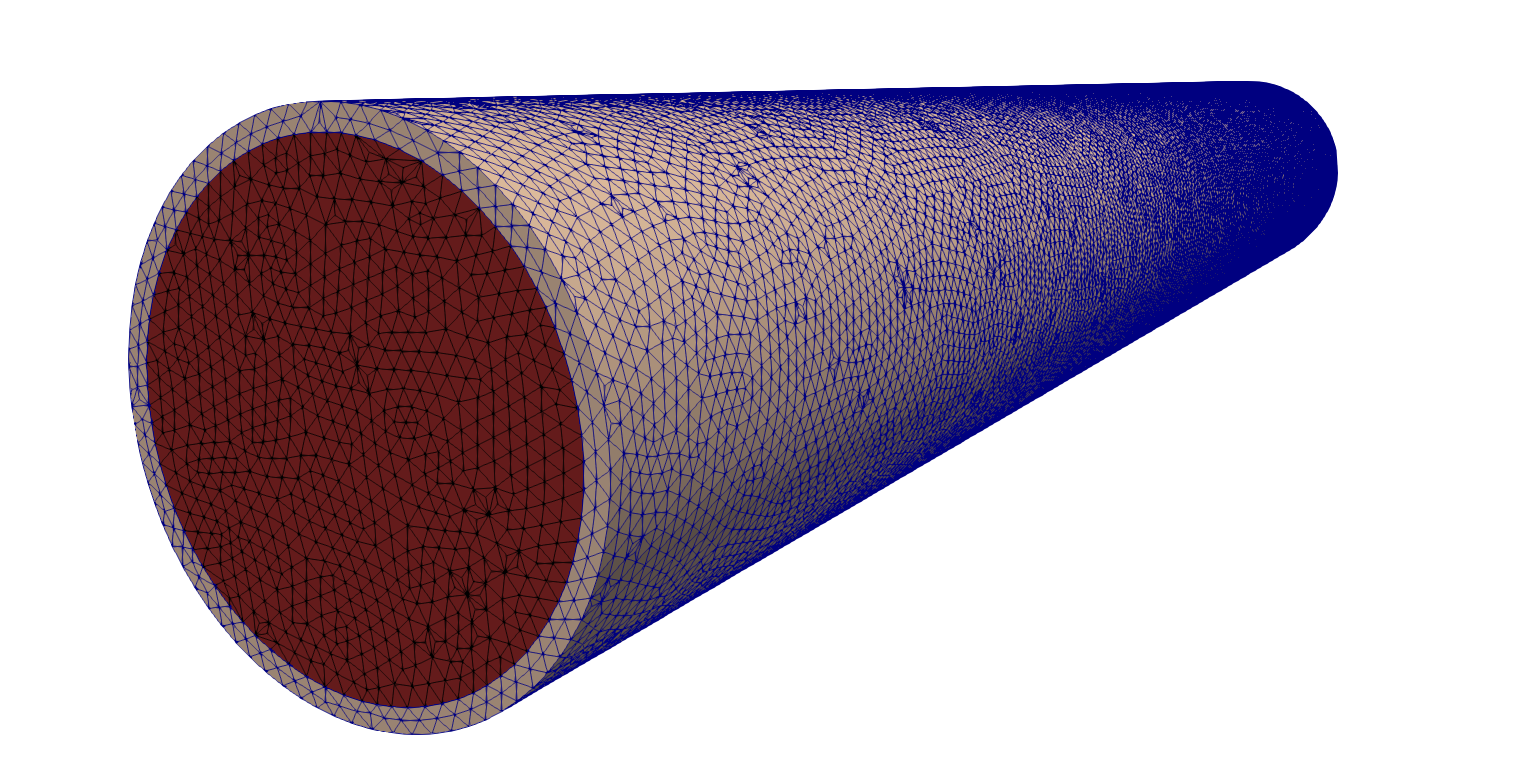}
\caption{Schematic representations and mesh for cylindrical pipe benchmark. (top) a rigid model with parabolic inflow and outlet resistance boundary condition. (bottom) an FSI model with same boundary conditions. For each model we show a magnified view of the tetrahedral finite element mesh. The fluid and solid domains are colored in red and gray, respectively. }\label{fig:cylinder}
\end{figure}

\noindent All tests discussed in this study are performed using the svFSI finite element solver from the SimVascular open source project, leveraging the message passing interface (MPI) library and optimized to run efficiently on large computational clusters~\cite{Esmaily2015a}. 
We note this implementation assigns fluid and solid elements to separate cores and assumes unique interface nodes, i.e., common nodes on the fluid and solid mesh are expected to match. An FSI mesh with matching interface nodes is generated through the freely-available Meshmixer software, with details reported in~\cite{Vedula2017}.

We consider a simple pipe benchmark whose size and boundary conditions are chosen to represent the ascending aorta of a healthy subject, having $4$ cm diameter, $30$ cm length and $0.2$ cm thickness, assuming a radius/thickness ratio of 10\%.
The inflow is steady with a parabolic velocity profile, having a mean flow rate of $Q=83$ mL/s ($5$ L/min). 
A resistance boundary condition is used at the outlet equal to $R=1600$ g/cm$^4$/s, which produces a mean pressure of approximately $\Delta P=100$ mmHg, typical of the systemic circulation of a healthy subject.
Simulations are performed with rigid and deformable walls (see Figure~\ref{fig:cylinder}), using 1,002,436 tetrahedral elements for the fluid domain and 192,668 tetrahedral elements for the wall, generated in SimVascular with the TetGen mesh generator plug-in~\cite{Updergrove2016,Hongzhi2017}.
We measure the wall clock time on the XSEDE Comet cluster for simulations consisting of 10 time steps of 1 millisecond each, using 38 and 48 cores for rigid and deformable simulations, respectively. The XSEDE Comet cluster has 1,944 compute nodes with the Intel Xeon E5-2680v3 cores, 24 cores/node, 2.5GHz clock speed, 960 GFlop/s flop speed, and 120GB/s memory bandwidth. For more information about the Comet cluster, please refer to the XSEDE user portal. We use the restarting scheme for GMRES. We set the restart number of 200 which showed superior performance against smaller restart numbers (See Appendix \ref{app:restart}). We set the ILUT drop tolerance to ${10^{-2}}$ and the fill-in level to 2. In our test, changing the drop tolerance to ${10^{-4}}$ and ${10^{-6}}$ did not significantly change the performance of the linear solver reported here. For the multigrid preconditioner we selected a maximum level of four, a Gauss-Seidel smoother, and the symmetric Gauss-Seidel for the subsmoother. We confirmed that this setting was superior to other choices for smoother and subsmoother (See Appendix \ref{app:ML}).
Finally, as the node ordering affects the amount of fill-in produced by an ILU decomposition, we apply Reverse-Cuthill-McKee (RCM) reordering prior to the incomplete factorization. RCM has been shown to be effective among many reordering schemes in the solution of non-symmetric sparse linear systems (see, e.g.,~\cite{Benzi1999}). From our testing, ILUT with RCM reordering provided superior performance against ILUT without reordering and ILUT with METIS reordering (See Appendix \ref{app:reordering}). 

\subsection{Rigid wall benchmark}\label{sec:Rigid}
%
\begin{table}[]
\centering
\begin{tabular}{ c c c c c  }
\hline
& $\epsilon=10^{-1}$& $\epsilon=10^{-3}$& $\epsilon=10^{-6}$& $\epsilon=10^{-9}$\\
\hline
$Err$ & 0.2235 & 0.0028 &1.618$\times10^{-7}$& 1.107$\times 10^{-9}$ \\
\hline
\end{tabular}
\caption{$l^2$-norm of velocity errors from different residual norm tolerances. The solution error is obtained from nodal velocity errors after 10 time steps, using a reference simulation with a tolerance up to the machine precision, $\epsilon=10^{-12}$.}
\label{table:error}
\end{table}

\begin{figure}[t]
\centering
{\includegraphics[width=.37\textwidth, keepaspectratio]{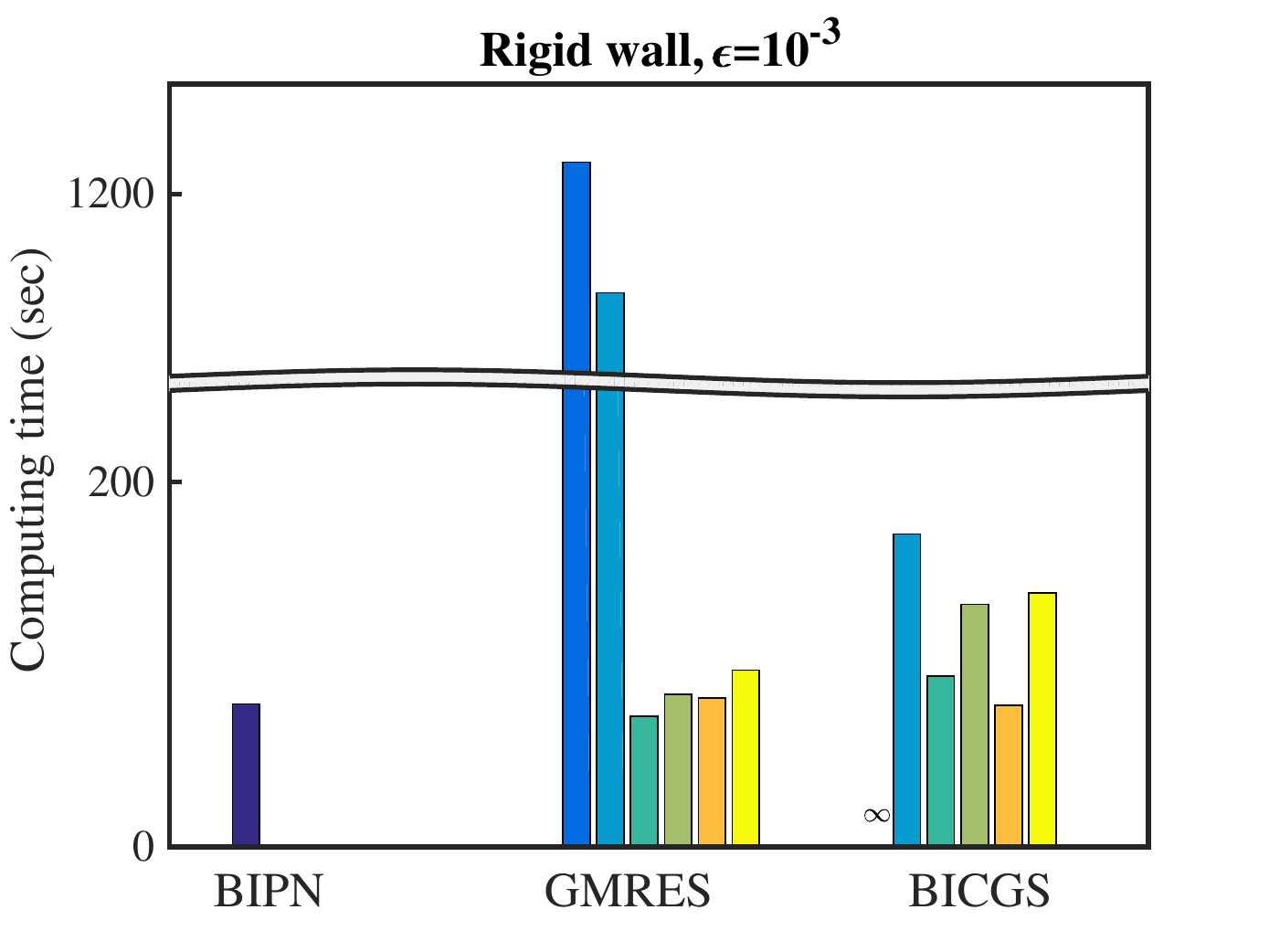}}
{\includegraphics[width=.37\textwidth, keepaspectratio]{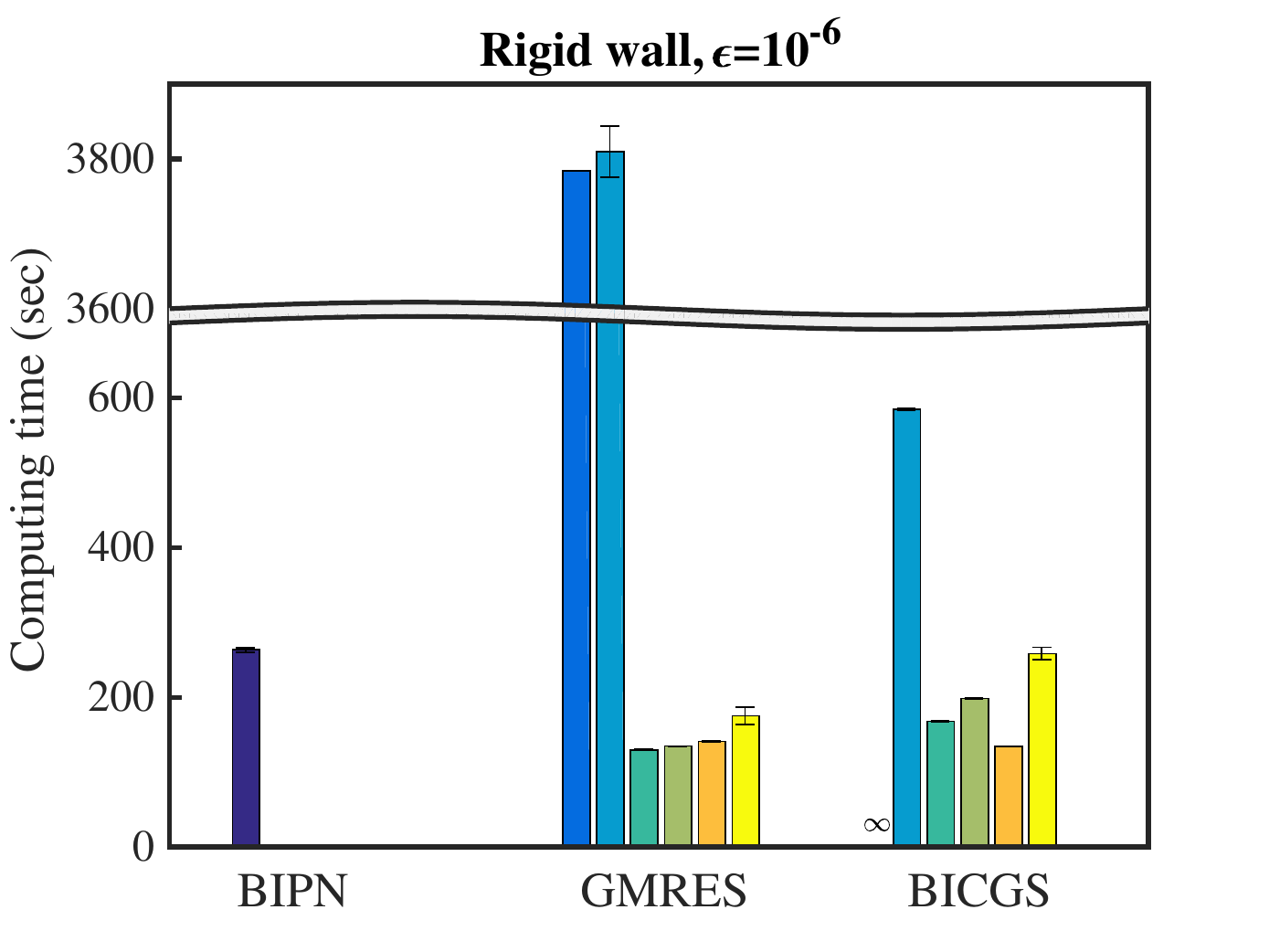}}
{\includegraphics[width=.35\textwidth, keepaspectratio]{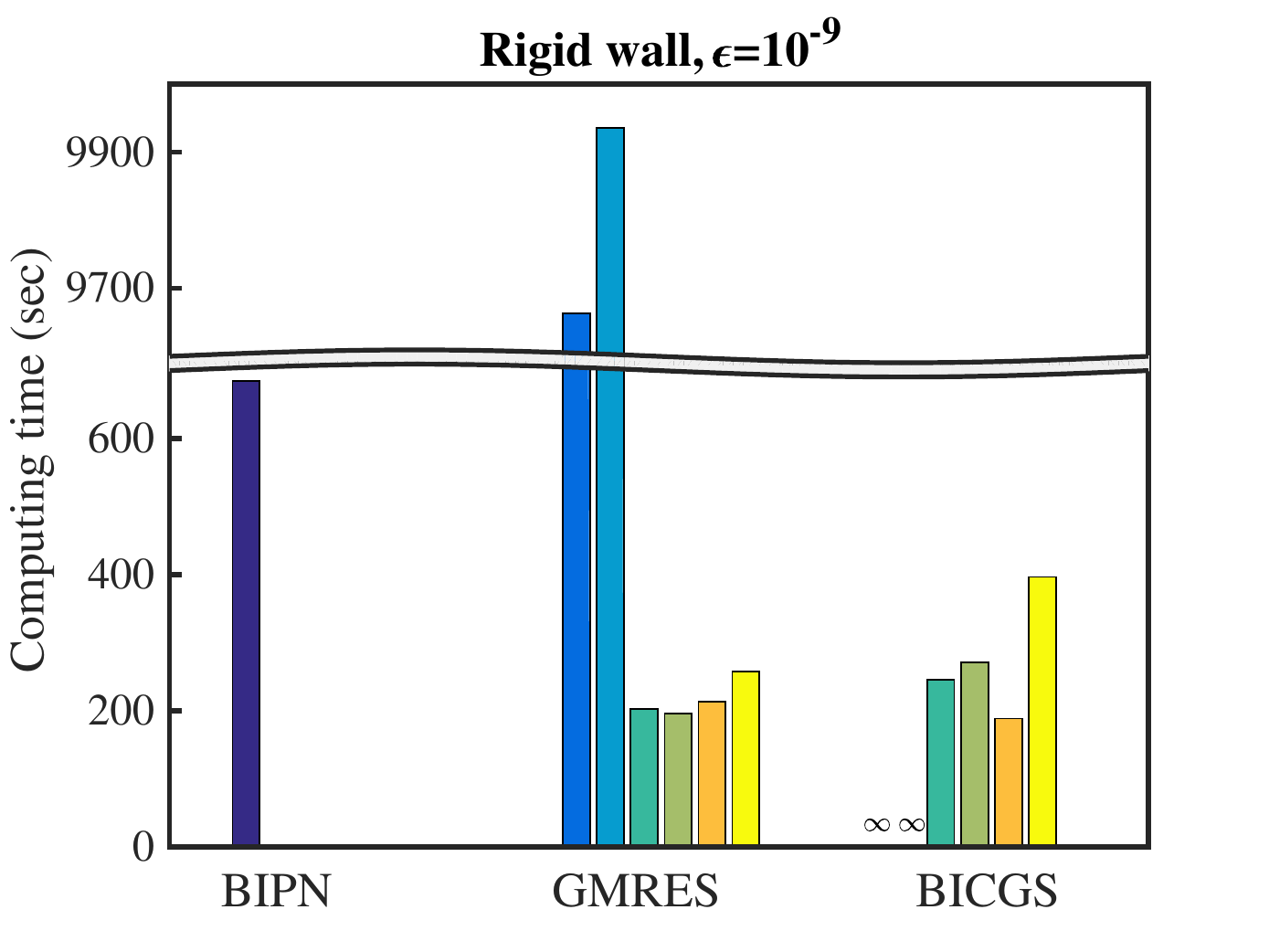}}
{\includegraphics[width=.075\textwidth, keepaspectratio]{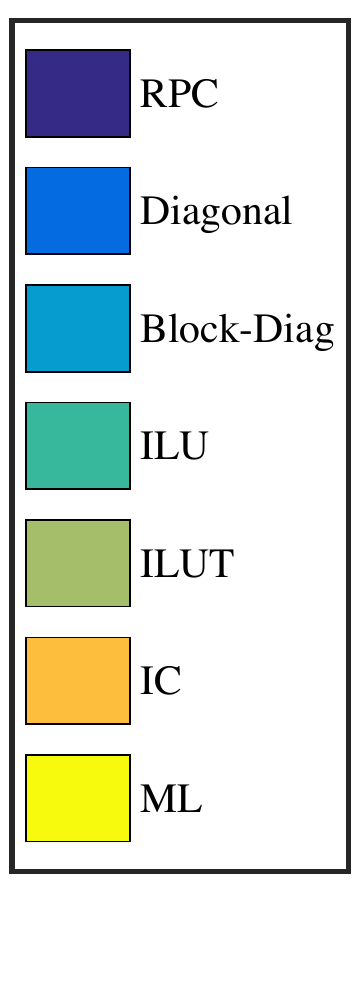}}
\caption{Compute times for linear solvers and preconditioners using a rigid pipe model with tolerances (top) $\epsilon=10^{-3}$, (center) $\epsilon=10^{-6}$, (bottom) $\epsilon=10^{-9}$. For $\epsilon=10^{-6}$, error bars are plotted by taking standard deviations from two repeated simulations. Differences between repeated simulations are caused by different computing nodes assigned by the scheduler on the Comet cluster.}
\label{fig:pipe_performance}
\end{figure}

\noindent  We plot the linear solver performance measured by wall clock time in Figure~\ref{fig:pipe_performance}. In Table \ref{table:2}, we report the number of iterations and the portion of total compute time consumed by solving the linear system. Three iterative solver tolerances, $\epsilon=10^{-3}, 10^{-6}, 10^{-9}$, are tested and compared. 
Table~\ref{table:error} shows the effect of tolerance on the velocities at $t=10$ ms, suggesting the velocity error norm is of the same order of the selected tolerance.

Figure~\ref{fig:pipe_performance} shows that incomplete factorization preconditioners are fast and exhibit robust performance across all tolerances, irrespective of the underlying iterative linear solver.
\begin{table}
\centering
\begin{tabular}{ |c |c |c  | c| c | c|}
 \hline
 \multicolumn{6}{|c|}{BIPN-RPC /$\epsilon=10^{-3}$} \\
 \hline
$\text{N}_\text{bipn}$ & $\text{t}_\text{LS}/\text{t}_\text{T}$& $\text{N}_\text{gmres}$  & $\text{t}_\text{gmres}/\text{t}_\text{LS}$&$\text{N}_\text{cg}$ & $\text{t}_\text{cg}/\text{t}_\text{LS}$\\
\hline
129& 89.5$\%$&1475& 5.3$\%$&38582& 77.6$\%$\\
\hline
\end{tabular}

\begin{tabular}{ |c |c |c  | c| c | c|}
 \hline
 \multicolumn{6}{|c|}{BIPN-RPC /$\epsilon=10^{-6}$} \\
 \hline
$\text{N}_\text{bipn}$ & $\text{t}_\text{LS}/\text{t}_\text{T}$& $\text{N}_\text{gmres}$  & $\text{t}_\text{gmres}/\text{t}_\text{LS}$&$\text{N}_\text{cg}$ & $\text{t}_\text{cg}/\text{t}_\text{LS}$\\
\hline
847& 95.1$\%$&22699& 17.25$\%$&91470& 55.7$\%$\\
\hline
\end{tabular}

\begin{tabular}{ |c |c |c  | c| c | c|}
 \hline
 \multicolumn{6}{|c|}{BIPN-RPC /$\epsilon=10^{-9}$} \\
 \hline
$\text{N}_\text{bipn}$ & $\text{t}_\text{LS}/\text{t}_\text{T}$& $\text{N}_\text{gmres}$  & $\text{t}_\text{gmres}/\text{t}_\text{LS}$&$\text{N}_\text{cg}$ & $\text{t}_\text{cg}/\text{t}_\text{LS}$\\
\hline
1923& 95.0$\%$&247142& 43.2$\%$&791817& 49.4$\%$\\
\hline
\end{tabular}

\begin{tabular}{| c ||c |c  | c| c |}
 \hline
$\epsilon=10^{-3}$ &\multicolumn{2}{c|}{GMRES} &\multicolumn{2}{c|}{BICG} \\
\hline
PC & $\text{N}_\text{gmres}$ & $\text{t}_\text{LS}/\text{t}_\text{T}$ &  $\text{N}_\text{bicgs}$ & $\text{t}_\text{LS}/\text{t}_\text{T}$\\
\hline
Diag  & 155544 & 98.8$\%$ & N/A & N/A\\
\hline
Block-D  & 147768 & 98.7$\%$ & 30027 & 94.4$\%$\\
\hline
ILU  & 2768 & 85.3$\%$ & 3104 & 89.6$\%$\\
\hline
ILUT  & 1493 & 88.0$\%$ & 2360 & 92.1$\%$ \\
\hline
IC  & 3636 & 40.5$\%$ & 3290 & 41.3$\%$ \\
\hline
ML  & 1080 & 51.9$\%$ & 1054 & 66.7$\%$\\
\hline
\end{tabular}

\begin{tabular}{ |c ||c |c  | c| c |}
\hline
$\epsilon=10^{-6}$ &\multicolumn{2}{c|}{GMRES} &\multicolumn{2}{c|}{BICGS} \\
\hline
PC & $\text{N}_\text{gmres}$ & $\text{t}_\text{LS}/\text{t}_\text{T}$ &  $\text{N}_\text{bicgs}$ & $\text{t}_\text{LS}/\text{t}_\text{T}$\\
\hline
Diag  & 492752 & 99.9$\%$ & N/A & N/A\\
\hline
Block-D  & 491185 & 99.9$\%$ & 110237 & 96.3$\%$\\
\hline
ILU  & 6632 & 89.6$\%$ & 6770 & 91.7$\%$\\
\hline
ILUT  & 3382 & 89.8$\%$ & 4045 & 92.9$\%$ \\
\hline
IC  & 8191 & 54.3$\%$ & 7710 & 51.3$\%$ \\
\hline
ML  & 2526 & 64.6$\%$ & 2376 & 76.6$\%$\\
\hline
\end{tabular}

\begin{tabular}{ |c ||c |c  | c| c |}
\hline
$\epsilon=10^{-9}$ &\multicolumn{2}{c|}{GMRES} &\multicolumn{2}{c|}{BICGS} \\
\hline
PC & $\text{N}_\text{gmres}$ & $\text{t}_\text{LS}/\text{t}_\text{T}$ &  $\text{N}_\text{bicgs}$ & $\text{t}_\text{LS}/\text{t}_\text{T}$\\
\hline
Diag  & 1252448 & 100$\%$ & N/A & N/A\\
\hline
Block-D  & 1251025 & 100$\%$ & N/A & N/A\\
\hline
ILU  & 10066 & 91.0$\%$ & 9766 & 92.2$\%$\\
\hline
ILUT  & 5132 & 90.6$\%$ & 5468 & 93.0$\%$ \\
\hline
IC  & 13728 & 59.8$\%$ & 11022 & 53.4$\%$ \\
\hline
ML  & 3992 & 69.6$\%$ & 3711 & 80.4$\%$\\
\hline
\end{tabular}
\caption{The number of iterations and the portion of compute time consumed by the linear solver for the pipe benchmark problem with rigid walls. The number of linear solver iterations is counted for 10 time step calculations. t$_\text{T}$ is the total compute time, t$_\text{LS}$ is the compute time consumed by solving the linear system. }
\label{table:2}
\end{table}
Despite similar performance for ILU, ILUT and IC preconditioners across all cases, the slightly worse performance of ILUT with respect to ILU seems to suggest that constructing a more accurate factor may lead to smaller run times than the savings in the factorization cost acheived by dropping additional fill-ins.
GMRES with diagonal preconditioners (either diagonal or block-diagonal) appears to be significantly slower than other schemes, particularly as the tolerance $\epsilon$ becomes smaller.
This degrading performance of standard GMRES for cardiovascular modeling is consistent with previous studies, showing that resistance boundary conditions are responsible for an increase in the condition number~\cite{Esmaily2013a,Esmaily2015}.
While BICGS seems to perform better than GMRES with diagonal preconditioners, its 
performance becomes increasingly unstable with smaller tolerance $\epsilon$. 
Furthermore, while algebraic multigrid preconditioners appear to be superior to diagonal preconditioning, they are inferior to BIPN or GMRES/BICGS with ILU.
Finally, the performance of BIPN with RPC preconditioning is comparable to ILU for small tolerances (i.e., $\epsilon=10^{-3}$) but significantly degrades as the tolerance value decreases.

{\color{black} From Table \ref{table:2}, we see that the time consumed by the linear system solvers constitutes the majority of the total compute time.} In BIPN, the compute time of GMRES is significantly smaller than the compute time of CG. As the tolerance value decreases, the relative percentage of compute time for GMRES solve becomes larger. All Trilinos preconditioners show larger iteration numbers with decreasing tolerance. The relatively small percentage of linear solver compute time against the total compute time with IC and ML implies that building such preconditioners is expensive. This suggests that storing and reusing a preconditioner for several time steps could increase efficiency. 

\subsection{Deformable wall benchmark (FSI)}\label{sec:FSI}
%
\begin{figure}
\centering
{\includegraphics[width=.37\textwidth, keepaspectratio]{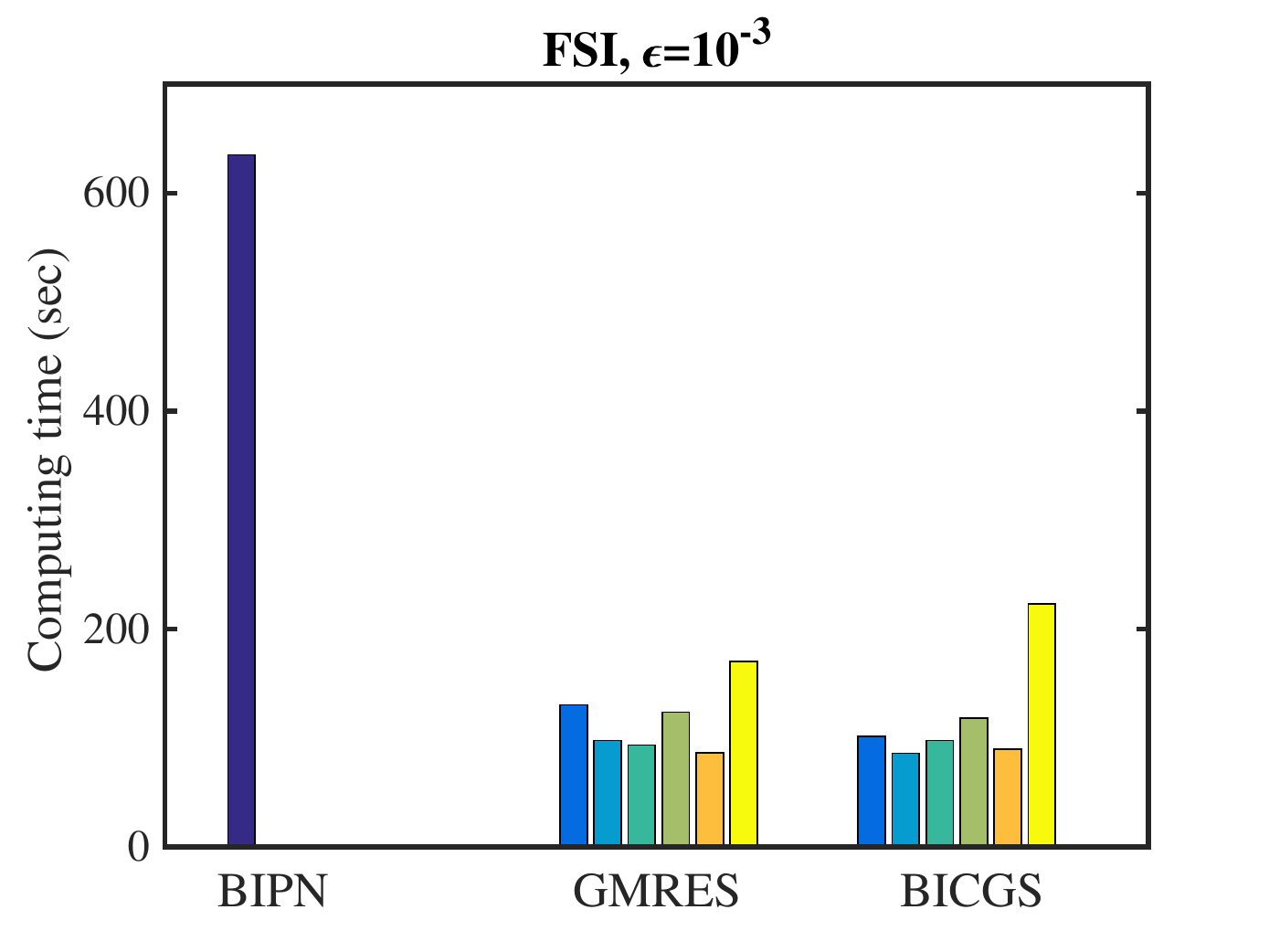}}
{\includegraphics[width=.37\textwidth, keepaspectratio]{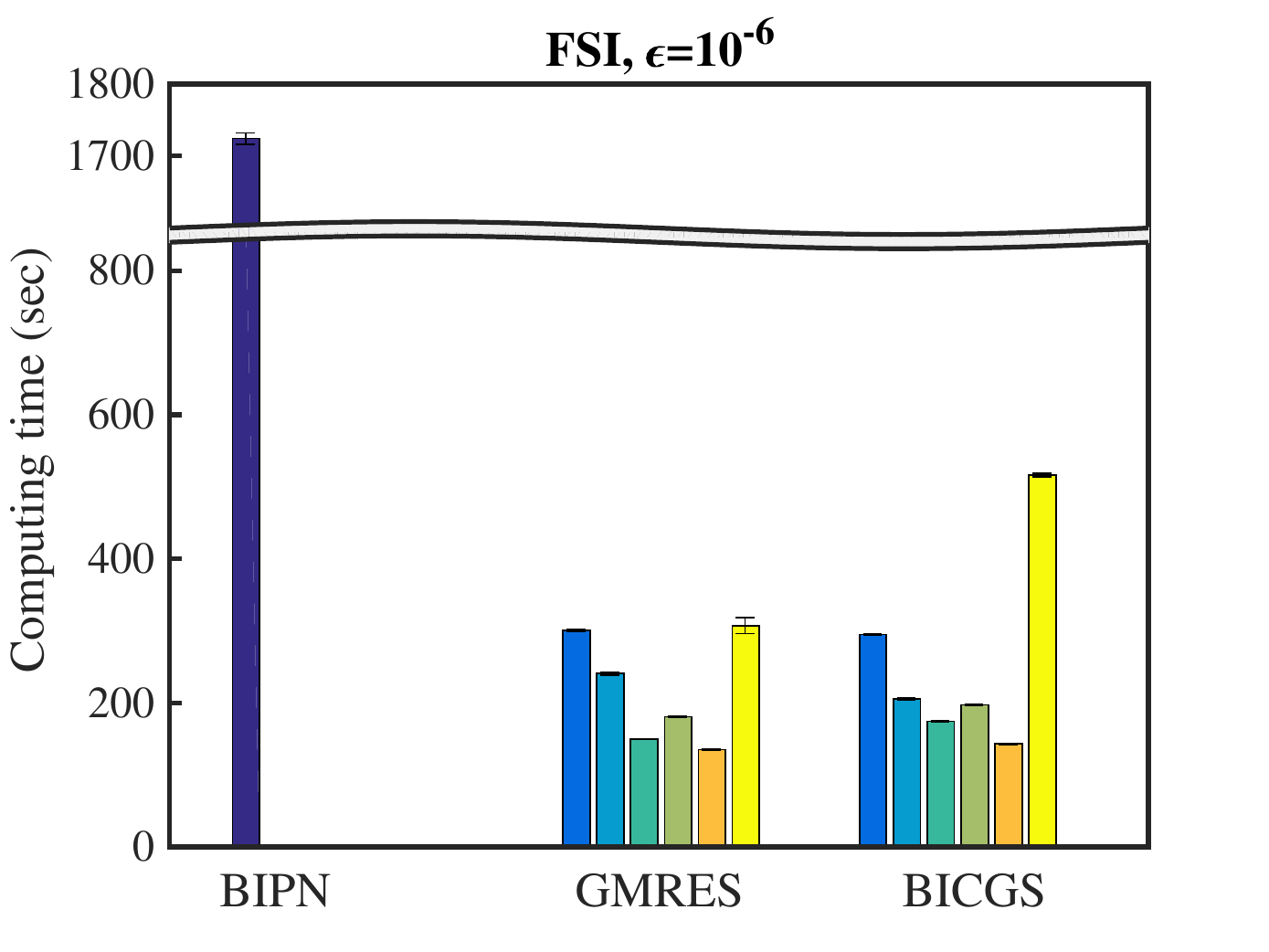}}
{\includegraphics[width=.35\textwidth, keepaspectratio]{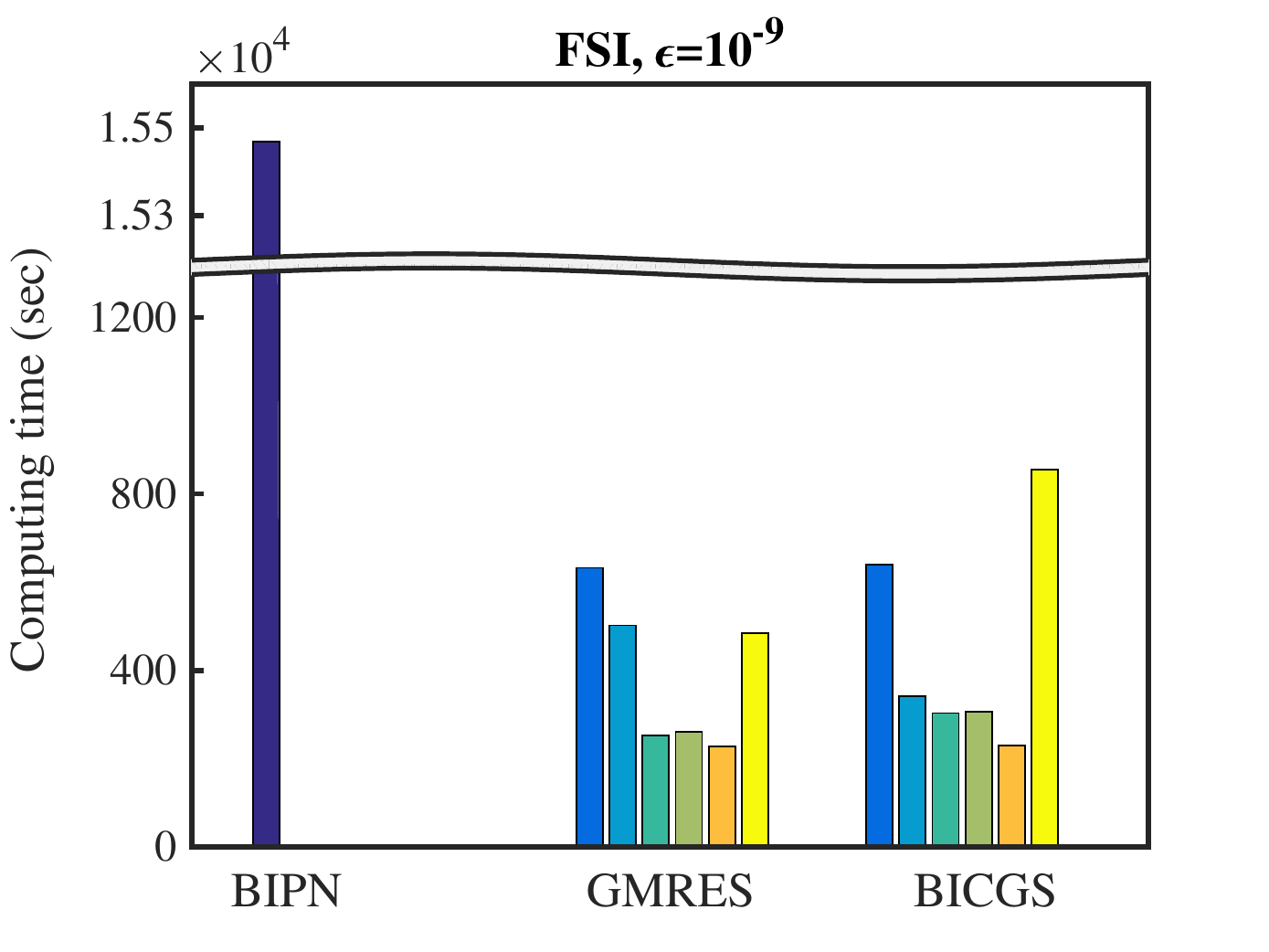}}
{\includegraphics[width=.075\textwidth, keepaspectratio]{legend.pdf}}
\caption{Compute times for linear solvers and preconditioners using a deformable wall pipe model (FSI) with tolerances (top) $\epsilon=10^{-3}$, (middle) $\epsilon=10^{-6}$, (bottom) $\epsilon=10^{-9}$. For $\epsilon=10^{-6}$, error bars are plotted by taking standard deviations from two repeated simulations.}
\label{fig:pipe_performance_fsi}
\end{figure} 

\noindent We illustrate the compute times for the deformable wall case in Figure~\ref{fig:pipe_performance_fsi} and summarize the number of iterations and percentage compute time of the linear solvers in Table \ref{table:3}. 
For FSI simulations with a tolerance $\epsilon=10^{-3}$, BIPN with RPC shows more than an 8 fold increase in compute time compared to the rigid wall case, which also suggests the limitations of this approach for the deformable wall case. The increase of iteration number for GMRES as tolerance values decrease is notably higher than in the rigid wall case, and the percentage of compute time for the GMRES solve in BIPN is significantly higher, about 80$\%$ of the total compute time while the CG part continues to make up a smaller percentage. This suggests directions for future improvement of BIPN in the GMRES part for FSI simulations. 
Conversely, incomplete factorization preconditioners (both for GMRES and BICGS), exhibit good performance across all tolerances and a limited increase in compute time compared to the rigid case.
Diagonal preconditioners show a comparable performance to incomplete factorization schemes at large tolerances, but the performance degrades for smaller tolerances.
Among all algorithms implemented in Trilinos, the algebraic multigrid preconditioners is the slowest, while BICG with ILU appears to be the best solution scheme overall. 

\begin{table}
\centering
\begin{tabular}{ |c |c |c  | c| c | c|}
 \hline
 \multicolumn{6}{|c|}{BIPN-RPC /$\epsilon=10^{-3}$} \\
 \hline
$\text{N}_\text{bipn}$ & $\text{t}_\text{LS}/\text{t}_\text{T}$& $\text{N}_\text{gmres}$  & $\text{t}_\text{gmres}/\text{t}_\text{LS}$&$\text{N}_\text{cg}$ & $\text{t}_\text{cg}/\text{t}_\text{LS}$\\
\hline
608& 78.6$\%$&117377& 77$\%$&64865& 15.6$\%$\\
\hline
\end{tabular}

\begin{tabular}{ |c |c |c  | c| c | c|}
 \hline
 \multicolumn{6}{|c|}{BIPN-RPC /$\epsilon=10^{-6}$} \\
 \hline
$\text{N}_\text{bipn}$ & $\text{t}_\text{LS}/\text{t}_\text{T}$& $\text{N}_\text{gmres}$  & $\text{t}_\text{gmres}/\text{t}_\text{LS}$&$\text{N}_\text{cg}$ & $\text{t}_\text{cg}/\text{t}_\text{LS}$\\
\hline
1580& 83.9$\%$&318272& 79.4$\%$&145183& 13.4$\%$\\
\hline
\end{tabular}

\begin{tabular}{ |c |c |c  | c| c | c|}
 \hline
 \multicolumn{6}{|c|}{BIPN-RPC /$\epsilon=10^{-9}$} \\
 \hline
$\text{N}_\text{bipn}$ & $\text{t}_\text{LS}/\text{t}_\text{T}$& $\text{N}_\text{gmres}$  & $\text{t}_\text{gmres}/\text{t}_\text{LS}$&$\text{N}_\text{cg}$ & $\text{t}_\text{cg}/\text{t}_\text{LS}$\\
\hline
14872& 83.3$\%$&4087258& 79.6$\%$&874018& 11.0$\%$\\
\hline
\end{tabular}

\begin{tabular}{ |c ||c |c  | c| c |}
 \hline
$\epsilon=10^{-3}$ &\multicolumn{2}{c|}{GMRES} &\multicolumn{2}{c|}{BICGS} \\
\hline
PC & $\text{N}_\text{gmres}$ & $\text{t}_\text{LS}/\text{t}_\text{T}$ &  $\text{N}_\text{bicgs}$ & $\text{t}_\text{LS}/\text{t}_\text{T}$\\
\hline
Diag  & 11283 & 71.9$\%$ & 12094 & 66.3$\%$\\
\hline
Block-D  & 9127 & 65.1$\%$ & 9438 & 59.9$\%$\\
\hline
ILU  & 3900 & 61.2$\%$ & 3655 & 64.2$\%$\\
\hline
ILUT  & 3493 & 70.6$\%$ & 3333 & 70.9$\%$ \\
\hline
IC  & 4089 & 17.2$\%$ & 3873 & 20.02$\%$ \\
\hline
ML  & 3735 & 31.74$\%$ & 3799 & 44.43$\%$\\
\hline
\end{tabular}

\begin{tabular}{ |c ||c |c  | c| c |}
\hline
$\epsilon=10^{-6}$ &\multicolumn{2}{c|}{GMRES} &\multicolumn{2}{c|}{BICGS} \\
\hline
PC & $\text{N}_\text{gmres}$ & $\text{t}_\text{LS}/\text{t}_\text{T}$ &  $\text{N}_\text{bicgs}$ & $\text{t}_\text{LS}/\text{t}_\text{T}$\\
\hline
Diag  & 33455 & 86.2$\%$ & 45623 & 85.3$\%$\\
\hline
Block-D  & 26311 & 81.7$\%$ & 29589 & 77.3$\%$\\
\hline
ILU  & 7699 & 70.1$\%$ & 7338 & 74.9$\%$\\
\hline
ILUT  & 5904 & 75.1$\%$ & 5537 & 77.1$\%$ \\
\hline
IC  & 8241 & 28.7$\%$ & 7819 & 32.0$\%$ \\
\hline
ML  & 7109 & 49.9$\%$ & 7610 & 63.9$\%$\\
\hline
\end{tabular}

\begin{tabular}{ |c ||c |c  | c| c |}
\hline
$\epsilon=10^{-9}$ &\multicolumn{2}{c|}{GMRES} &\multicolumn{2}{c|}{BICGS} \\
\hline
PC & $\text{N}_\text{gmres}$ & $\text{t}_\text{LS}/\text{t}_\text{T}$ &  $\text{N}_\text{bicgs}$ & $\text{t}_\text{LS}/\text{t}_\text{T}$\\
\hline
Diag  & 71884 &90.25$\%$ & 100431 & 90.6$\%$\\
\hline
Block-D  & 55392 & 88.2$\%$ & 49695 & 82.1$\%$\\
\hline
ILU  & 13801 & 75.1$\%$ & 13093 & 80.0$\%$\\
\hline
ILUT  & 9848 & 76.7$\%$ & 9404 & 79.3$\%$ \\
\hline
IC  & 14615 & 42.2$\%$ & 13858 & 43.4$\%$ \\
\hline
ML  & 12178 & 59.8$\%$ & 9587 & 78.4$\%$\\
\hline
\end{tabular}
\caption{The number of iterations and the portion of compute time consumed by the linear solver for the pipe benchmark problem with deformable walls. The number of linear solver iterations is counted for 10 time step calculations. t$_\text{T}$ is the total compute time, t$_\text{LS}$ is the compute time consumed by solving the linear system. }
\label{table:3}
\end{table}

\section{Parallel scalability}\label{sec:Scaling}
%
\noindent Parallel scalability is investigated for two preconditioned linear solvers, BIPN-RPC and BICG-ILU, in terms of speedup (see, e.g.,~\cite{Chen2006}), defined as the computing speed of multiple cores compared to a {\color{black}single core} calculation, i.e., $S_p=T_1/T_p$, where $T_p$ is the compute time on $N_p$ cores. The ideal strong scalability performance between $S_p$ and $N_p$ is linear, however, sublinear scaling is expected due to communication cost. 

\subsection{Strong scaling}\label{sec:Strong}
%
\begin{figure}
\centering
\includegraphics[width=0.33\textwidth, keepaspectratio]{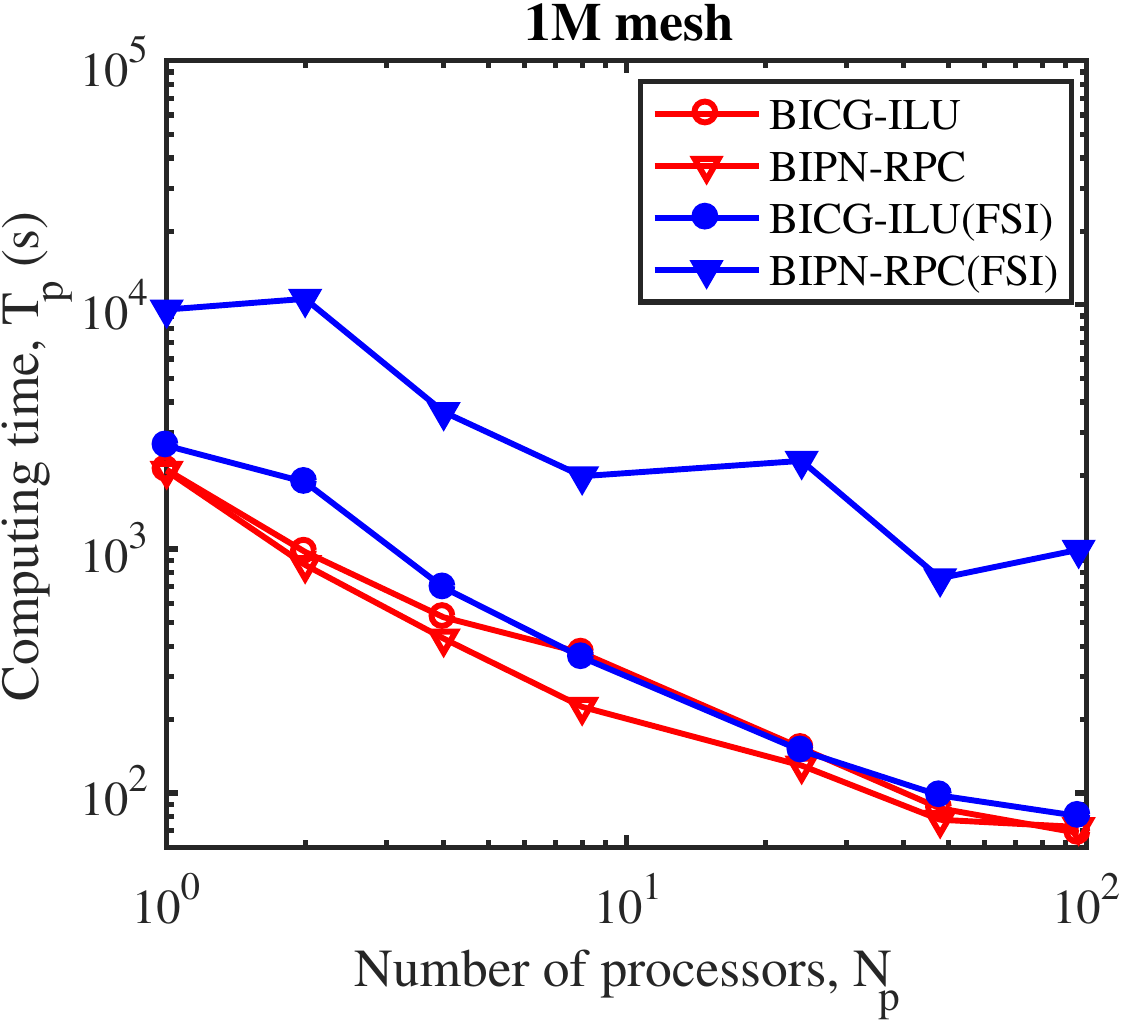}
\includegraphics[width=0.33\textwidth, keepaspectratio]{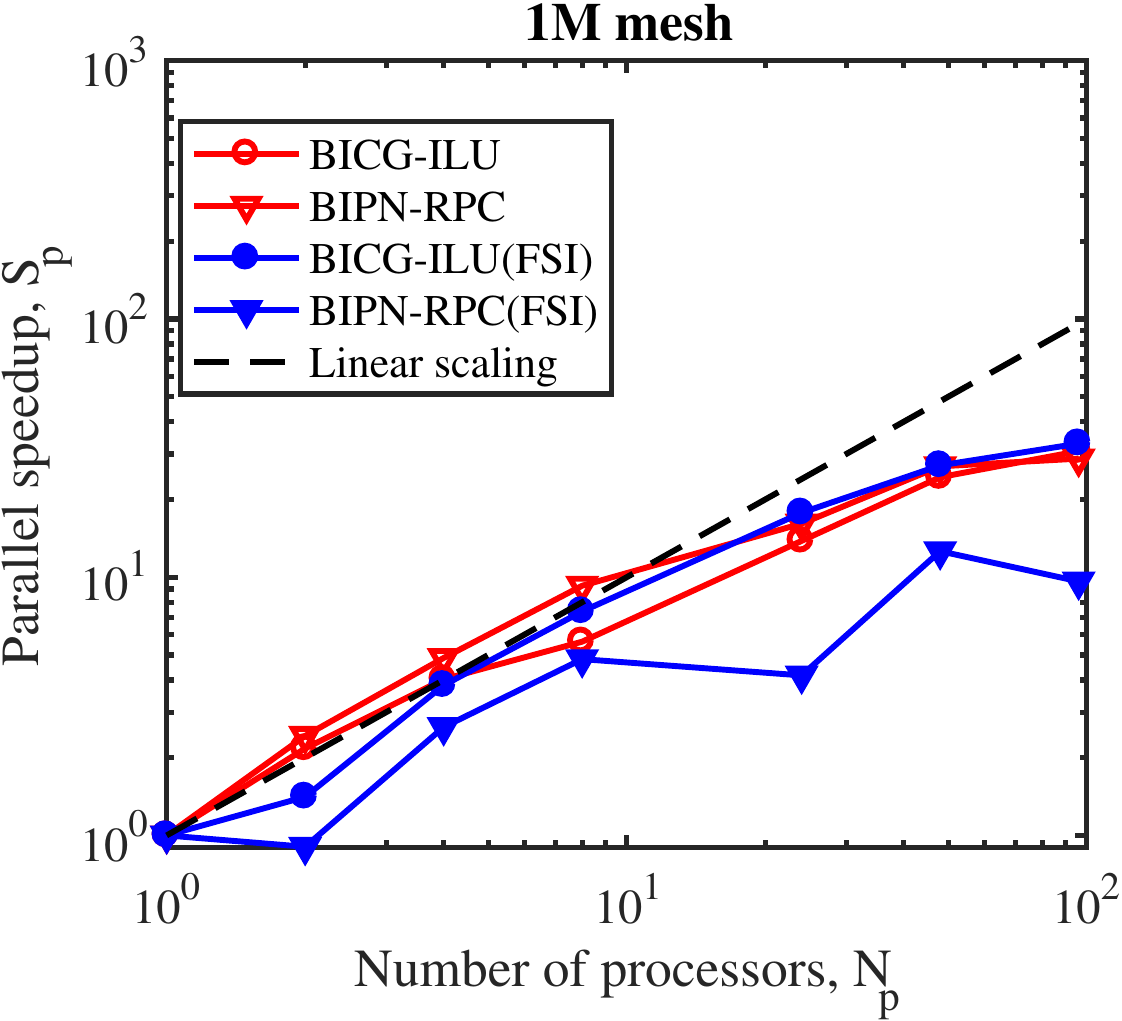}
\caption{Strong scaling of BIPN-RPC and BICG-ILU for pipe benchmark with rigid and deformable walls on the 1M lumen mesh model (Mesh2). (top) compute time, $T_p$, versus number of cores, $N_p$, (bottom) speedup, $S_p=T_1/T_p$, versus $N_p$.}
\label{fig:strong}
\end{figure} 

\begin{figure}
\centering
\includegraphics[width=0.33\textwidth, keepaspectratio]{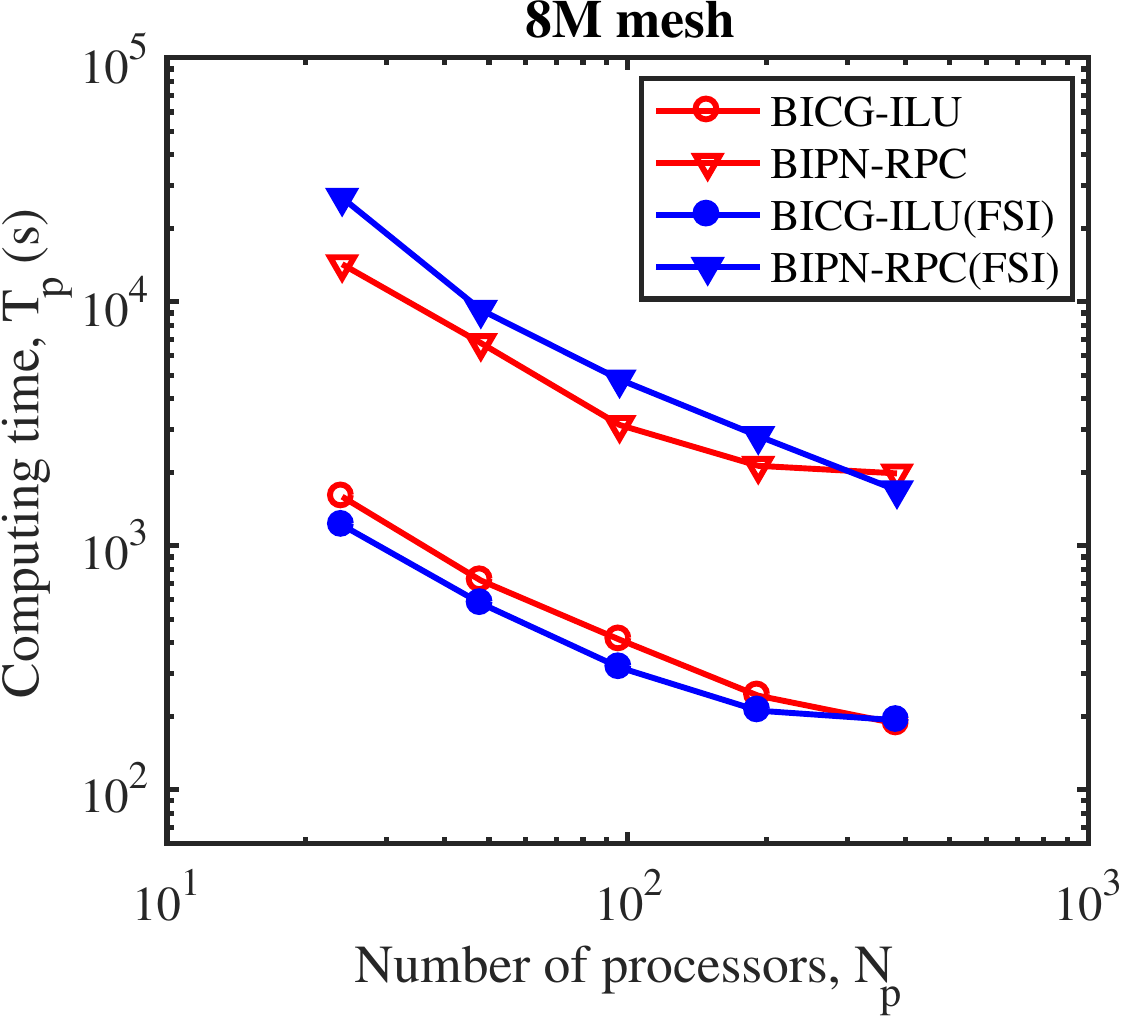}
\includegraphics[width=0.33\textwidth, keepaspectratio]{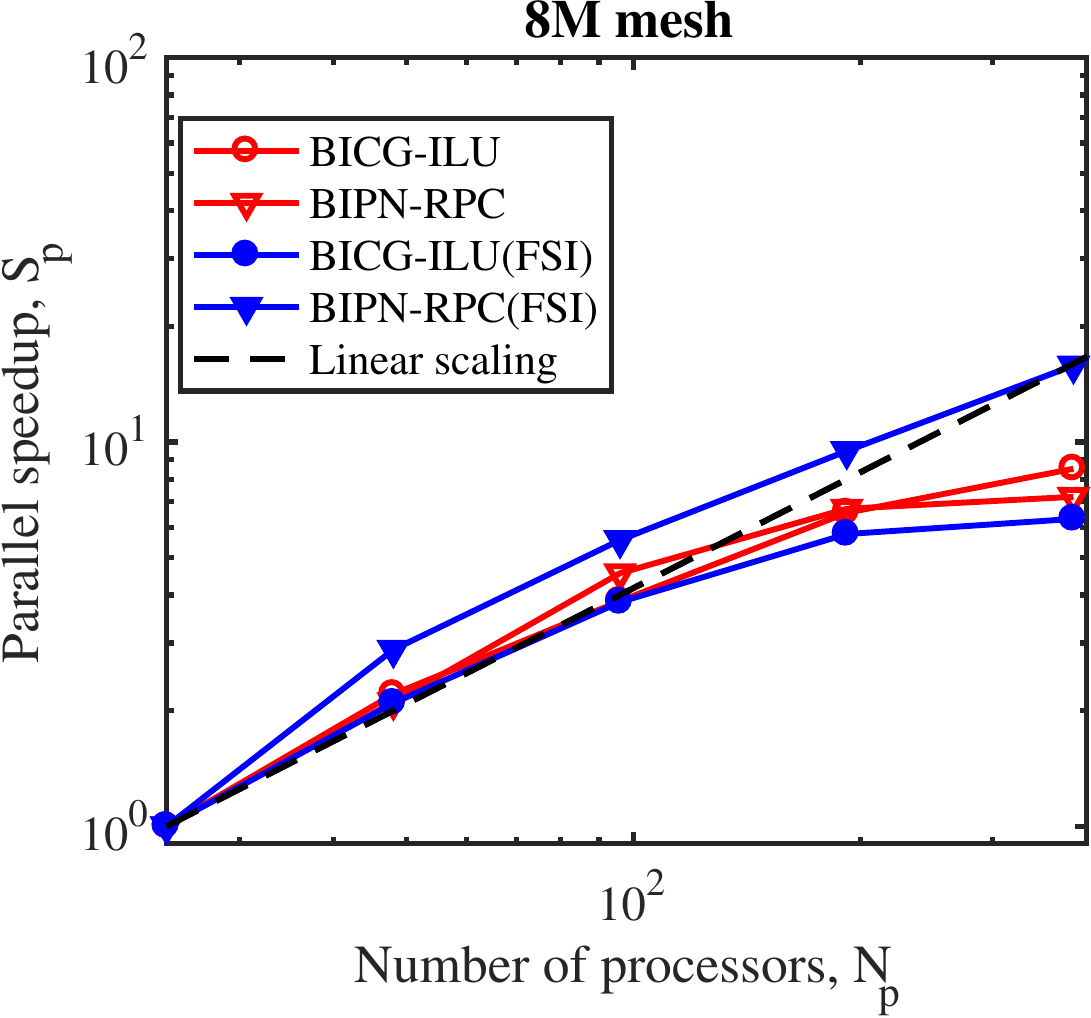}
\caption{Strong scaling of BIPN-RPC and BICG-ILU for pipe benchmark on the 8M lumen mesh model (Mesh5). (top) $T_p$ versus $N_p$, (bottom) $S_p=T_{24}/T_p$ versus $N_p$.}
\label{fig:strong2}
\end{figure} 

\noindent In this section, we monitor the compute time for a model with a fixed number of degrees of freedom, while progressively  increasing the number of cores. We first test the strong scalability by varying the number of cores, testing 1, 2, 4, 8, 24, 48, and 96 cores for an $\approx$1 million element mesh (Mesh1 in Table~\ref{table:weakScaling}) (Figure~\ref{fig:strong}). We chose the number of cores as multiples of 24, to use all cores in any given node. {\color{black}We note, however, that one should use only between 2/3 and 3/4 of the total number of cores in a given node since the local memory bandwidth is often a bottleneck resulting in higher overhead and reduced speed improvement.} In the rigid wall model, BIPN-RPC and BICG-ILU show similar performance across all core numbers as shown in Figure~\ref{fig:pipe_performance}. The parallel speedup shows that both methods scale well up to 24 cores, leading to about 40,000 elements per core, while their parallel performance is reduced when running on more than 48 cores. In Figure ~\ref{fig:pipe_performance}, the zig-zag behavior of BIPN-RPC (FSI) reveals excessive inter-core communications and memory references. In the FSI problem, the total compute time of BIPN-RPC is larger, and the speed up is worse than BICG-ILU. BICG-ILU shows consistently good scalability for both rigid and FSI models whereas the scalability of BIPN-RPC degrades significantly in the FSI case. 

We conduct an additional scaling study on a refined mesh with 8M elements (Mesh5 in Table~\ref{table:weakScaling}). We use 24, 48, 96, 192, and 384 cores and we use the $N_p=24$ case as a reference for $S_p$. As shown in Figure \ref{fig:strong2}, the BICG-ILU speedup scales almost linearly up to 192 cores, i.e., $\sim$40,000 elements per core. In the 8M mesh, the compute time of BIPN-RPC becomes significantly larger due to a poor weak scalability, as discussed in the next section. 

\subsection{Weak scaling}\label{sec:Weak}
%
\noindent In this section, we solve models of increasing size and report the simulation time by keeping approximately the same number of elements per core. The number of nodes, elements and non-zeros in the coefficient matrix is summarized for each mesh in Table~\ref{table:weakScaling}.
We use 20, 40, 80, 160, 320 cores for meshes 1 to 5, resulting in 27,000 elements and about 4,500 {\color{black}nodes per a core}. 

\begin{table}
\centering
\begin{tabular}{ c c c c  }
\hline
&\multicolumn{3}{c}{Fluid}\\
\cline{2-4}  
&{$\#$ Nodes}&{$\#$ Tetrahedra}& $\#$ Non-zeros \\
\hline
Mesh1 & 93,944 & 551,819 &1,496,375  \\
Mesh2 & 166,899 & 1,002,436 & 2,718,385\\
Mesh3 & 349,469 & 2,131,986 & 5,784,902\\
Mesh4 & 716,298 & 4,415,305 & 11,990,977 \\
Mesh5 & 1,314,307 & 8,117,892 & 22,144,741 \\
\end{tabular}

\begin{tabular}{ c c c c  c c }
\hline
& \multicolumn{3}{c}{Structure} & Total\\
\cline{2-3}   \cline{5-5}  
&{$\#$ Nodes}&{$\#$ Tetrahedra}&&{$\#$ Non-zeros}\\
\hline
Mesh1 & 31,223 & 96,147&&1,834,624  \\
Mesh2 & 50,909 & 192,668 &&3,331,744\\
Mesh3 & 100,759 & 412,408 && 7,028,673  \\
Mesh4 & 97,206 & 893,452 && 14,565,414 \\
Mesh5 & 378,089 & 1,752,270 && 27,188,079  \\
\hline
\end{tabular}

\caption{Number of nodes, elements and non-zero entries in tangent stiffness matrix for selected meshes in scaling studies.}
\label{table:weakScaling}
\end{table}

\begin{figure}
\centering
\includegraphics[width=0.32\textwidth, keepaspectratio]{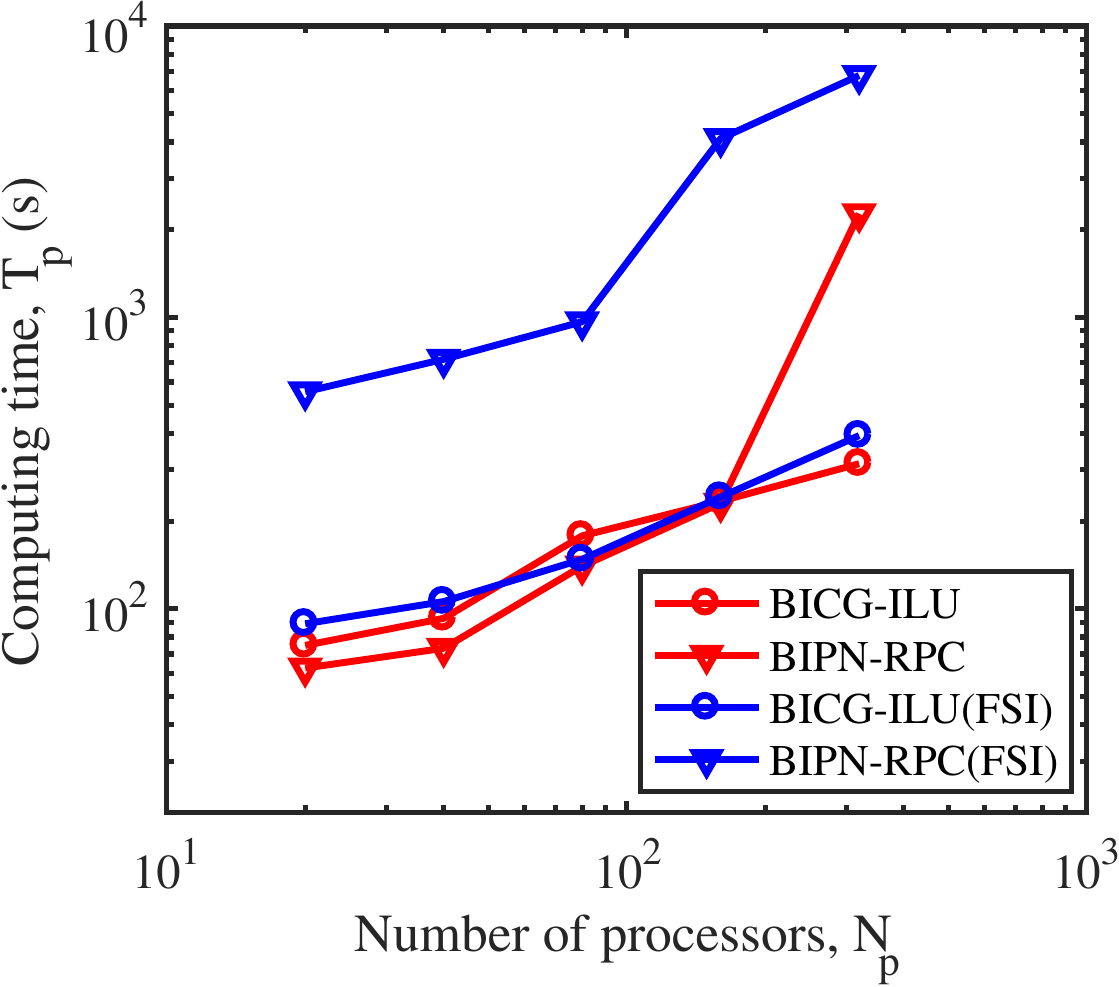}
\caption{Weak scaling of BICG-ILU and BIPN-RPC for pipe benchmark with rigid and deformable walls.}\label{fig:weak}
\end{figure}
BICG-ILU shows increased compute time but does not show any significant changes in scaling when increasing the number of cores. For the rigid model, BIPN-RPC shows poor scalability as the number of cores exceeds 160. This explains the loss of superior performance of BIPN-RPC against BICG-ILU in the strong scalability study with 8M mesh. Similar to the rigid case, BIPN-RPC shows performance loss after 160 cores for the FSI model. 

\section{Characteristics of unpreconditioned matrices}\label{sec:CharMat}
%
\noindent In this section, we investigate the properties of the coefficient matrices discussed in section~\ref{sec:LSf} to better understand the performance results.
With reference to the pipe benchmark, we visualize the matrix sparsity pattern and investigate its properties including bandwidth, symmetry, diagonal dominance and spectrum.
We also investigate the characteristics of both {\it global} and {\it local} matrices. The {\it global} matrix contains all mesh nodes from all cores, while a {\it local} matrix contains only a subset of the nodes in the {\it global} matrix assigned to a single core upon partitioning.
We report single-core, {\it local} matrix characteristics with $\text{N}_{lnd}\sim 5000$ nodes to represent the typical case of distributed discretizations consisting of $\sim$ 25000 tetrahedral elements per core.
This way we focus on detailed {\it local} information in a region of specific interest (e.g. resistance boundary), and also calculate properties of the matrix such as eigenvalues and condition numbers in a cost-effective way. 
%
We discuss properties for two groups of coefficient matrices, i.e., matrices associated with fluid flow and matrices associated with solid mechanics.

\subsection{Matrix properties for fluid flows in rigid vessels}
\begin{figure}
\centering
{\includegraphics[width=0.47\textwidth, keepaspectratio]{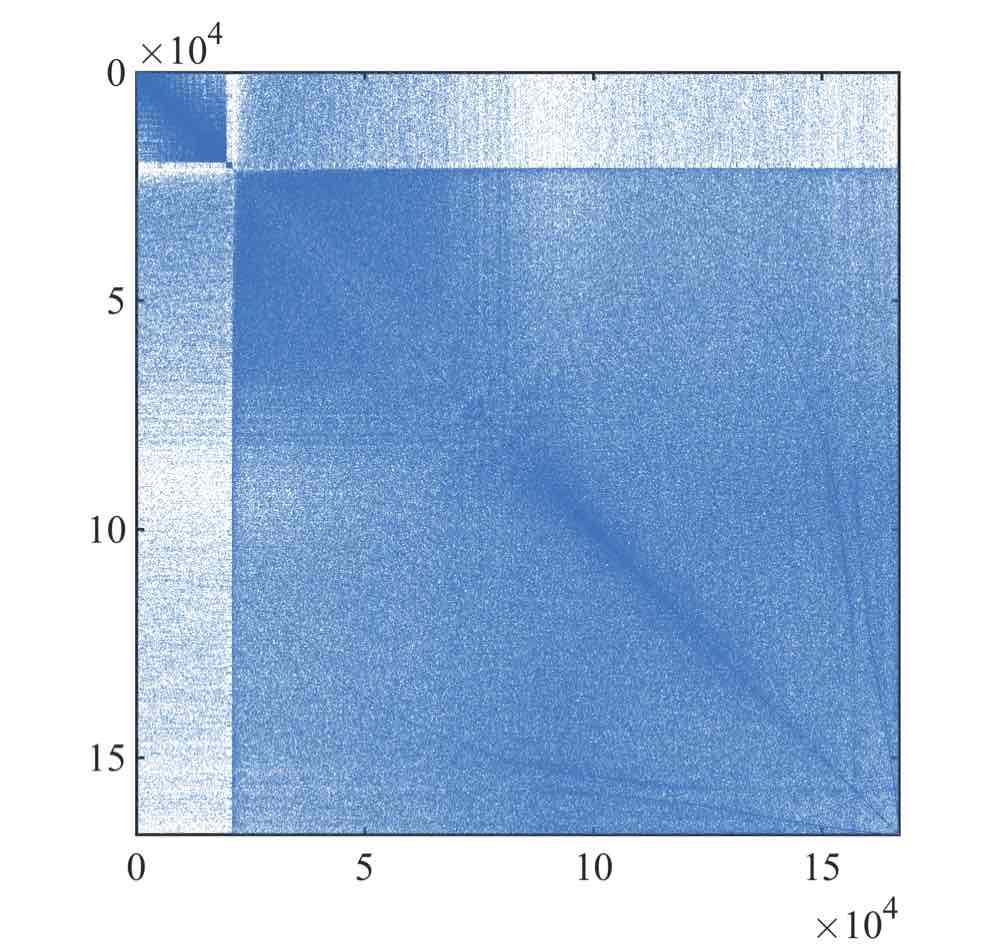}}
{\includegraphics[width=0.44\textwidth, keepaspectratio]{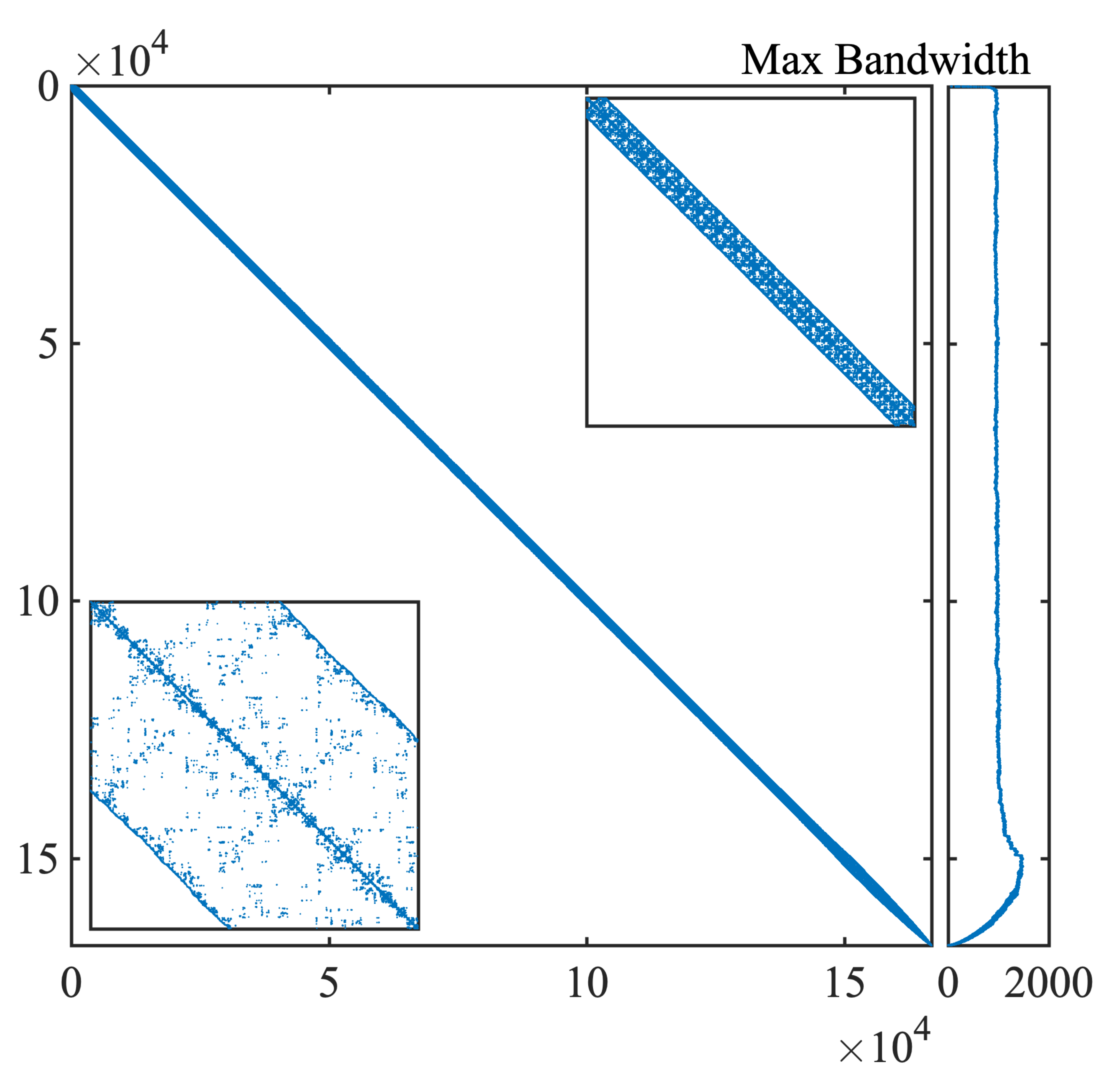}}
{\includegraphics[width=0.205\textwidth, keepaspectratio]{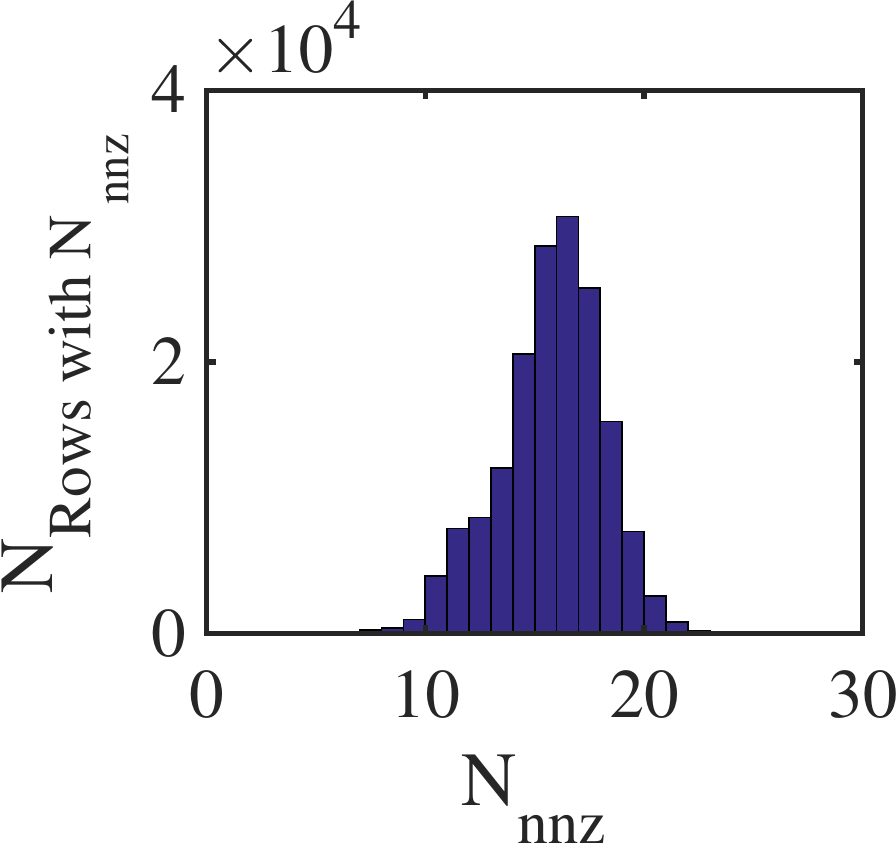}}
{\includegraphics[width=0.265\textwidth, keepaspectratio]{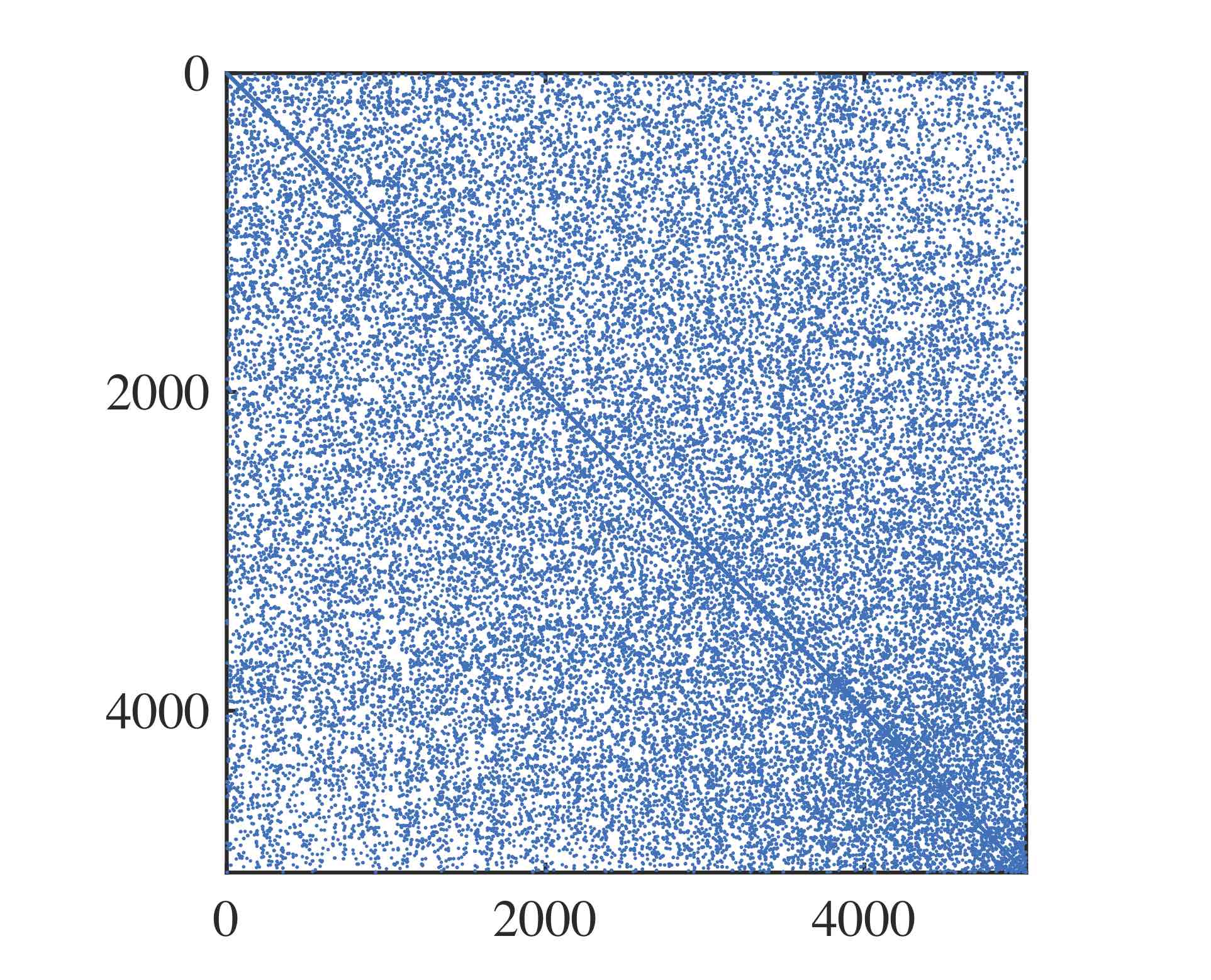}}
\caption{Sparsity pattern of a connectivity matrix for the 1M pipe mesh. Among all matrix elements, only non-zero values are colored in the plot. (top) The {\it global} sparsity pattern for the whole pipe model with $N_{nd}=166,899$. (center) Reverse Cuthill-Mckee reordering of the {\it global} connectivity matrix. The upper-right inset is 100 magnification of the diagonal of the reordered matrix. The lower-left inset is 10,000 magnification. For each row, the left exterior plot shows the maximum bandwidth in that row. (bottom left) Frequencies for the number of non-zero elements in a row (N$_{nnz}$). (bottom right) A {\it local } sparsity pattern from a core with $N_{lnd}=5014$.}
\label{fig:Connectivity}
\vspace{-4pt}
\end{figure} 
\noindent {\bf Sparsity pattern.} The structure of $\mathbf{A}_f$ is determined by element nodal connectivity, i.e., the non-zero columns for node $a$ are from nodes belonging to the element \emph{star} associated to $a$. The \emph{star} of elements associated to a given node $a$ is the set of all elements connected to $a$.
An example of {\it global} sparsity pattern for the pipe benchmark unstructured mesh is reported in Figure~\ref{fig:Connectivity}. The node ordering starts from the outer surface and goes to the interior of the pipe, as seen from the reduced connectivity between the upper-left block and the remaining inner nodes. The density is less than 1 percent, and sparsity is more than 99.99 percent (Table~\ref{table:weakScaling}).

A Reverse-Cuthill-McKee (RCM) bandwidth minimizing permutation of the same raw matrix shows a banded sparse structure illustrated in Figure~\ref{fig:Connectivity}. We report quantitative estimates for bandwidth and number of non-zeros, showing that bandwidth is approximately 1000 with a maximum of about 1500, and most nodes are connected to 15-16 other nodes.

\begin{figure}
\centering
{\includegraphics[width=0.45\textwidth, keepaspectratio]{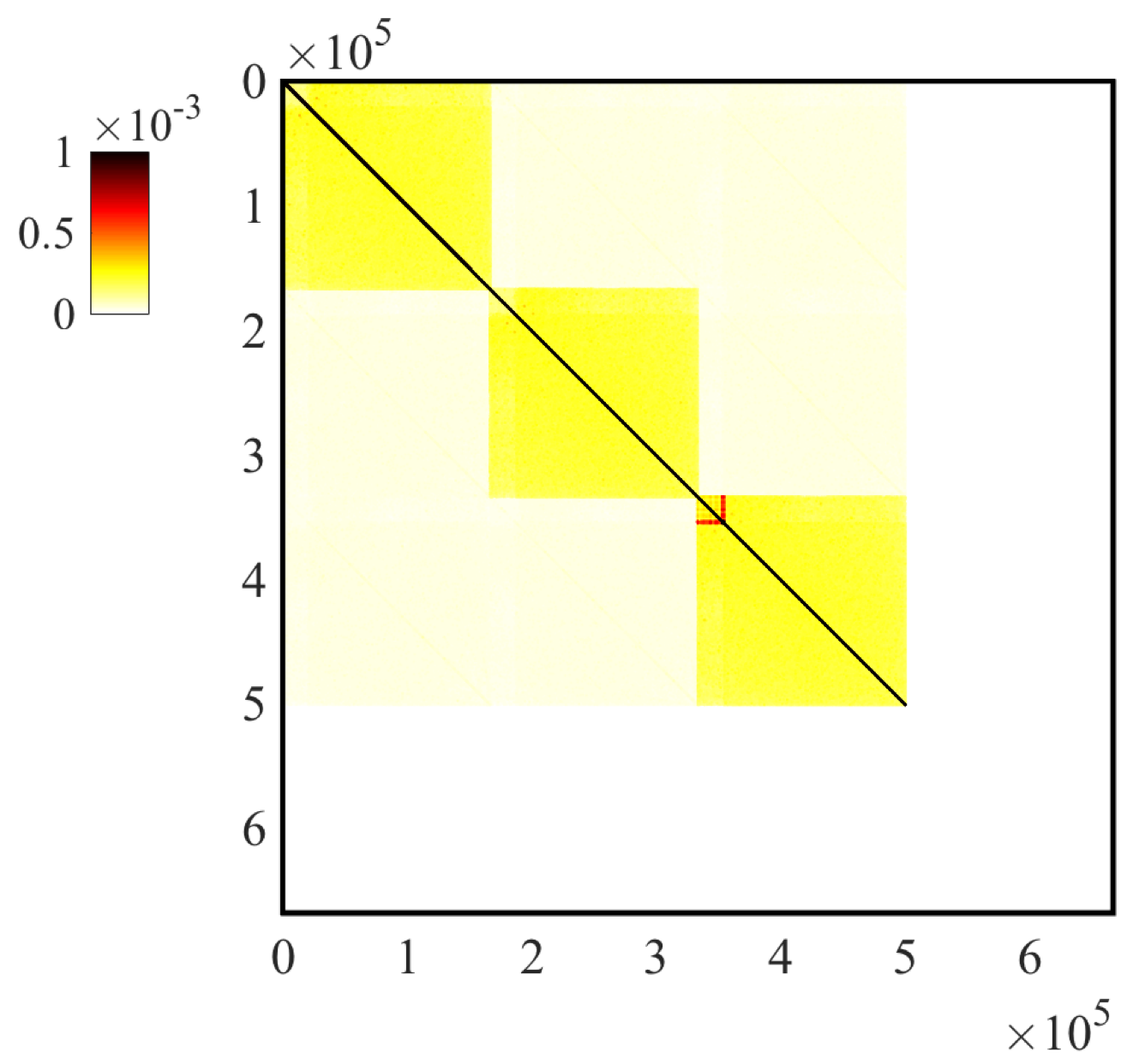}}
{\includegraphics[width=0.47\textwidth, keepaspectratio]{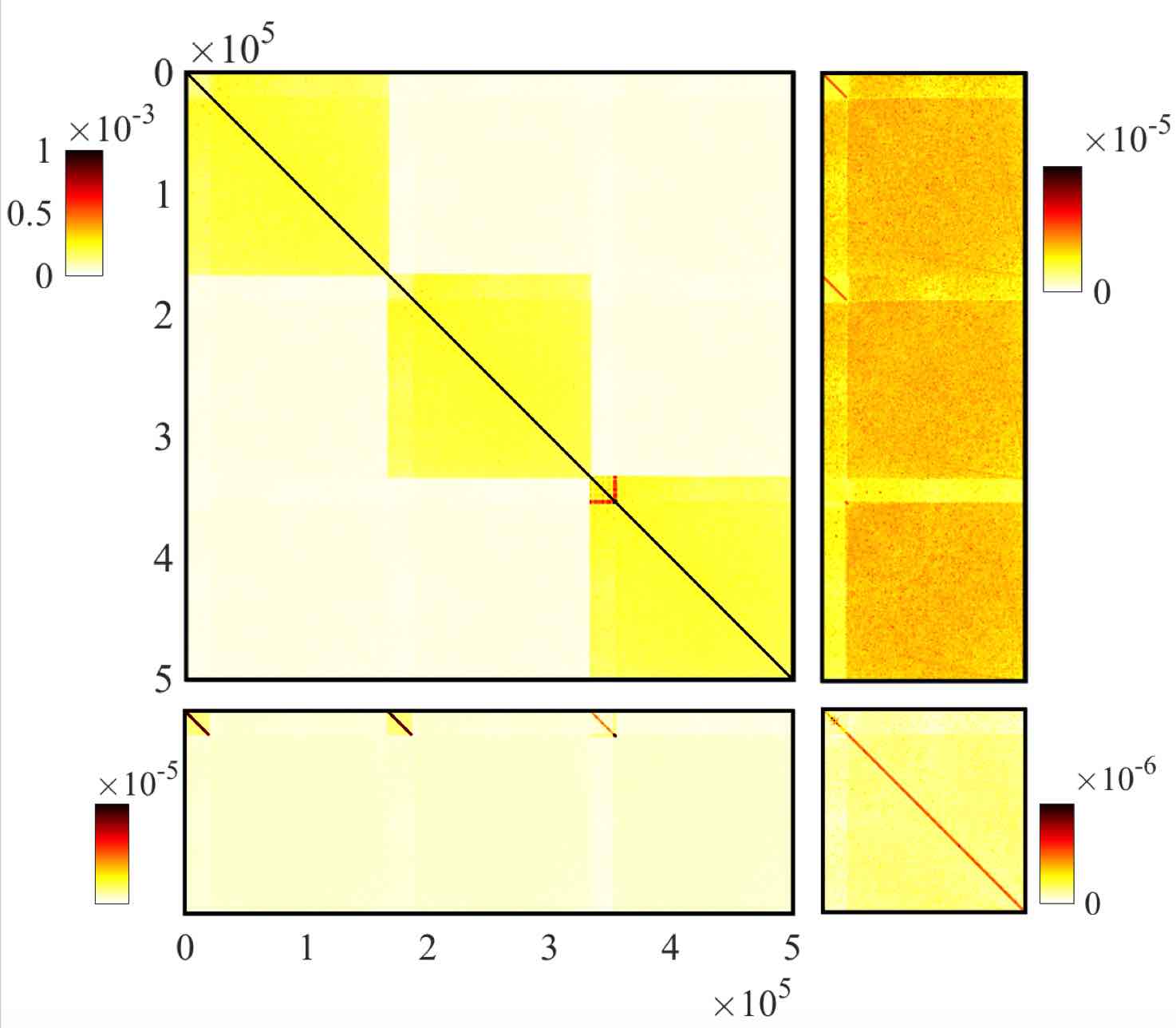}}
\caption{A visual representation of the {\it global} sparse matrix $\mathbf{A}_f$ with entries colored by absolute magnitude. (top) The full matrix  $\mathbf{A}_f$ colored by magnitudes ranging 0 to $10^{-3}$. (bottom) A decomposed matrix colored by magnitudes of each sub-block. In $\mathbf{K}$ in the upper-left block, the colorbar ranges from 0 to $10^{-3}$. In $\mathbf{G}$ in the upper-right block and $\mathbf{D}$, in the lower-left block, the colorbar ranges from 0 to $10^{-5}$. In $\mathbf{L}$ in the lower-right block the colorbar ranges from 0 to $10^{-6}$.
}\label{fig:matrix}
\end{figure} 
\vspace{3pt}

\begin{figure}
\centering
{\includegraphics[width=0.53\textwidth, keepaspectratio]{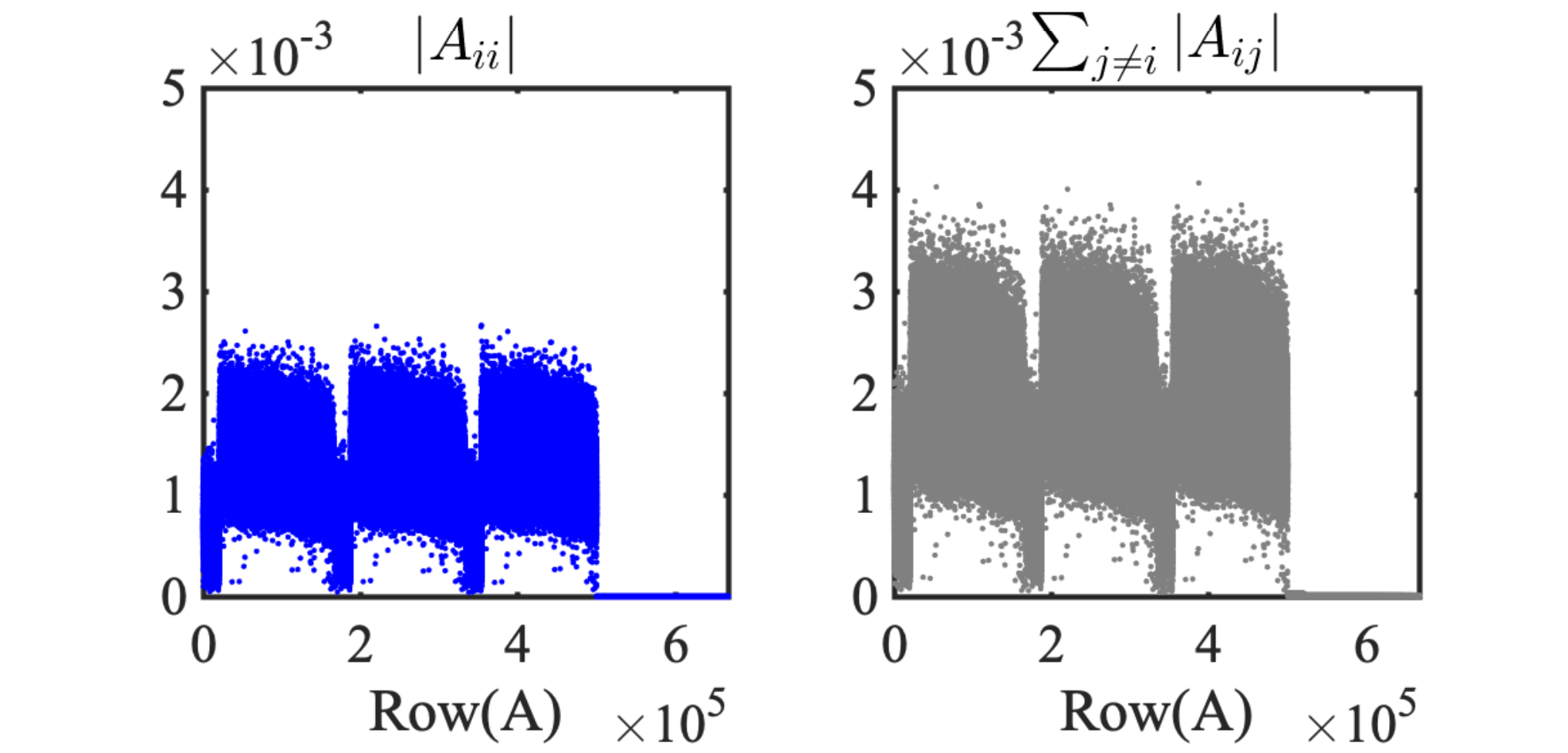}}
{\includegraphics[width=0.4\textwidth, keepaspectratio]{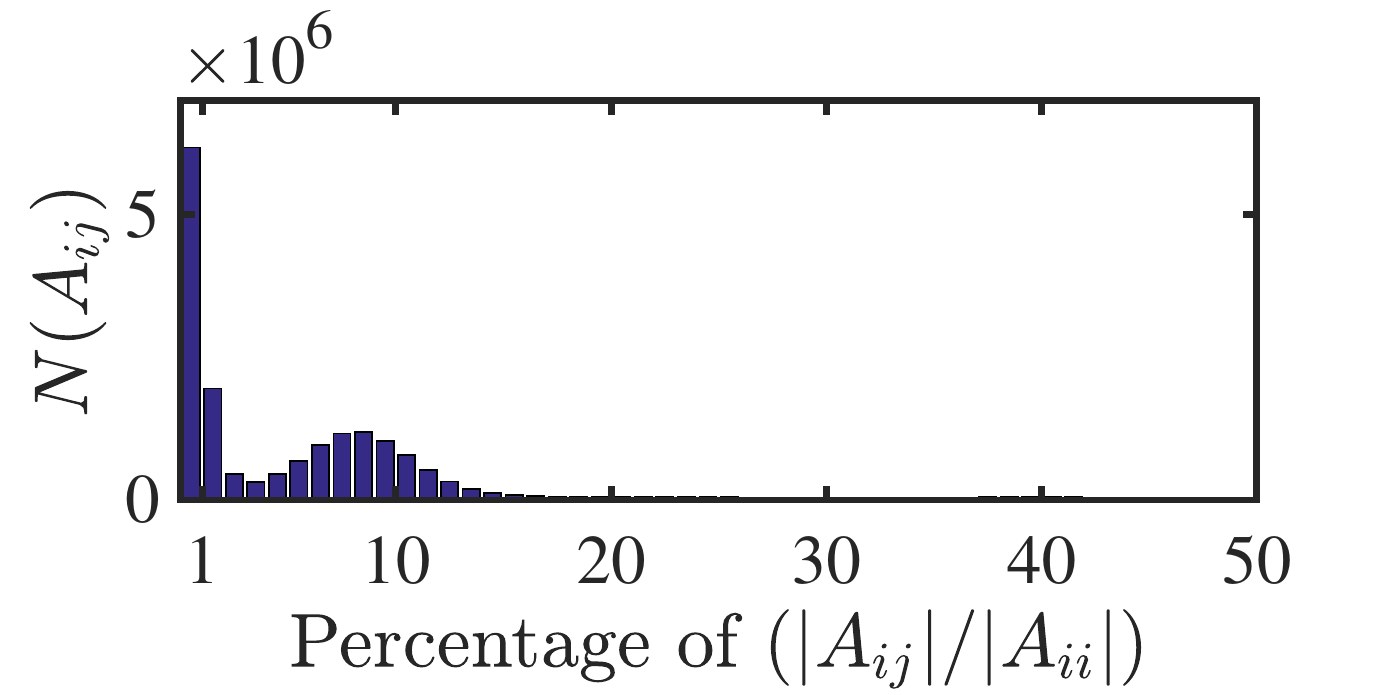}}
\caption{Measures of diagonal dominance. (top left) The magnitude of diagonal values in each row of $\mathbf{A}_f$. (top right) The sum of absolute magnitude of off-diagonal values in each row of  $\mathbf{A}_f$. (bottom) Histogram of the number of matrix elements $N(A_{ij}$) with magnitudes that correspond to the certain percentage of the associated diagonal entry. Only elements that are more than $1\%$ of diagonal entries are counted.}\label{fig:diag}
\end{figure}

\noindent {\bf Diagonal dominance.} A closer look at the magnitudes of entries in the {\it global} matrix $\mathbf{A}_{f}$ reveals a clear block-structure (Figure~\ref{fig:matrix}). The 4-by-4 block structure corresponds to the matrix blocks in equation (\ref{eq:blockmatrix}). This shows that diagonal entries are larger than the off-diagonals in each block of $\mathbf{K}$, suggesting that magnitudes of entries in the blocks of $\mathbf{G}, \mathbf{D}, \mathbf{L}$ are small compared to diagonals in $\mathbf{K}$. 
This relates to the dominant contribution of the acceleration and advection terms, over the stabilization and viscous terms in \eqref{eq:Kblock} (see, e.g.,~\cite{Esmaily2015}).
The matrix $\mathbf{A}_{f}$ is, however, not diagonally dominant (i.e. $|A_{ii}|<\sum_{j=1} |A_{ij}|$). To show this, we quantified the relative magnitudes of off-diagonal and diagonal entries (Figure~\ref{fig:diag}). 
Specifically, we counted the number of elements that have a certain percentage of absolute magnitudes compared to the diagonal values, showing that most off-diagonal values are less than $20\%$ of the associated diagonal. To report quantitative estimates of diagonal dominance, we measure the mean of the relative magnitude of the sum of off-diagonal values to the diagonal, 
\beq
D(\mathbf{K})=\frac{1}{N_{r}}\sum_{i=1}^{N_{r}}\Big[|K_{ii}| / \Big(\sum_{\substack{j=1 \\ i\neq j}}^{N_{c}}|K_{ij}| \Big)\Big],  
\eeq
in which $N_{r}$ and $N_{c}$ is the number of rows and columns in $\mathbf{K}$ respectively. $D$ increases with the diagonal dominance. D($\mathbf{A}_f$) for the {\it global} matrix is 0.514 and D($\mathbf{K}$) for the {\it global} matrix is 0.678.

\vspace{3pt}

\noindent {\bf Symmetry.}  We use an index, $S$, to quantify how close a matrix is to symmetric. We first obtain off-diagonal elements of $\mathbf{A}$ by subtracting its diagonal as $\tilde{\mathbf{A}} = (\mathbf{A}-\text{diag}(\mathbf{A}))$. We then decompose the $\tilde{\mathbf{A}}$ into the symmetric part, $\tilde{\mathbf{A}}_{sym}$=$(\tilde{\mathbf{A}}+\tilde{\mathbf{A}}^T)/2$, and the skew-symmetric part, $\tilde{\mathbf{A}}_{skew}=(\tilde{\mathbf{A}}-\tilde{\mathbf{A}}^T)/2$. The index $S$ is defined as
\begin{equation}
S(\mathbf{A})=\frac{|\tilde{\mathbf{A}}_{sym}|-|\tilde{\mathbf{A}}_{skew}|}{|\tilde{\mathbf{A}}_{sym}|+|\tilde{\mathbf{A}}_{skew}|},  
\end{equation}
in which we use the 2-norm for $|\mathbf{A}|$. The index equals $-1$ for a perfectly skew-symmetric matrix and $1$ for a perfectly symmetric matrix. 
As shown in Section~\ref{sec:LSf}, the matrix is nonsymmetric in the $\mathbf{K}$, $\mathbf{G}$, $\mathbf{D}$ blocks due to stabilization and convective terms, with $S(\mathbf{A}_f)$ equal to 0.9859.
That is, $S(\mathbf{A}_f)$ is a nearly symmetric matrix in the analyzed regime (ideal aortic flow).
Finally, $\mathbf{L}$ is a symmetric and semi-positive definite matrix with $S(\mathbf{L})=1$.

\vspace{3pt}

\noindent {\bf Eigenvalues.} Spectral properties are widely used to characterize the convergence and robustness of iterative solvers.
It is well known, for example, that the rate of convergence of CG depends on the spectral radius of a left-hand-side SPD matrix. 
Despite eigenvalues clustered around 1 leading to rapid convergence of iterative solvers for well-conditioned SPD matrices, the eigenvalues may not be solely responsible for the convergence rate of these solvers and other matrix characteristics may play a role~\cite{Saad2003}. Calculation of all eigenvalues ($\lambda_i$) of the {\it global} matrix for a typical cardiovascular model with order 1 million mesh elements is prohibitively expensive. In this paper we therefore report the spectrum of {\it local} matrices instead of the {\it global} matrix. For a smaller size system, we also demonstrate that the distribution of eigenvalues from {\it local} matrices is a good approximation to the distribution of eigenvalue of the {\it global} matrix (See appendix \ref{app:eigen}). 

\begin{figure}
\centering
{\includegraphics[width=0.50\textwidth, keepaspectratio]{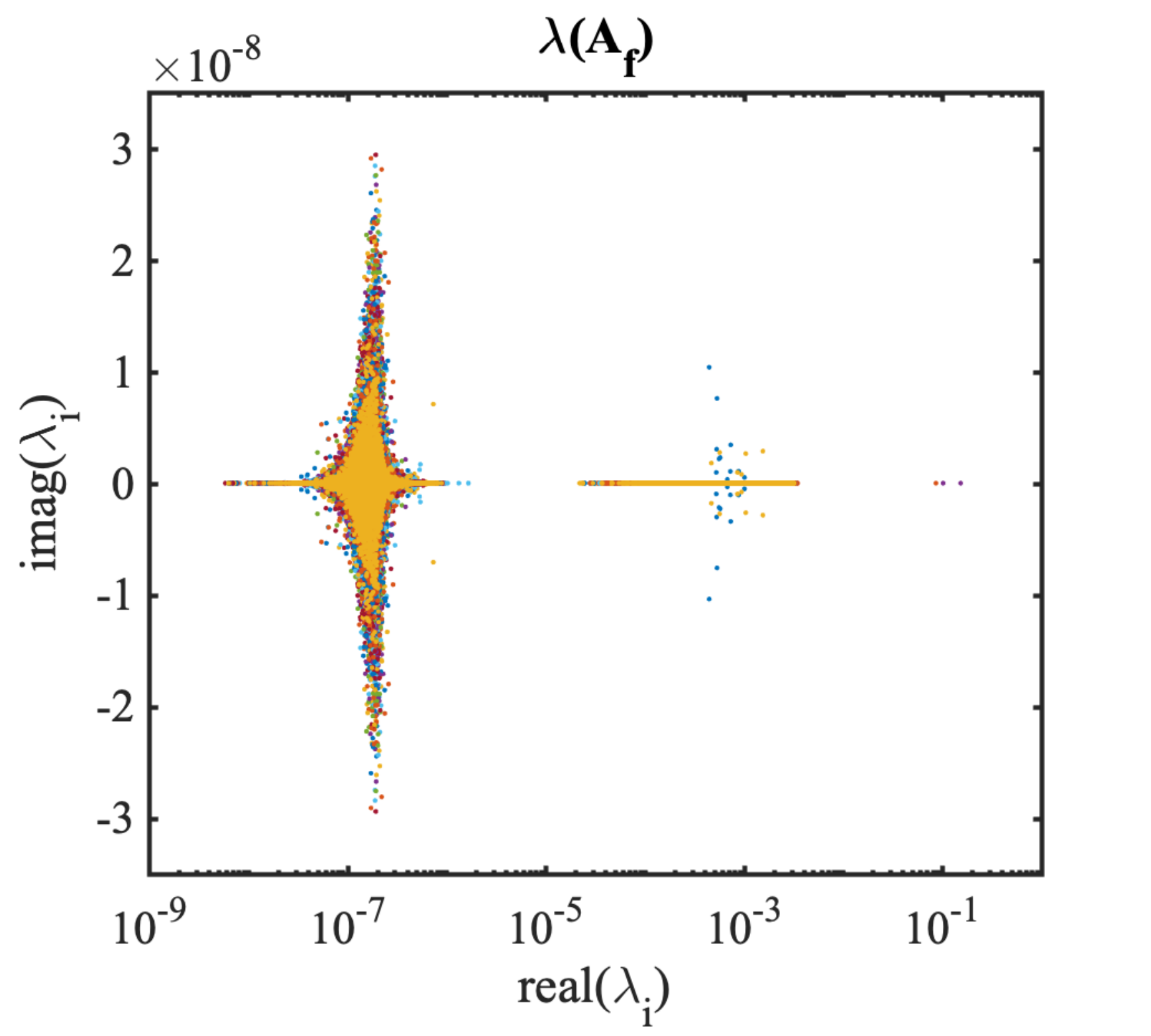}}
{\includegraphics[width=0.54\textwidth, keepaspectratio]{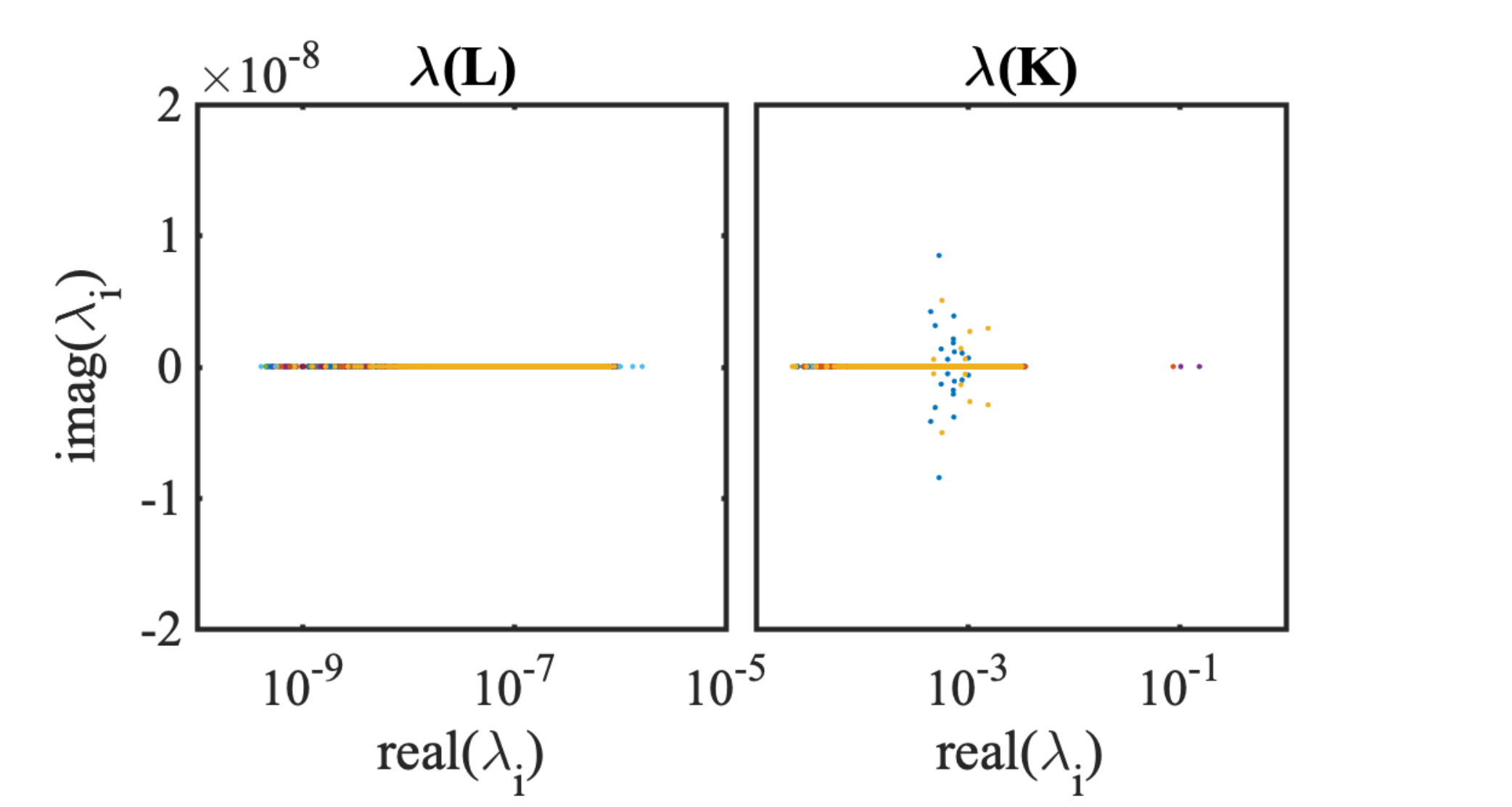}}
\caption{ Spectrum of (top) {\it local} fluid matrices $\mathbf{A}_{f}$, (bottom left) $\mathbf{L}$ blocks, (bottom right) $\mathbf{K}$ blocks. All {\it local} eigenvalue spectrums from 38 cores are plotted together with different colors.}
\label{fig:spectrum}
\vspace{-5pt}
\end{figure} 

In Figure \ref{fig:spectrum}, we plotted the spectrum of {\it local} $\mathbf{A}_f$ matrices from the pipe benchmark with rigid walls. The eigenvalues of $\mathbf{A}_f$ are complex with small magnitudes of the imaginary part up to $O(10^{-8})$, while the magnitudes of the real part ranges from $O(10^{-9})$ to $O(10^{-1})$. In Figure \ref{fig:spectrum}, there are three distinct groups of eigenvalues with different ranges of the real part. The first group contains those with the real part less than $10^{-5}$. The second group contains those with real part ranging between $10^{-5}$ and $10^{-2}$. The spectrums of $\mathbf{K}$ and $\mathbf{L}$ in Figure \ref{fig:spectrum} show that the third group, with eigenvalues larger than $10^{-5}$ in $\mathbf{A}_f$, is attributed to the $\mathbf{K}$ block, while the eigenvalues of $\mathbf{L}$, $\mathbf{G}$ and $\mathbf{D}$ are responsible for the group of smallest real eigenvalues in $\mathbf{A}_f$. We list several minimum and maximum eigenvalues of a {\it local} matrix without resistance {\color{black} boundary condition (BC)} in Table~\ref{table:3}. The maximum eigenvalues of $\mathbf{K}$ and $\mathbf{A}_f$ appear to be the same, suggesting that block $\mathbf{K}$ dominates the high portion of the spectrum, with the smallest eigenvalues provided by the blocks $\mathbf{L}$, $\mathbf{G}$ and $\mathbf{D}$. This suggests that the large condition number of $\mathbf{A}_f\sim O(10^6)$, obtained from MATLAB {\it condest}, relates to the inhomogeneous eigenvalue spectrum observed in the momentum, continuity and coupling blocks.
Additionally, the small condition number of $\mathbf{K}$ justifies the idea behind the BIPN approach, i.e., to solve the $\mathbf{K}$ block separately, while expressing the other blocks in Schur complement form~\cite{Esmaily2015}.
$\mathbf{L}$ is singular, and thus has zero eigenvalue and an extremely large condition number. Lastly, the resistance boundary condition is responsible for the few largest eigenvalues order of $O(10^{-1})$ in Figure \ref{fig:spectrum}, as we discuss in the next section. 

\begin{table}
\centering
\begin{tabular}{ c |c c c c c  }
\hline
\multicolumn{6}{c}{{$\mathbf{A}_f$}: Condition number=1.293$\times10^{6}$} \\
\hline
{$\lambda_i$}& 1${^{st}}$ & 2${^{nd}}$ & 3${^{rd}}$ & 4${^{th}}$ & 5${^{th}}$ \\
\hline
Max$(\times10^{-3})$ & 3.304 & 3.301 & 3.298 & 3.282 & 3.270\\
\hline
Min$(\times10^{-8})$& 0.748 & 1.042 & 1.348 & 1.524 & 1.672\\
\hline
\hline
\multicolumn{6}{c}{{$\mathbf{K}$}: Condition number=163} \\
\hline
{$\lambda_i$}& 1${^{st}}$ & 2${^{nd}}$ & 3${^{rd}}$ & 4${^{th}}$ & 5${^{th}}$ \\
\hline
Max$(\times10^{-3})$ & 3.304 & 3.301 & 3.298 & 3.282 & 3.270\\
\hline
Min$(\times10^{-5})$& 4.615 & 4.621 & 5.958 & 6.381 & 6.614\\
\hline
\hline
\multicolumn{6}{c}{{$\mathbf{L}$}: Condition number=$1.555\times 10^{18}$} \\
\hline
$\lambda_i$& 1${^{st}}$ & 2${^{nd}}$ & 3${^{rd}}$ & 4${^{th}}$ & 5${^{th}}$ \\
\hline
Max$(\times10^{-7})$ & 9.053 & 8.180 & 8.085 & 8.083 & 7.983\\
\hline
Min$(\times10^{-9})$& 0.000 & 0.456 & 1.224 & 1.765 & 2.214\\
\hline
\end{tabular}
\caption{The 1-norm condition number estimates and five maximum and minimum eigenvalues ($\lambda_i$) of a {\it local} matrix $\mathbf{A}_f$, and blocks $\mathbf{K}$, and $\mathbf{L}$ without resistance BC.}
\label{table:5}
\end{table}

\vspace{3pt}


\noindent {\bf Effects of the resistance boundary condition.}
%
\noindent A resistance boundary condition perturbs the condition number of the coefficient matrix $\mathbf{A}_f$, and may be responsible for a significant increase in the solution time for the tangent linear system~\cite{Esmaily2013a}.
The boundary traction $\mathbf{h}$, is given, in this case by 
\begin{equation}
\mathbf{h}(\mathbf{u}, p, \mathbf{x}, t)=-P_i\,\mathbf{n},\,\,\mathbf{x}\in \Gamma_h,
\end{equation}
in which $P_i$ is the pressure at surface $i$, evaluated as $P_i=R_i\,Q_i$, i.e., proportional to the flow rate $Q_i$ across the surface
\begin{equation}
Q_i(t)=\int_{\Gamma_i} \mathbf{u} \cdot \mathbf{n}\,d\Gamma, 
\end{equation}
through the prescribed resistance $R_i$. Thus, the contribution of the resistance boundary condition to the coefficient matrix is 
\begin{equation}
\mathbf{K}^{bc}=\sum_{i=1}^{n^{bc}}\tilde{R}_i\,\mathbf{S}_i \otimes \mathbf{S}_i,\,\,\mathbf{S}_i =  \int_{\Gamma_i}N_a\,\mathbf{n}\,d\Gamma, 
\end{equation}
where $n^{bc}$ is the number of resistance boundaries, and $\tilde{R}_i= \gamma\,\Delta t\,R_i$. 
$\mathbf{K}^{bc}$ is finally added to the $\mathbf{K}$ sub-matrix, resulting in the coefficient matrix $\mathbf{\tilde{K}}=\mathbf{K}+\mathbf{K}^{bc}$. 
Generalization from an outlet resistance to a coupled lumped parameter network model is accomplished using a slightly more general expression for $\mathbf{K}^{bc}$, i.e.
\begin{multline}
\mathbf{K}^{bc}=\sum_{k=1}^{n^{bc}}\sum_{l=1}^{n^{bc}}\gamma\,\Delta t\,M_{kl} \int_{\Gamma_k} N^{a}\,\mathbf{n}_{i}\,d\Gamma \int_{\Gamma_l} N^{b}\,\mathbf{n}_{j}\,d\Gamma,\\
M_{kl}=\frac{\partial P_k^{n+1}}{\partial Q_l^{n+1}}, 
\end{multline}
where the resistance matrix $M_{kl}$ is obtained by coupling pressures and flow rates at different outlets~\cite{Esmaily2012}.

\vspace{3pt}

\begin{figure}
\centering
{\includegraphics[width=0.3\textwidth, keepaspectratio]{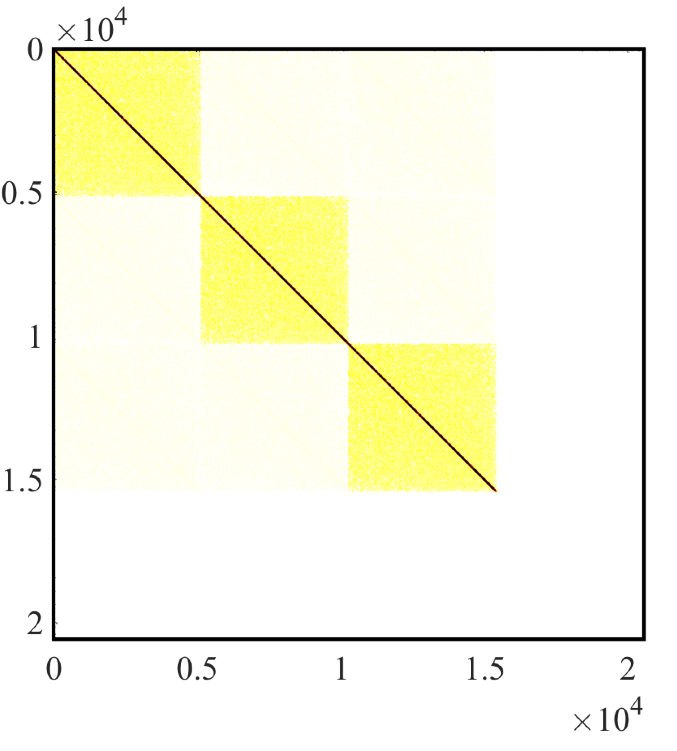}}
{\includegraphics[width=0.35\textwidth, keepaspectratio]{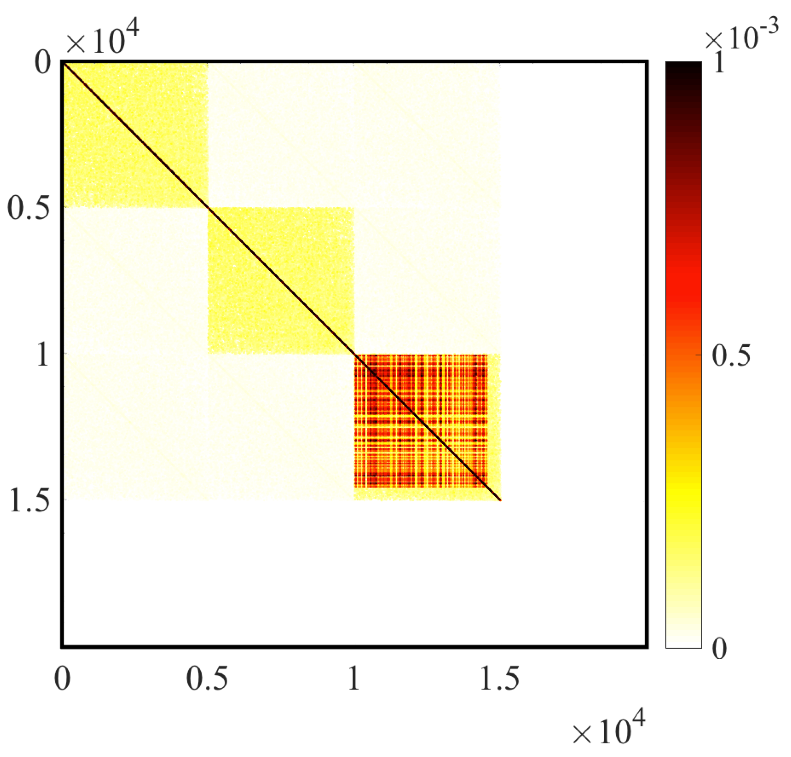}}
\caption{A visual representation of (top) a {\it local} matrix $\mathbf{A}_f$ without resistance BC (bottom) a {\it local} matrix $\mathbf{A}_{f,res}$ with a resistance BC, R=1600$g/cm^4/s$. Matrix elements are colored by their absolute magnitude. For both figures color ranges from 0 to $10^{-3}$.}
\label{fig:matrix_res}
\vspace{-5pt}
\end{figure}

Addition of a resistance boundary condition alters the topology of the coefficient matrix due to the rank one contribution $\mathbf{S}_{i}\otimes\mathbf{S}_{i}$. 
This, in practice, couples all velocity degrees of freedom on a given outlet, significantly affecting the performance of matrix multiplication for the $\mathbf{K}$ block and the fill-in generated by its LU decomposition.
Thus, the vector $\mathbf{S}_{i}$ is stored separately, to improve the efficiency of matrix multiplication and for RPC preconditioning.
Figure~\ref{fig:matrix} shows how the {\it global} matrix entries are affected by the presence of a resistance boundary condition, i.e., large magnitude components in the z-directional velocity block (lower-right block in $\mathbf{K}$) arise. To better highlight this effect, we show two {\it local} matrices with and without a resistance boundary condition in Figure~\ref{fig:matrix_res}. 
Addition of $\mathbf{K}^{bc}$ increases the contribution of off-diagonal entries moving the matrix further away from diagonal dominance.

\vspace{3pt}

The resistance BC perturbs the spectrum and increases the condition number of $\mathbf{A}_f$. The largest few eigenvalues in the spectrum of $\mathbf{A}_f$ and $\mathbf{K}$ in Figure \ref{fig:spectrum} are calculated from {\it local} matrices in partitioned domains interfacing the resistance boundary. The change of spectral properties of $\mathbf{\tilde{K}}$ due to rank one contribution $\mathbf{K}^{bc}$ (see, e.g.,~\cite{Esmaily2015}) is measured with maximum and minimum eigenvalues reported in Table~\ref{table:6}.
The maximum eigenvalue of $\mathbf{\tilde{K}}$ is significantly larger than $\mathbf{K}$, leading to a $\sim O(10)$-fold increase in the condition number of $\mathbf{\tilde{K}}$, thus increasing the spectral radius of the whole spectrum as shown in Figure \ref{fig:spectrum}. 
Additionally, our tests confirm that the first maximum eigenvalue $\mathbf{A}_{f}$ increases linearly with the assigned resistance.
Thus, in a more general case, we expect several large eigenvalues to be added to $\mathbf{A}_{f}$ for models with multiple outlet resistances.

\begin{table}
\centering
\begin{tabular}{ c |c c c c c  }
\hline
\multicolumn{6}{c}{{$\mathbf{\tilde{A}}_f$}: Condition number=3.299$\times10^{7}$} \\
\hline
$\lambda_i$ & 1${^{st}}$ & 2${^{nd}}$ & 3${^{rd}}$ & 4${^{th}}$ & 5${^{th}}$ \\
\hline
Max$(\times10^{-3})$ & 86.525 & 3.253 & 3.244 & 3.239 & 3.185\\
\hline
Min$(\times10^{-8})$& 0.671 & 1.036 & 1.350 & 1.427 & 1.504\\
\hline
\hline
\multicolumn{6}{c}{{$\mathbf{\tilde{K}}$}: Condition number=5.883$\times10^{3}$} \\
\hline
$\lambda_i$ & 1${^{st}}$ & 2${^{nd}}$ & 3${^{rd}}$ & 4${^{th}}$ & 5${^{th}}$ \\
\hline
Max$(\times10^{-3})$ & 86.525 & 3.253 & 3.244 & 3.239 & 3.185\\
\hline
Min$(\times10^{-5})$& 2.878 & 2.886 & 3.457 & 3.981 & 4.002\\
\hline
\end{tabular}
\caption{The 1-norm condition number estimates and five maximum and minimum eigenvalues ($\lambda_i$) of a {\it local} matrix $\mathbf{\tilde{A}}_{f}$ and $\mathbf{\tilde{K}}$ with a resistance BC.}
\label{table:6}
\end{table}

\subsection{Matrix properties for fluid flow in deformable vessels}

\begin{figure}
\centering
{\includegraphics[width=0.46\textwidth, keepaspectratio]{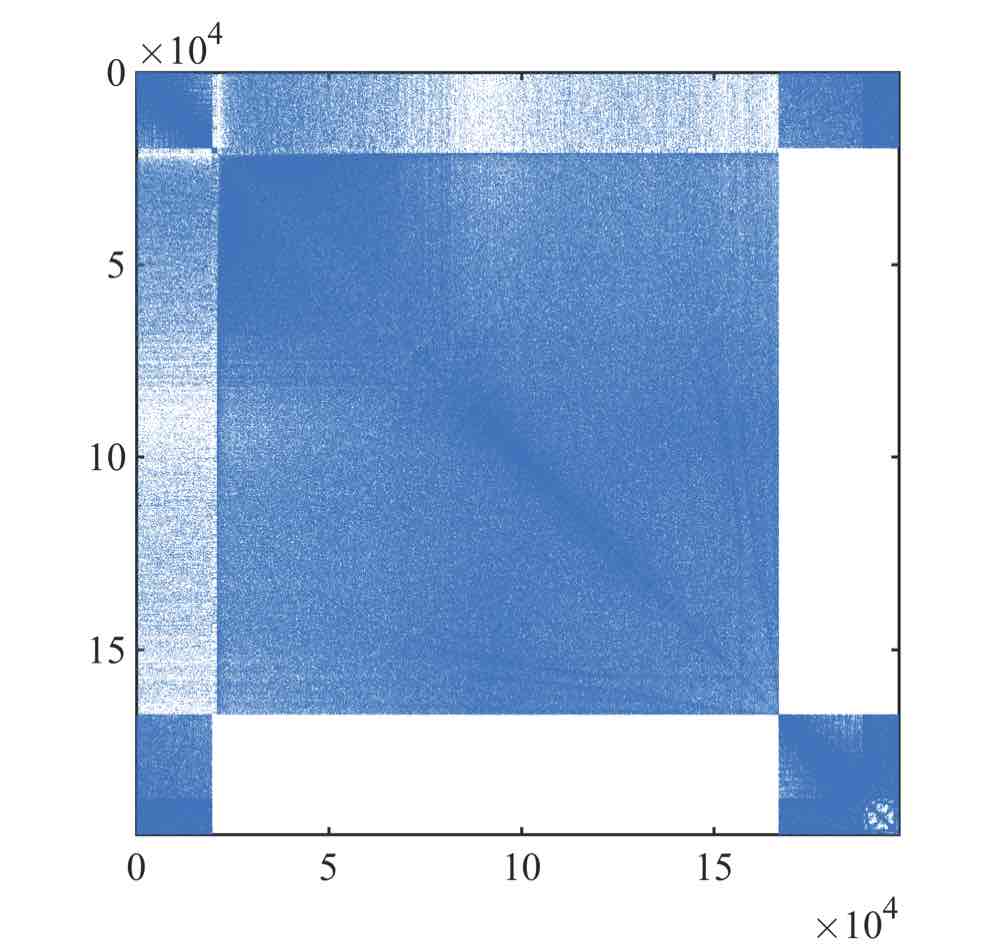}}
{\includegraphics[width=0.43\textwidth, keepaspectratio]{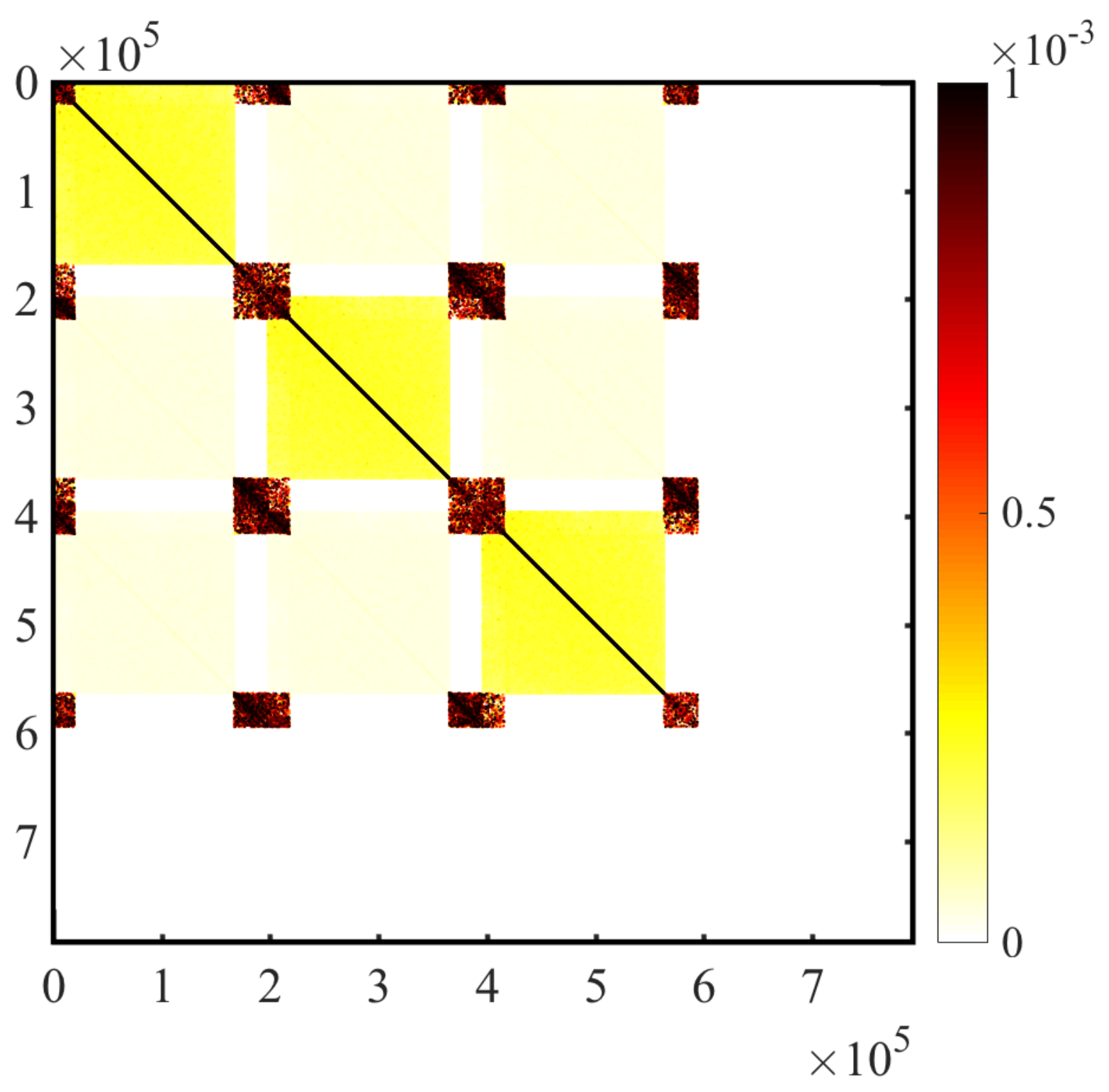}}
\caption{(Top) {\it Global} sparsity pattern for the FSI pipe benchmark with $N_{nd}=198,128$. (bottom) Visual representation of the entry magnitudes for the {\it global} matrix $\mathbf{A}_{\text{FSI}}$.
}\label{fig:RCMs}
\vspace{-10pt}
\end{figure}

\begin{figure}
\centering
{\includegraphics[width=0.45\textwidth, keepaspectratio]{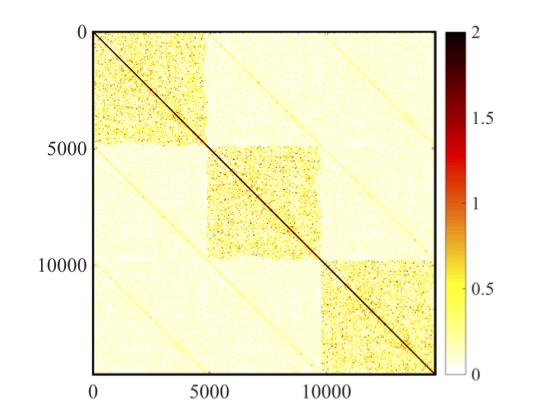}}
{\includegraphics[width=0.45\textwidth, keepaspectratio]{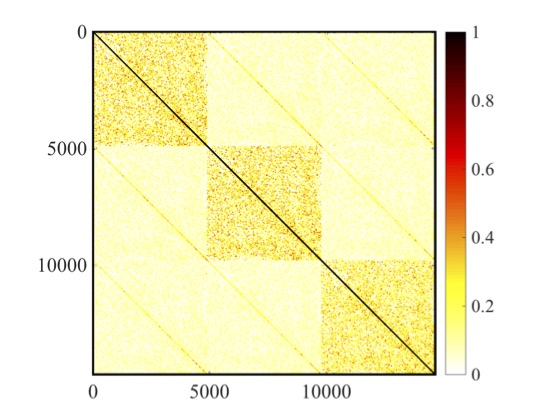}}
{\includegraphics[width=0.45\textwidth, keepaspectratio]{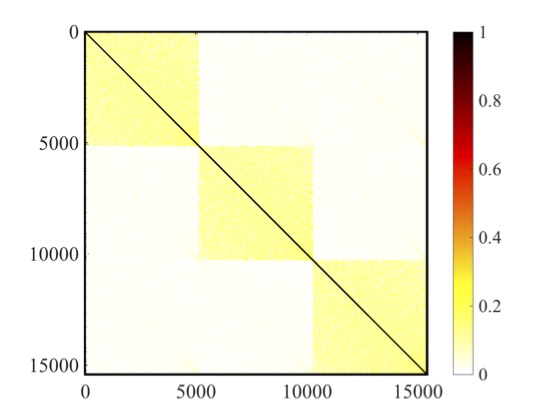}}
\caption{Sparsity pattern of a {\it local} $\mathbf{K}_{s}$ colored by absolute magnitude of each entry, for $N_{lnd}=4892$. (top) A raw matrix, $\mathbf{K}_{s}$, in the solid domain (center) A scaled matrix with its diagonal $\mathbf{K}_s^*$ in the solid domain (bottom) A scaled {\it local} $\mathbf{K}^*$ with its diagonal in the fluid domain.}
\label{fig:matrix_s}
\end{figure}

\noindent {\bf Sparsity pattern.} The {\it global} sparsity pattern for the FSI mesh is illustrated in Figure~\ref{fig:RCMs}, where nodes in the solid-fluid interface are ordered first, followed by nodes in the fluid region next to the interface, and nodes in the solid domain. As shown in Figure~\ref{fig:Connectivity} when the connectivity matrix is reordered by RCM, the global sparsity pattern has a banded sparse matrix structure with the larger maximum bandwidth of $\approx1750$ compared to the rigid case, while the number of non-zeros in a row is mostly clustered around 15 to 16, similar to the rigid case. Magnitudes of entries associated with solid nodes appear to be significantly larger than those in the fluid domain. For example, the magnitude of $\mathbf{K}_s$ is order one (Figure~\ref{fig:matrix_s}), whereas the magnitude of $\mathbf{K}$ is order $10^{-3}$ (Figure~\ref{fig:matrix_res}). 
In what follows, we focus on the {\it local} matrix $\mathbf{K}_s$ from Eq. \eqref{eq:solid} since the characteristics of $\mathbf{A}_f$ in FSI are similar to the rigid wall case. 

\noindent {\bf Diagonal dominance.}  The metric introduced above to quantify diagonal dominance drops to $D(\mathbf{A}_{\text{FSI}})=0.4775$ for the {\it global} matrix $\mathbf{A}_{\text{FSI}}$, i.e., the additional FSI terms reduce the diagonal dominance of the system. 
The magnitudes of a {\it local} $\mathbf{K}_s$ matrix before and after diagonal scaling is compared, in Figure~\ref{fig:matrix_s}, to a {\it local} $\mathbf{K}$ from the fluid domain.
The diagonally scaled block $\mathbf{K}_s$ qualitatively shows how the off-diagonals in $\mathbf{K}_s$ are larger than those in $\mathbf{K}$. 
The diagonal dominance metric for the {\it local} blocks $\mathbf{K}_{s}$ and $\mathbf{K}$ are found to be $D(\mathbf{K}_{s})=0.379$ and $D(\mathbf{K})=0.678$, respectively. 
\vspace{3pt}

\noindent {\bf Symmetry and positive definiteness.}  The symmetry metric for matrix $\mathbf{K}_s$ is one, i.e., $S(\mathbf{K_s})=1$ and positive definite as expected from its properties and confirmed numerically. 

\vspace{3pt}

\begin{figure}
\centering
{\includegraphics[width=0.50\textwidth, keepaspectratio]{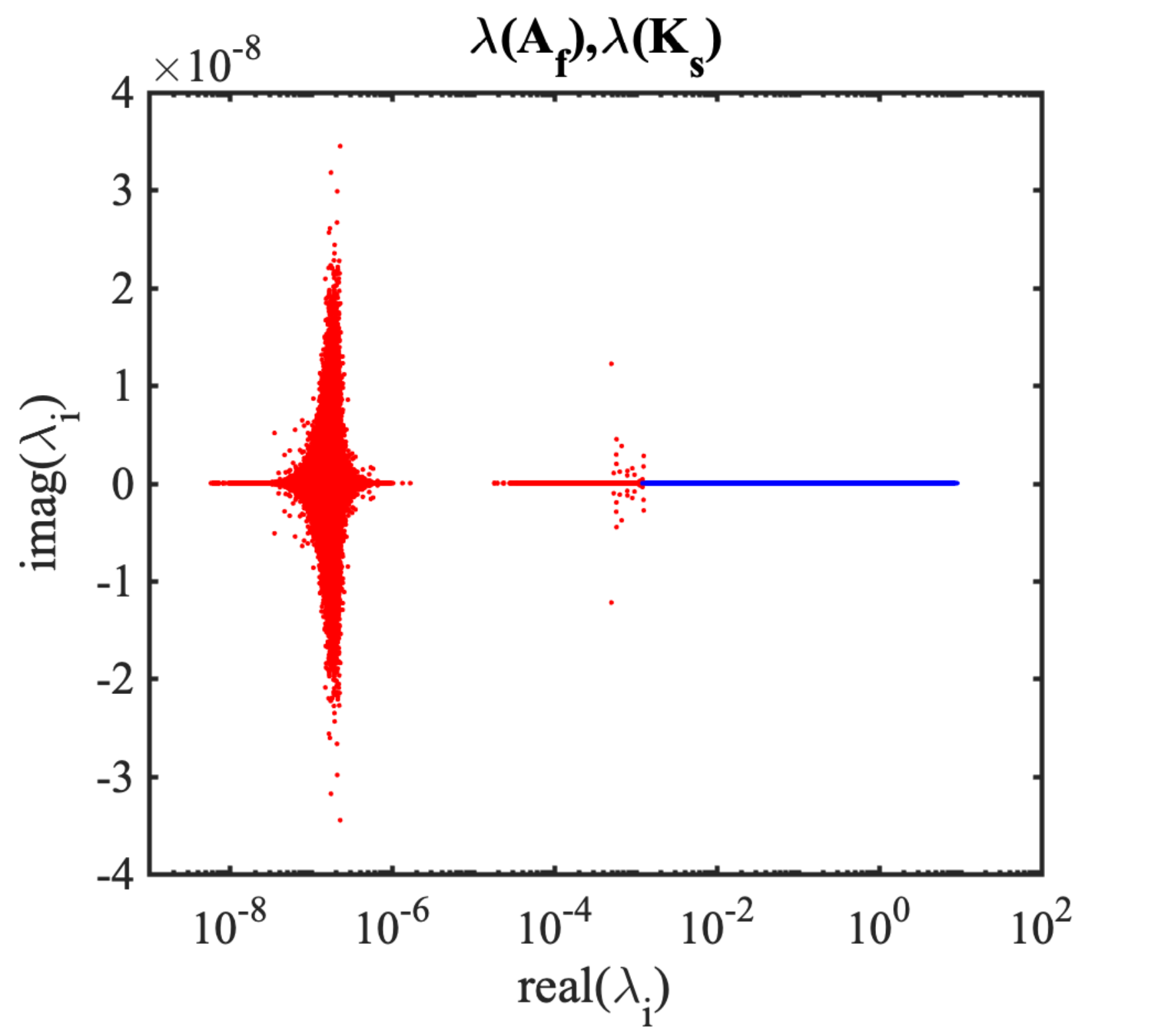}}
\caption{ Spectrum of (red) {\it local} $A_{f}$ in the fluid domain and (blue) $K_s$ in the solid domain.  All {\it local} eigenvalue spectrums from 48 cores are plotted together.}
\label{fig:spectrum_solid}
\vspace{-5pt}
\end{figure}

\noindent {\bf Eigenvalues.}  We calculated and plotted the eigenvalue spectrum of {\it local} matrices from the FSI benchmark in Figure~\ref{fig:spectrum_solid}. All eigenvalues of $\mathbf{K}_s$ are real due to the symmetry of $\mathbf{K}_s$. As listed in Table~\ref{table:7}, the magnitudes of eigenvalues in $\mathbf{K}_s$ is significantly larger than the magnitudes of eigenvalues from $\mathbf{A}_f$. 

\begin{table}
\centering
\begin{tabular}{ c |c c c c c }
\hline
\multicolumn{6}{c}{{$\mathbf{K}_s$}: Condition number=3.169$\times10^{4}$} \\
\hline
$\lambda_i$ & 1${^{st}}$ & 2${^{nd}}$ & 3${^{rd}}$ & 4${^{th}}$ & 5${^{th}}$ \\\hline
Max & 7.706& 7.469 & 7.389 & 7.325 & 7.294\\
\hline
Min$(\times10^{-3})$& 1.374 & 1.404 & 1.468 & 1.510 & 1.529\\
\hline
\end{tabular}
\caption{The 1-norm condition number estimate and five maximum and minimum eigenvalues ($\lambda_i$) of the raw {\it local} $\mathbf{K}_s$ matrix.}
\label{table:7}
\end{table}

\subsection{Discussion}
%
\noindent Results from the previous sections suggest the following conclusions. 
First, the condition number of both the fluid and solid tangent matrices $\mathbf{A}_f$ and $\mathbf{K}_s$ appears to be large, and therefore preconditioning is necessary. 
Second, the fluid matrix $\mathbf{A}_f$ is more diagonally dominant than the solid matrix $\mathbf{K}_{s}$.
This suggests that diagonal preconditioning is expected to be more effective for rigid wall simulations, but incomplete factorization preconditioners are expected to work better under fluid-structure interaction, consistent with the results obtained in the pipe benchmark.
Third, resistance and coupled multidomain boundary conditions need a special treatment for preconditioning, due to their effect on the maximum eigenvalue and condition number.

\section{Effect of preconditioning }\label{sec:PC}

\noindent In this section we investigate how application of various preconditioners affects the spectral properties of the coefficient matrix in both the rigid and deformable case by explicitly computing the preconditioned matrix $\mathbf{M}_l^{-1}\mathbf{A}\mathbf{M}_r^{-1}$, where $\mathbf{M}_l^{-1}$ is a left preconditioner and $\mathbf{M}_r^{-1}$ is a right preconditioner. 

\begin{figure}
\centering
{\includegraphics[width=0.50\textwidth, keepaspectratio]{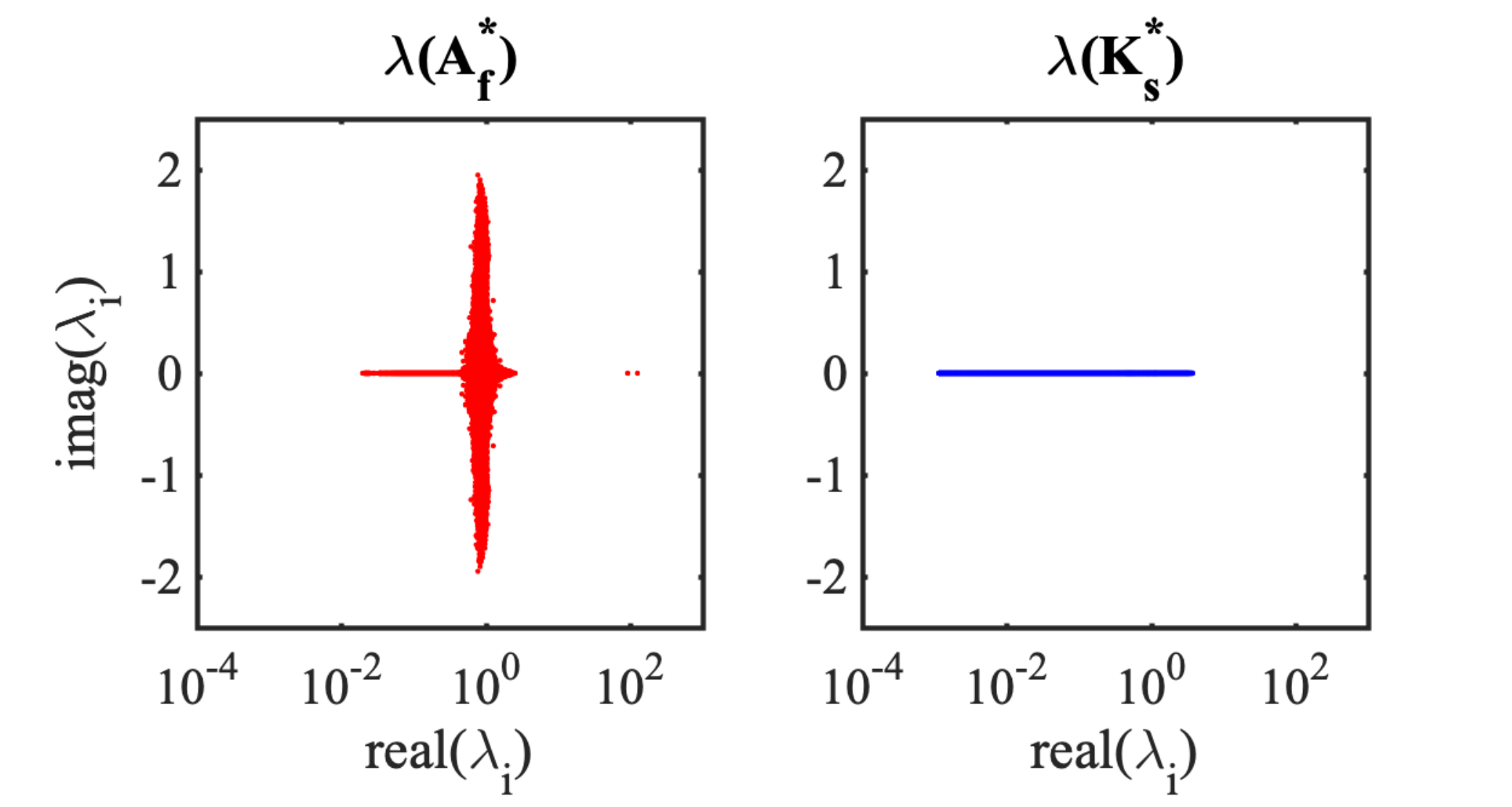}}
\caption{ Spectrum of (red) {\it local} $\mathbf{A}_{f}^*$ in the fluid domain and (blue) $\mathbf{K}_s^*$ in the solid domain.  All {\it local} eigenvalue spectrums from 48 cores are plotted together.}
\label{fig:spectrum_pc}
\vspace{-5pt}
\end{figure}

Consider a left and right Jacobi preconditioning for $\mathbf{A}_f$:
\beq
\begin{split}
\mathbf{W}^m=\text{diag}(\mathbf{K})^{-1/2},
\mathbf{W}^c=\text{diag}(\mathbf{L})^{-1/2},\\
\mathbf{K}^*\leftarrow \mathbf{W}^m\mathbf{K}\mathbf{W}^m,
\mathbf{G}^*\leftarrow \mathbf{W}^m\mathbf{G}\mathbf{W}^c,\\
\mathbf{D}^* \leftarrow \mathbf{W}^c\mathbf{D}\mathbf{W}^m,
\mathbf{L}^*\leftarrow \mathbf{W}^c\mathbf{L}\mathbf{W}^c,\\
\Delta \mathbf{u}^*\leftarrow \mathbf{W}^m\Delta\mathbf{u},
\Delta \mathbf{p}^*\leftarrow \mathbf{W}^c\Delta \mathbf{p},\\
\end{split}
\eeq
resulting in the linear system $\mathbf{A}_f^*\mathbf{y}^*=-\mathbf{R}^*$. Spectral properties of the preconditioned matrix $\mathbf{A}_f^*$ and $\mathbf{K}_s^*$ are reported in Figure \ref{fig:spectrum_pc} and Table~\ref{table:8}.

Table~\ref{table:8} shows how diagonal preconditioning is effective in improving the conditioning of $\mathbf{A}_f$, particularly for $\mathbf{K}$, without resistance BC.
The condition number of $\mathbf{K}^*$ is reduced to $\sim$10, and only a few linear solver iterations are expected to be sufficient to substantially reduce the approximated residual.
This again justifies the approach followed by the BIPN solver, where the linear system involving the momentum block $\mathbf{K}$ is solved separately, thus shifting the computational cost to the iterative solution of its Schur-complement.
The condition number of the approximated Schur complement block $\mathbf{L}^*+{\mathbf{G}^*}^T{\mathbf{G}^*}$ is, in this example, equal to 120, consistent with previous findings in~\cite{Esmaily2015}. 
This also justifies the fact that $\sim$80 percent of total compute time in BIPN is dedicated to the solution of the Schur complement linear system.
Conversely, a symmetric Jacobi preconditioning does not seem to significantly reduce the condition number of the solid block $\mathbf{K}_s$.
This is attributed to the presence of large off-diagonal values in $\mathbf{K}_{s}$ that are only marginally affected by diagonal preconditioning. 
As a result, the eigenvalues of $\mathbf{K}_{s}^*$ range from $O(10^{-3})$ to $O(1)$ while the eigenvalues of $\mathbf{A}_f^*$ range $O(10^{-2})$ to $O(1)$, shown in Figure \ref{fig:spectrum_pc}, with the exception of a few eigenvalues from the resistance BC. This, in turn, justifies the superiority of incomplete factorization preconditioners for FSI simulations. 

\begin{table}
\centering
\begin{tabular}{ c |c c c c c }
\hline
\multicolumn{6}{c}{{$\mathbf{A}^*_f$}: Condition number=645} \\\hline
$\lambda_i$ & 1${^{st}}$ & 2${^{nd}}$ & 3${^{rd}}$ & 4${^{th}}$ & 5${^{th}}$ \\\hline
Max& 2.271 & 2.248 & 2.248 & 2.247&2.244\\\hline
Min$(\times10^{-2})$& 2.557 & 3.558 & 4.623 & 5.183&5.795\\\hline\hline
\multicolumn{6}{c}{{$\mathbf{K}^*$}: Condition number=14} \\\hline
$\lambda_i$ & 1${^{st}}$ & 2${^{nd}}$ & 3${^{rd}}$ & 4${^{th}}$ & 5${^{th}}$ \\\hline
Max & 2.249 & 2.248 & 2.248 & 2.244&2.244 \\\hline
Min& 0.579 & 0.580 & 0.582 & 0.582&0.583 \\\hline\hline
\multicolumn{6}{c}{{$\mathbf{L}^*$}: Condition number=$1.662\times 10^{18}$} \\\hline
$\lambda_i$ & 1${^{st}}$ & 2${^{nd}}$ & 3${^{rd}}$ & 4${^{th}}$ & 5${^{th}}$ \\\hline
Max & 2.290 & 2.224 & 2.216 & 2.213&2.201 \\\hline
Min$(\times10^{-2})$& 0.000 & 0.200 & 0.555 & 0.798&1.019\\\hline\hline
\multicolumn{6}{c}{$\mathbf{K}^*_s$: Condition number=1.517$\times 10^{4}$} \\\hline
$\lambda_i$ & 1${^{st}}$ & 2${^{nd}}$ & 3${^{rd}}$ & 4${^{th}}$ & 5${^{th}}$ \\\hline
Max & 3.567 & 3.556 & 3.214 & 3.181&3.161 \\\hline
Min$(\times10^{-3})$& 1.173 & 1.193 & 1.205 & 1.222&1.238 \\\hline
\hline
\end{tabular}
\caption{The 1-norm condition number estimates and five maximum and minimum eigenvalues ($\lambda_i$) of the preconditioned {\it local} matrices $\mathbf{A}^*_f$, $\mathbf{K}^*$, $\mathbf{L}^*$ without resistance BC and $\mathbf{K}^*_s$. The 5th maximum eigenvalue of {\it local} $\mathbf{A}_f^*$ is complex, with imaginary part of $-3.257\times10^{-5}$.}
\label{table:8}
\end{table}
 
\begin{figure}
\centering
 {\includegraphics[width=0.30\textwidth, keepaspectratio]{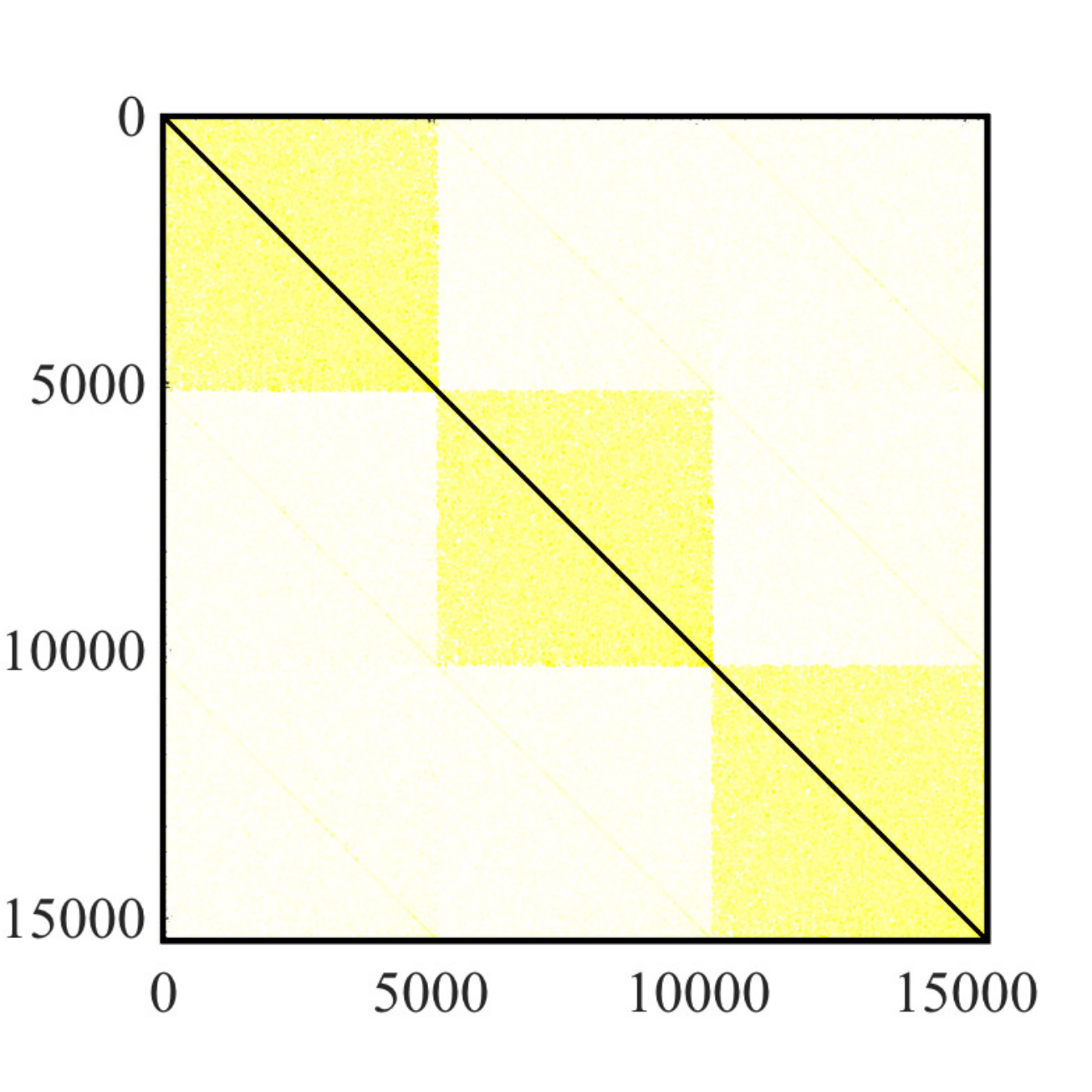}}
 {\includegraphics[width=0.30\textwidth, keepaspectratio]{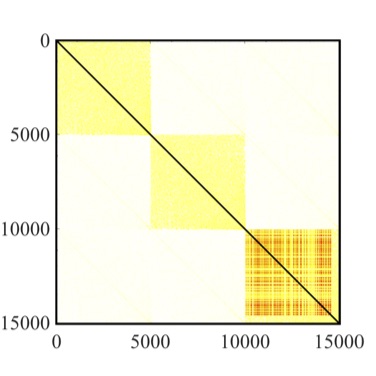}}
 {\includegraphics[width=0.34\textwidth, keepaspectratio]{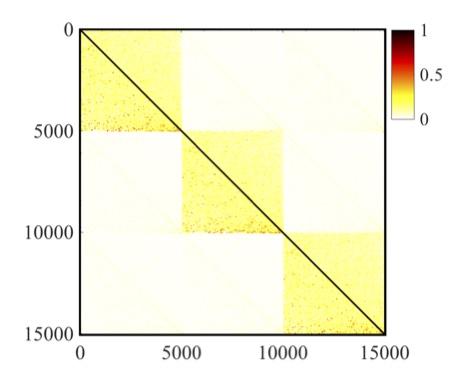}}
\caption{{\it local} sparse matrix structure of preconditioned $\mathbf{K}$ block. (top) $\mathbf{\tilde{K}^*}$ without resistance BC, (center) $\mathbf{\tilde{K}^*}$ with a resistance BC, (bottom) $\mathbf{H\tilde{K}}$. The colorbar is same for all figures.}\label{fig:prec_res}
\end{figure}

Application of a symmetric Jacobi preconditioner to a {\it local} coefficient with resistance boundary condition leads to the eigenvalues reported in Table~\ref{table:9}. 
Despite a reduction by three orders of magnitude, the condition number is still one order of magnitude larger than in the case without resistance BC. 
Note that the maximum eigenvalue of the preconditioned matrix is one order of magnitude larger than the second maximum eigenvalue, consistent with previous observations.
Thus, the RPC preconditioning proposed in~\cite{Esmaily2013a} seeks a preconditioning matrix $\mathbf{H}$ such that $\mathbf{H}\simeq {(\mathbf{K}^{total})}^{-1}$. 
The idea is to construct $\mathbf{H}$ by combining the diagonal components of $\mathbf{K}$ with the resistance contributions stored in $\mathbf{S}_j$ as
\begin{equation}
\mathbf{H}=({\mathbf{K}}^d)^{-1} - \sum_{j=1}^{n^{bc}} \bigg[ \frac{R_j ({(\mathbf{K}^d)}^{-1}\mathbf{S}_j)\otimes({(\mathbf{K}^d)}^{-1}\mathbf{S}_j)}{1+R_j || {(\mathbf{K}^d)}^{-\frac{1}{2}}\mathbf{S}_j||^2}\bigg], 
\end{equation}
where $\mathbf{K}^d$ is $\text{diag}(\mathbf{K})$. 
The preconditioned matrix $\mathbf{H\tilde{K}}$ has a small condition number (Table~\ref{table:7}) and smaller off-diagonal entries (Figure~\ref{fig:prec_res}).

\begin{table}
\centering
\begin{tabular}{ c |c c c c c  }
\hline
\multicolumn{6}{c}{{$\mathbf{\tilde{A}}^*_f$}: Condition number=9243} \\
\hline
$\lambda_i$ & 1${^{st}}$ & 2${^{nd}}$ & 3${^{rd}}$ & 4${^{th}}$ & 5${^{th}}$ \\
\hline
Max& 55.15 & 2.262 & 2.247 & 2.247&2.243\\
\hline
Min$(\times10^{-1})$& 0.232 & 0.358 & 0.467 & 0.492&3.99\\
\hline
\hline
\multicolumn{6}{c}{{$\mathbf{\tilde{K}}^*$}: Condition number=1862} \\
\hline
$\lambda_i$ & 1${^{st}}$ & 2${^{nd}}$ & 3${^{rd}}$ & 4${^{th}}$ & 5${^{th}}$ \\
\hline
Max & 55.16 & 2.248 & 2.247 & 2.244&2.241 \\
\hline
Min& 0.217 & 0.384 & 0.392 & 0.396&0.399 \\
\hline
\hline
\multicolumn{6}{c}{{$\mathbf{H\tilde{K}}$}: Condition number=76} \\
\hline
$\lambda_i$ & 1${^{st}}$ & 2${^{nd}}$ & 3${^{rd}}$ & 4${^{th}}$ & 5${^{th}}$ \\
\hline
Max & 2.248 & 2.247 & 2.244 & 2.241&2.241 \\
\hline
Min& 0.217 & 0.384 & 0.392 & 0.396&0.399\\
\hline
\end{tabular}
\caption{The 1-norm condition number estimates and five maximum and minimum eigenvalues ($\lambda_i$) of the preconditioned {\it local} matrix $\mathbf{\tilde{A}}^*_{f}$, $\mathbf{\tilde{K}}^*$, and $\mathbf{H\tilde{K}}$.}
\label{table:9}
\end{table}

\section{Performance of linear solvers in patient-specific models}\label{sec:PS}

In this section we tested the performance of linear solvers on patient-specific cardiovascular models, in an effort to extrapolate the results obtained for the pipe benchmark to more realistic problems. 
We use three different models with a wide range of boundary conditions (i.e., resistance, RCR, coronary BC, closed-loop multidomain), with and without wall deformability and covering various patient-specific anatomies. All anatomic models were constructed from medical image data using SimVascular.  

\subsection {\bf Pulmonary hypertension}
\begin{figure}[h!]
\centering
{\includegraphics[width=0.43\textwidth, keepaspectratio]{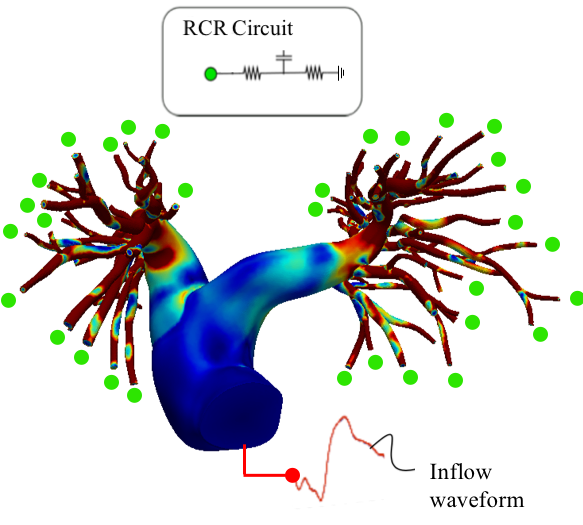}}
{\includegraphics[width=0.40\textwidth, keepaspectratio]{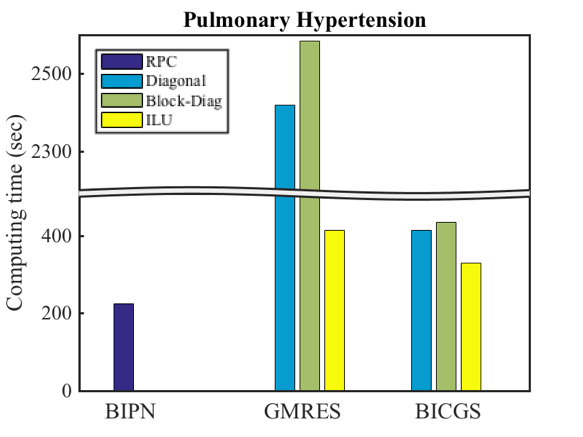}}
\caption{(Top) Patient-specific model for pulmonary hypertension with schematics of boundary conditions. The model is colored by instantaneous wall shear stress. (bottom) compute times for preconditioned iterative linear solvers.}\label{fig:PH}
\end{figure}

The first model represents the left and right pulmonary arteries with associated branches and is used to investigate the effects of pulmonary hypertension (PH). The finite element mesh contains 3,223,072 tetrahedral elements to represent the pulmonary lumen, has rigid walls and 88 outlets with Windkessel (RCR) boundary conditions, prescribed through a coupled 0-D multi-domain approach \cite{Esmaily2012,Yang2018} (Figure~\ref{fig:PH}). A pulsatile inflow waveform extracted from PC-MRI was imposed at the pulmonary artery inlet. This model is solved using a time step of 0.46 milliseconds and 120 cores ($\approx 25,000$ elements per core). The tolerance on the Newton-Raphson residual is set to $\epsilon =10^{-4}$.

We briefly report {\it global} matrix characteristics in the PH model. The diagonal dominance metric is $D(\mathbf{A}_f)=0.5598$, and $D(\mathbf{K})=0.7397$. The metric value for $D(\mathbf{A}_f)$ is similar to values from the pipe model with rigid walls. The matrix is nearly-symmetric, $S(\mathbf{A}_f)=0.9903$. 

Results in Figure~\ref{fig:PH}(c) compare the performance of diagonal, block-diagonal and ILU preconditioning. BIPN with RPC shows the best performance, followed by ILU-BICG. This is expected due to the large number of resistance boundary conditions (i.e., 88) at the model outlets. Diagonal preconditioners with GMRES instead perform poorly, consistent with our observations in the pipe benchmark. 

\subsection  {\bf Coronary artery bypass graft}
\begin{figure}[t!]
\vspace{-6pt}
\centering
{\includegraphics[width=0.38\textwidth, keepaspectratio]{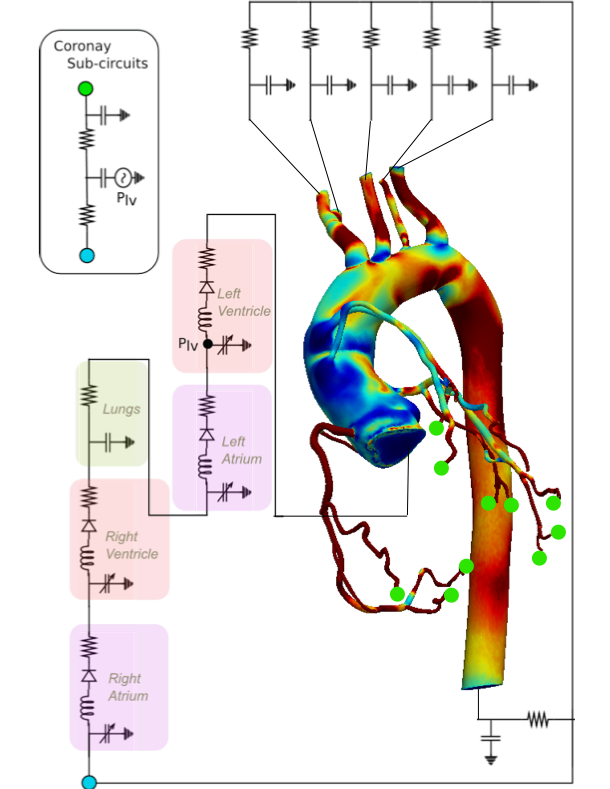}}
{\includegraphics[width=0.37\textwidth, keepaspectratio]{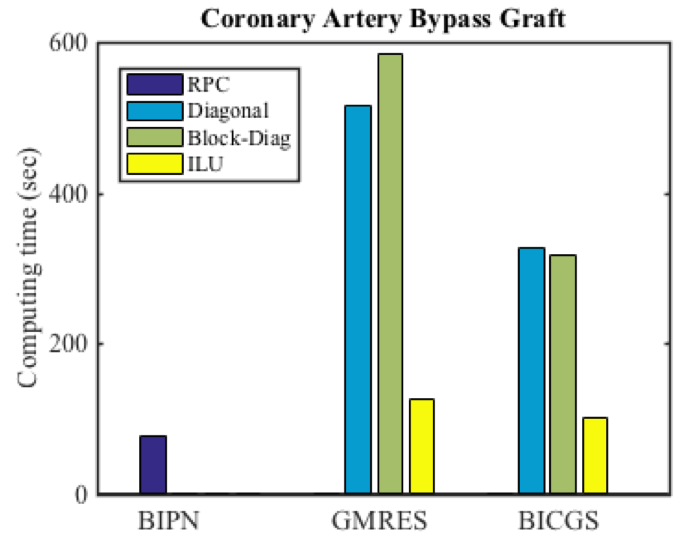}}
\caption{(Top) Patient-specific model for coronary bypass graft model with schematics of boundary conditions. The model is colored by instantaneous wall shear stress. (bottom) compute times for preconditioned iterative linear solvers.}\label{fig:CABG}
\end{figure}

Second, we consider a model of coronary artery bypass graft surgery (see, e.g.,~\cite{Ramachandra2017,Tran2017}) with 4,199,945 tetrahedral elements and rigid walls, coupled with a closed-loop 0D lumped parameter network (LPN), including coupled heart, coronary and systemic circulation models (Figure~\ref{fig:CABG}). Simulations were performed using 168 cores ($\sim24,000$ elements per core), with a time step of 0.87 millisecond and a non linear iteration tolerance of $\epsilon =10^{-3}$.

The diagonal dominance metric for the {\it global} matrix in the CABG model is $D(\mathbf{A}_f)=0.5200$, and $D(\mathbf{K})=0.6914$. The matrix in the CABG model is less diagonally dominant than the previous cylinder or pulmonary hypertension model. The matrix is also near symmetric, $S(\mathbf{A}_f)=0.9938$. 

As expected, due to the presence of coupled multi-domain boundary conditions~\cite{Esmaily2013a}, BIPN results in the best performance, followed closely by BICG with ILU, while GMRES with a diagonal preconditioner performs poorly. The relative performance of ILU against the diagonal preconditioner is better than seen in previous models. The smaller diagonal dominance metric in CABG model confirms superiority of ILU over the diagonal preconditioner. 

\subsection  {\bf Left coronary}
Next, we tested performance of linear solvers in a left coronary model. The left coronary model was extracted from a full coronary artery model used in Tran {\it et al}.\cite{Tran2017}. The pulsatile flow waveform at the inlet of the left coronary branch was extracted from the full model simulation and imposed at the inlet of the model. The model has six outlets, each with applied open-loop coronary outlet boundary conditions \cite{Kim2009}. All resistance and compliance values as well as inflow pulsatile waveforms are determined to produce a normal physiologic response of the left coronary artery following our prior work. We ran simulations with rigid and deformable walls with a lumen mesh containing 486,066 tetrahedral elements and a vessel wall mesh with 206,369 tetrahedral elements, time step of 1 millisecond and tolerance of $\epsilon=10^{-4}$. We used 20 cores for the rigid wall simulation and 24 cores for the deformable wall simulation. 

The diagonal dominance metric for the left coronary model with rigid wall is $D(\mathbf{A}_f)=0.5238$, and $D(\mathbf{K})=0.6972$, which are similar to the CABG model. The matrix is near symmetric, $S(\mathbf{A}_f)=0.9855$, however, this model is furthest from symmetric among all models considered.

For the ALE FSI model, the diagonal dominance is reduced as $D(\mathbf{A}_{FSI})=0.4323$, $D(\mathbf{K})=0.5320$. Note that the number of elements in a wall mesh is 40 percent of the fluid mesh, so the effect of adding solid mechanics in the linear system is more significant than the pipe case where only 20 percent of elements were in the in wall mesh. As a result, the symmetry metric is very close to 1, $S(\mathbf{A}_{FSI})=0.99998$, since $\mathbf{K}_s$ is symmetric. 

As shown in Figure \ref{fig:LC}, performance test results are consistent with previous findings. RPC-BIPN is the fastest method for the rigid wall simulation. In FSI, the performance of BIPN is poor, while diagonal and ILU preconditioners with GMRES perform better.  

\begin{figure}
\vspace{-6pt}
\centering
{\includegraphics[width=0.38\textwidth, keepaspectratio]{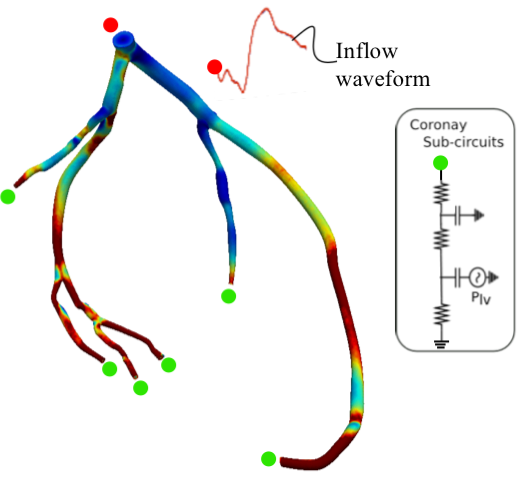}}
{\includegraphics[width=0.36\textwidth, keepaspectratio]{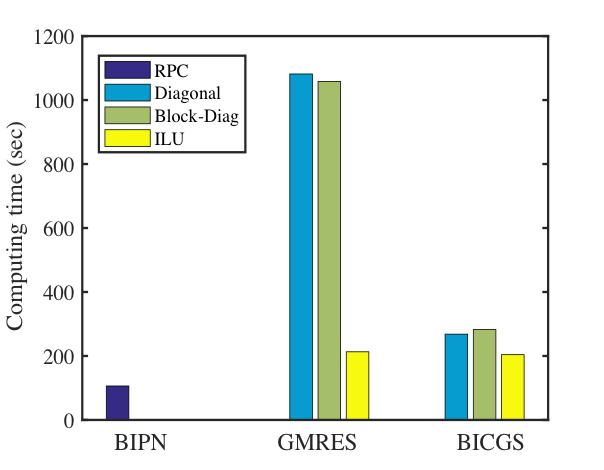}}
{\includegraphics[width=0.36\textwidth, keepaspectratio]{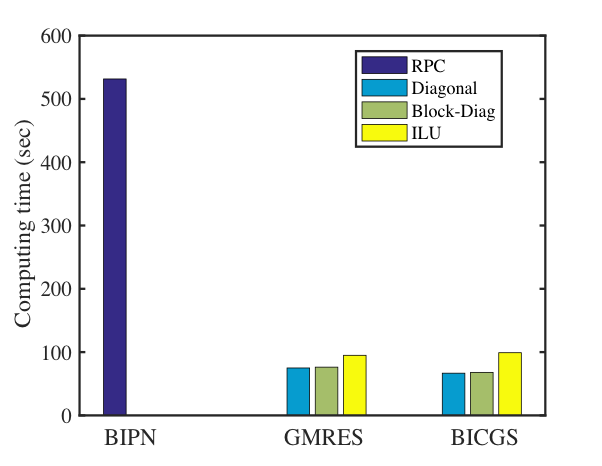}}
\caption{(Top) Patient-specific model for left coronary artery model with schematics of boundary conditions. The model is colored by instantaneous wall shear stress. (center) compute times for preconditioned iterative linear solvers in the rigid wall simulation. (bottom) compute times for preconditioned iterative linear solvers in the deformable wall simulation.}
\label{fig:LC}
\vspace{-4pt}
\end{figure} 

\subsection {Discussion}
From the performance results in the patient-specific models, we find that RPC-BIPN is the fastest method for rigid wall simulations with many resistance BCs in agreement with the pipe model. ILU-BICG is only slightly slower than RPC-BIPN while the standard diagonal scaled GMRES fails. The performance degradation of RPC-BIPN in FSI models is consistent with the pipe model and suggests the need for future improvements of BIPN for ALE FSI. 
\section{Summary and conclusions}\label{sec:Conclusions}
In this paper we study the performance of preconditioned iterative linear solvers for cardiovascular simulation in rigid and deformable walls. To this end, we implement various iterative linear solvers - GMRES, BICGS, and BIPN - and preconditioners - diagonal, block-diagonal, ILU, ILUT, ML, and RPC in a single flow solver. Standard iterative solvers and preconditioners are employed from the Trilinos library and compared against RPC-BIPN, implemented in our in house solver. Simulation wall clock time is measured and compared in a benchmark pipe flow with a resistance BC. 

ILU preconditioned BICG provides the best overall performance in both rigid and deformable wall simulations. RPC-BIPN in the FSI simulation shows $\sim 8$ fold increase in compute time compared to the rigid wall case. Strong and weak scalings of ILU-BICG and RPC-BIPN are reported. 

To better understand the observed performance, characteristics of the left-hand-side matrix in the linear system are examined. We report sparsity patterns, diagonal dominance, symmetry, eigenvalues and condition numbers in {\it global} and {\it local} matrices. Results show that the sparse matrix structure has a narrow banded structure after Reverse-Cuthill-McKee reordering. The matrix from the fluid domain has larger diagonal values than off-diagonals and is nearly symmetric. Eigenvalues and the condition number of the matrix from the fluid domain show that the $\mathbf{K}$ block has a significantly smaller condition number compared to the $\mathbf{A}_f$ matrix, supporting the main premise of BIPN. Effects of preconditioning on matrix characteristics are investigated by explicitly forming the preconditioned matrix. A diagonal preconditioner is shown to be effective to reduce the range of eigenvalues in the fluid domain, especially for the $\mathbf{K}$ matrix.

Adding wall deformability to the fluid simulation increases the bandwidth of the matrix and decreases the relative magnitudes of the diagonal values compared to the off-diagonal values. Due to the reduction of diagonal dominance, a diagonal preconditioner does not significantly reduce the condition number of the original matrix. 

The resistance boundary condition disturbs the sparsity and diagonal dominance of the original fluid matrix, and causes an ill-conditioned system by adding an eigenvalue which is larger than the maximum eigenvalue of matrix without resistance BC. The resistance based preconditioner successfully reduces the condition number of the system with a resistance boundary condition by four orders of magnitude, while a diagonal preconditioner only reduces the condition number by two orders of magnitudes. 

The performance of various preconditioned linear solvers is evaluated in four patient-specific models. In these models, RPC-BIPN is best for rigid wall models with multiple resistance or coupled LPN outlet boundary conditions. In the deformable wall simulation, RPC-BIPN shows significant performance degradation and diagonal preconditioners or ILU with BICG achieve the best performance. 

This study motivates several new research directions to develop new preconditioned linear solver strategies. The effectiveness of BIPN for the solution of fluids problems with rigid walls has been proven in the current study. Currently our in-house code (RPC-BIPN) uses the bi-partitioned approach for ALE FSI, forming a linear system from the momentum equations for the fluid and the solid domains together, and another system for the continuity equation for the fluid domain. However, since the characteristics of the matrix in the solid domain is different from the fluid domain, most notably diagonal dominance, linear systems from these two domains should be separately solved (i.e. Tri-partitioning). The inefficiency of solving FSI in BIPN stems from adding off-diagonal dominance to the left-hand-side matrix block $\mathbf{K}$. Since RPC is based on a simple diagonal preconditioner, solving the system $\mathbf{K}$ becomes less efficient. We suggest solving $\mathbf{K}_s$ separately with an Incomplete Cholesky preconditioner, exploiting its symmetric property, rather than a simple diagonal preconditioning. Exploration of this idea is the subject of future work.   

Additionally, we point out that the Schur complement block in BIPN is not preconditioned. Since a major portion of the computational cost in BIPN is consumed when solving the Schur complement block \cite{Esmaily2015}, acceleration of the linear solver performance by a proper preconditioning technique could significantly reduce the compute time. To form a preconditioner for the Schur complement block, one would need an efficient sparse matrix-matrix multiplication scheme as well as explicit formation of the Schur complement block. The open-source Trilinos library provides this option as well as various preconditioners so combining a partitioning approach with Trilinos is expected to provide consistent performance in both rigid and deformable wall simulations of cardiovascular hemodynamics. Implementation and testing of this approach is left for future investigation. Testing of linear solver performance on more complex patient-specific disease with large wall deformations (e.g. aortic dissection) are warranted, and would likely lead to further insights. Future studies are also warranted to further assess solver performance and matrix characteristics, towards development of new solver and preconditioner strategies.    

\section{Appendix}
\appendix
\section{GMRES restart}
\label{app:restart}
We tested different GMRES restart numbers in the pipe benchmark problems. In Figure \ref{fig:LC}, we plot compute times for preconditioned GMRES using a pipe model with rigid wall and a pipe with deformable wall. Our test shows that decreasing restart number increases the compute time of linear solver in the rigid pipe model. The FSI model does not show a notable difference. 

\begin{figure}
\vspace{-6pt}
\centering
{\includegraphics[width=0.38\textwidth, keepaspectratio]{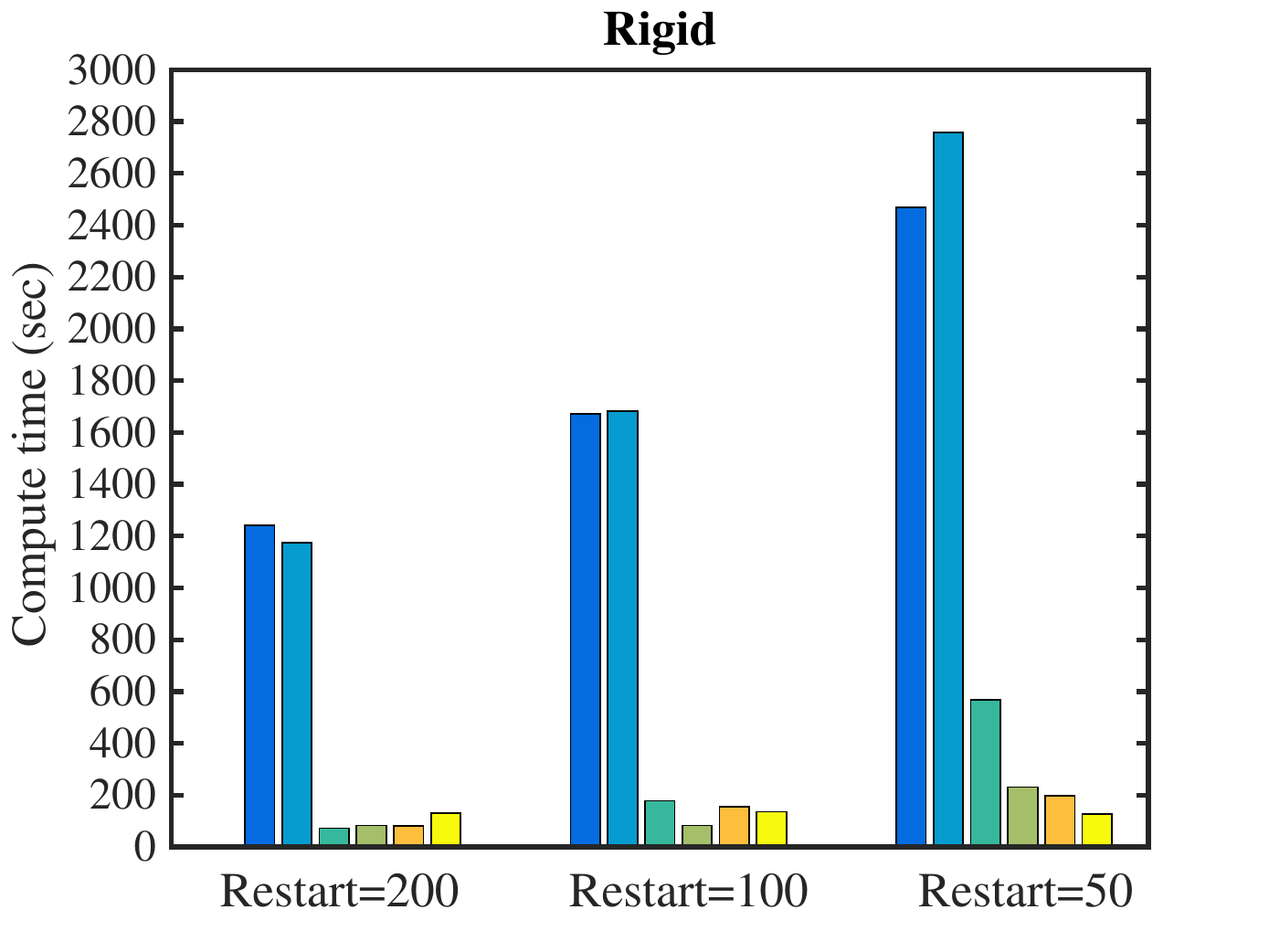}}
{\includegraphics[width=0.38\textwidth, keepaspectratio]{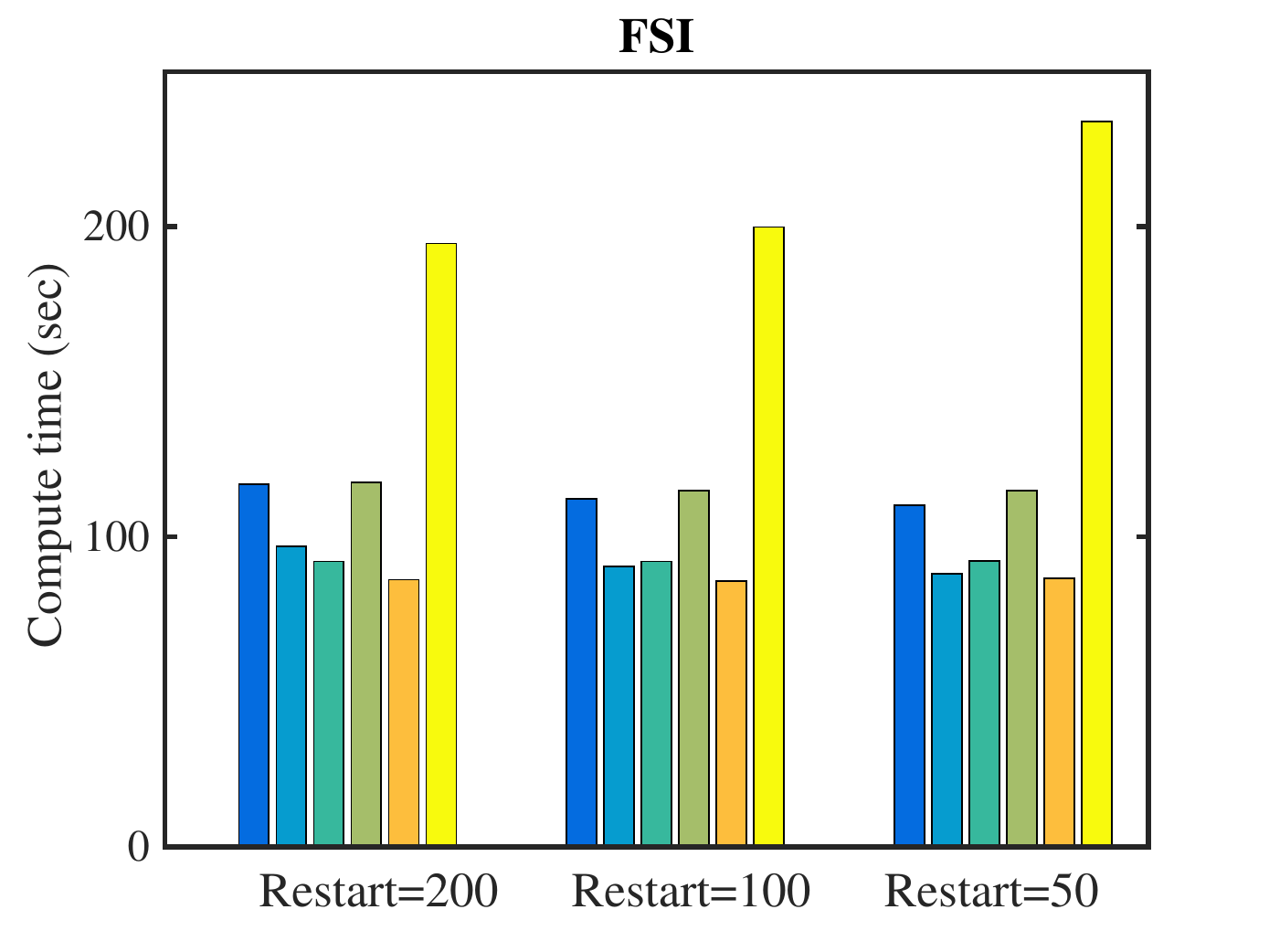}}
\caption{Compute times for GMRES and preconditioners using (top) a rigid pipe model, and (bottom) a pipe model with deformable wall using different the GMRES restart numbers.}
\label{fig:LC}
\vspace{-4pt}
\end{figure} 

\section{Choice of smoother and subsmoother for ML}
\label{app:ML}

\begin{figure}
\vspace{-6pt}
\centering
{\includegraphics[width=0.5\textwidth, keepaspectratio]{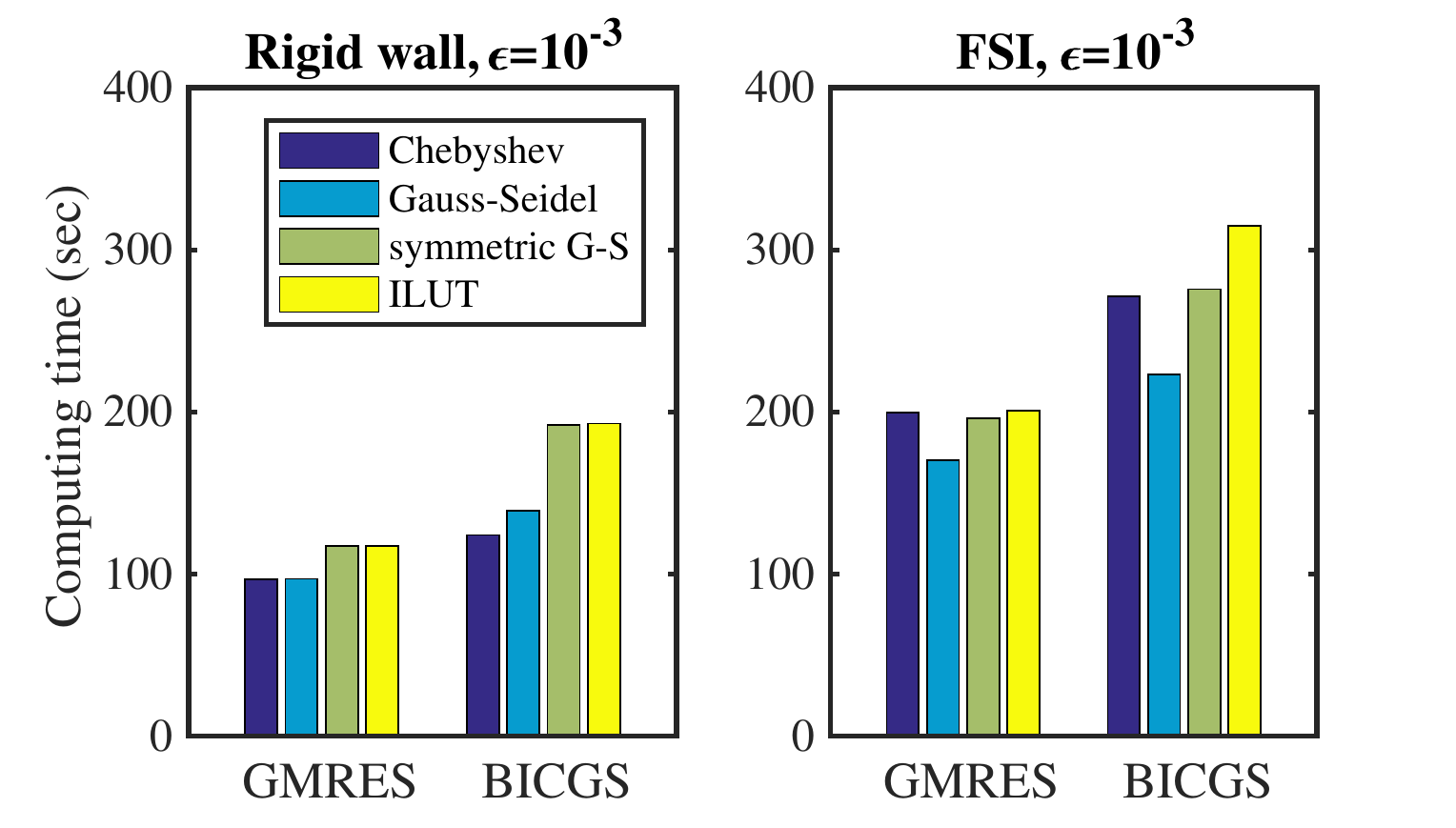}}
\caption{Compute times for linear solvers preconditioned with ML using a (a) rigid and (b) an FSI pipe model with tolerance $\epsilon=10^{-3}$, with different smoothers. The symmteric Gauss-Seidel subsmoother is used. For the rigid wall model, 38 cores are used. For the FSI model, 48 cores are used.}
\label{fig:smoother}
\vspace{-4pt}
\end{figure}

In the ML package, multiple options are available for the smoother and the subsmoother. As shown in Figure \ref{fig:smoother}, the Gauss-Seidel smoother works best among Chebyshev, symmetric Gauss-Seidel, and ILUT. For the subsmoother, the symmetric Gauss-Seidel is the best among Chebyshev and MLS. 

\begin{figure}
\vspace{-6pt}
\centering
{\includegraphics[width=0.5\textwidth, keepaspectratio]{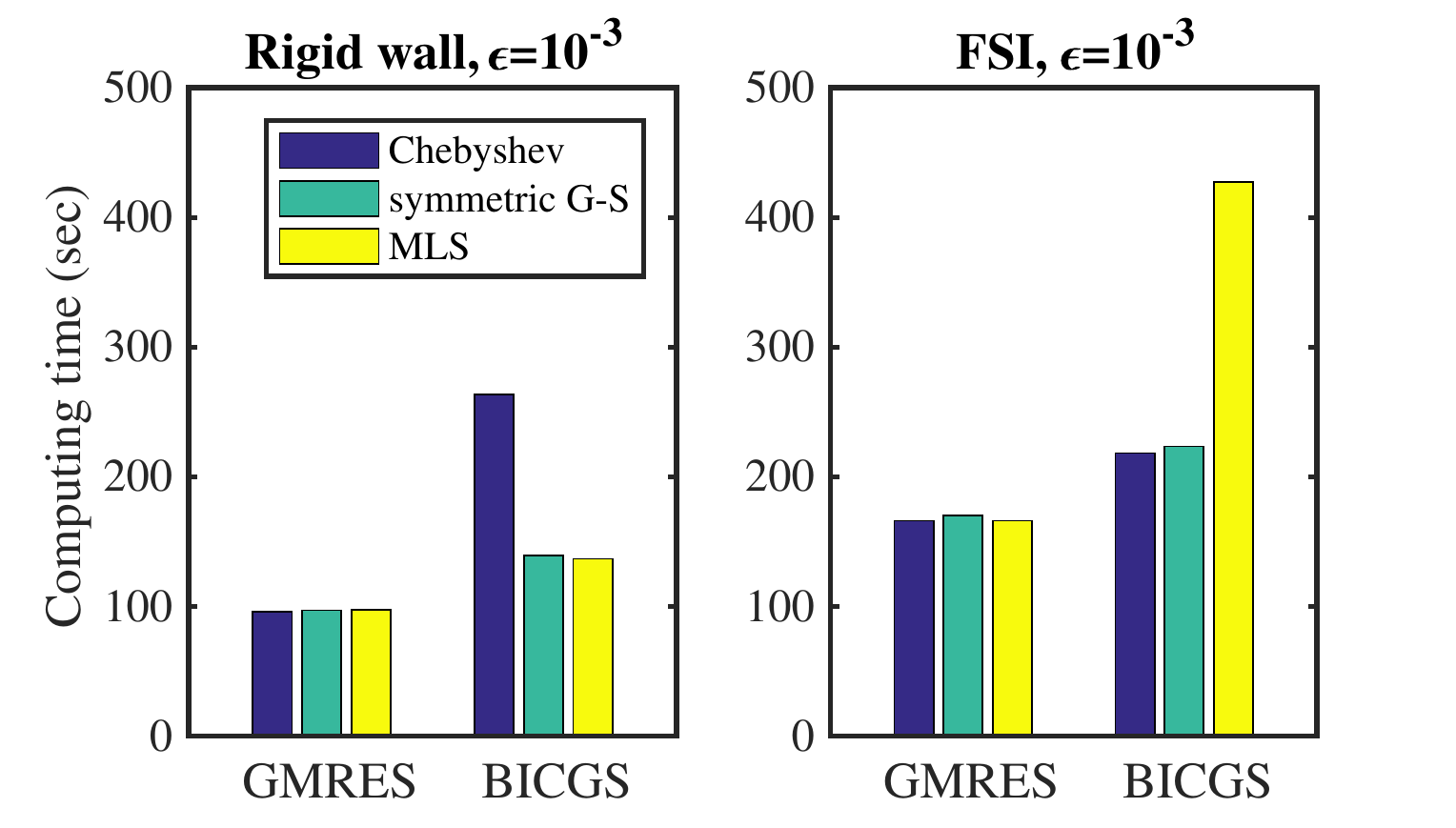}}
\caption{Compute times for linear solvers preconditioned with ML using a (a) rigid and (b) FSI pipe model with tolerance $\epsilon=10^{-3}$, with different subsmoothers. The Gauss-Seidel smoother is used. For the rigid wall model, 38 cores are used. For the FSI model, 48 cores are used.}
\label{fig:subsmoother}
\vspace{-4pt}
\end{figure}

\section{Effect of reordering in ILU}
\label{app:reordering}
\begin{figure}
\vspace{-6pt}
\centering
{\includegraphics[width=0.5\textwidth, keepaspectratio]{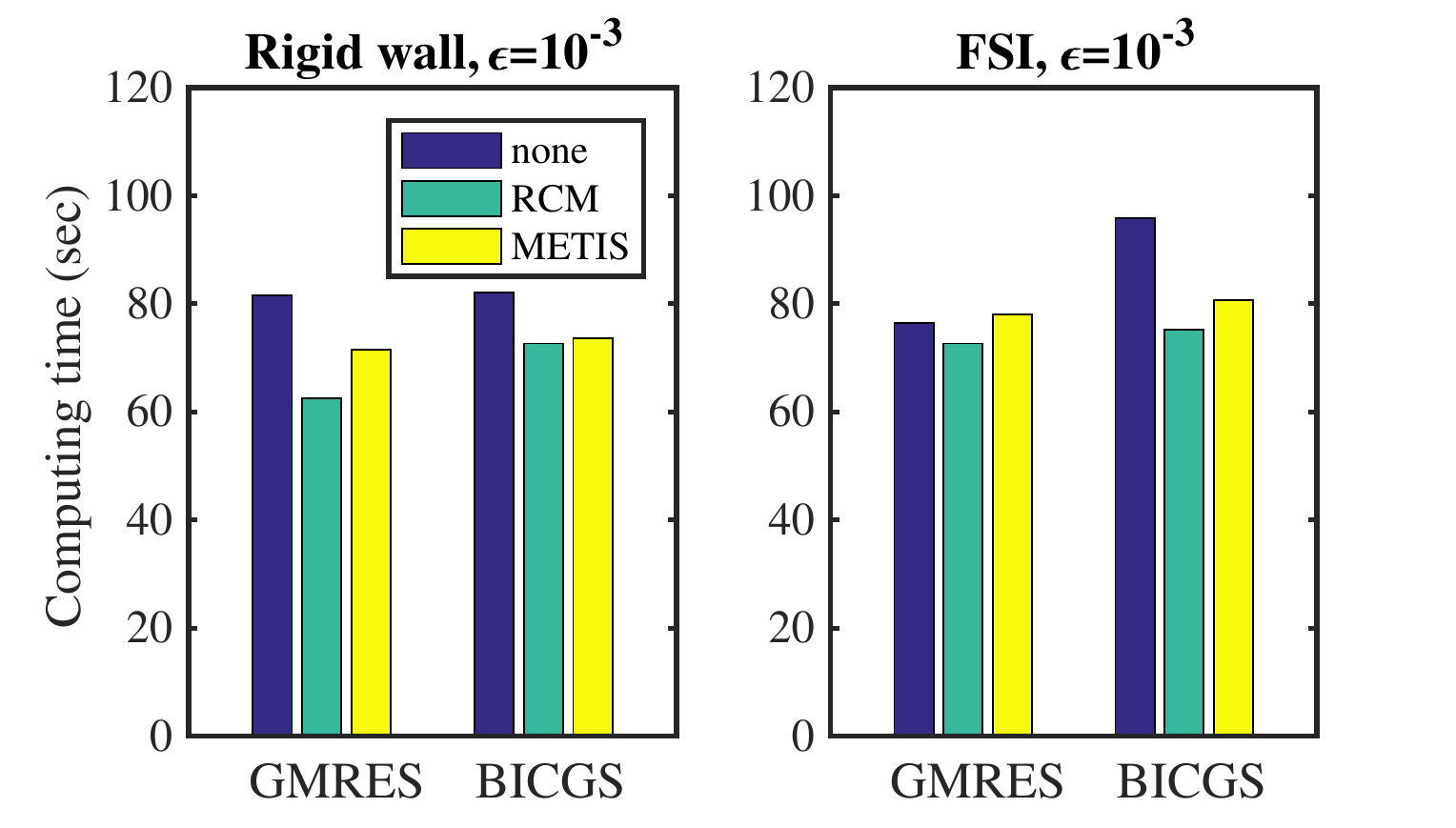}}
\caption{Compute times for linear solvers preconditioned with ILUT using a (a) rigid and (b) FSI pipe model with tolerance $\epsilon=10^{-3}$, with different reorderings. For the rigid wall, 38 cores are used. For the FSI, 48 cores are used.}
\label{fig:reordering}
\vspace{-4pt}
\end{figure}

We evaluated and compared compute times of linear solvers with different reordering methods. RCM and METIS reordering for ILUT via the Trilinos IFPACK are implemented. We use 2 level fill-in and $10^{-2}$ dropping tolerance for this test. Figure \ref{fig:reordering} shows performance differences between ILUT with different reordering schemes. From the testing, we confirm that the RCM is the fastest method against METIS and no reordering. The superior performance of RCM is notable when GMRES is used with ILUT. 

\section{Eigenvalue spectrums of the {\it local} and {\it global} matrix. }
\label{app:eigen}
\begin{figure}
\vspace{-6pt}
\centering
{\includegraphics[width=0.49\textwidth, keepaspectratio]{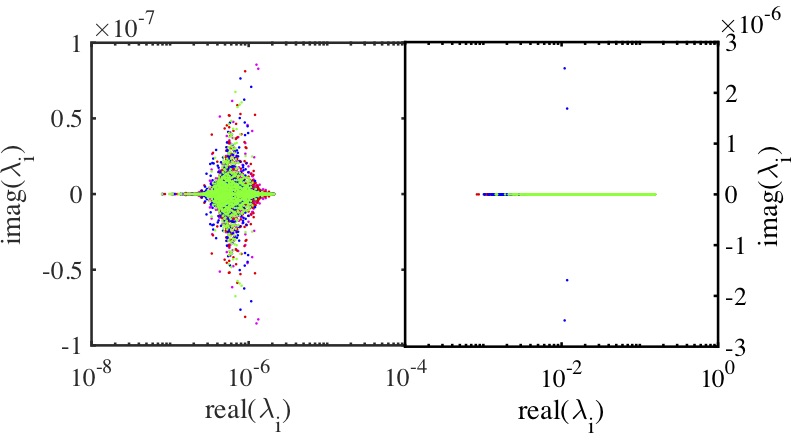}}
{\includegraphics[width=0.49\textwidth, keepaspectratio]{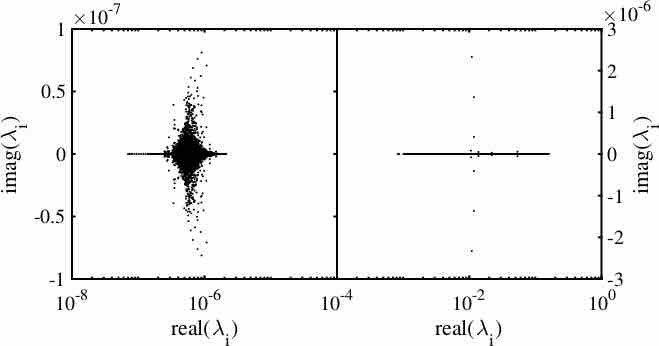}}
\caption{The spectrum of eigenvalues for a rigid pipe model with Neumann BC at the outlet. (top) eigenvalues obtained from four {\it local} matrices. Different colors are used to represent eigenvalues from different {\it local} matrices. (bottom) eigenvalues obtained from the {\it global} matrix.}
\label{fig:globalocal}
\vspace{-4pt}
\end{figure}

In this section we compare the spectrum of eigenvalues in the {\it local} and {\it global} matrices and investigate how our analysis on {\it local} eigenvalues can be generalized to the {\it global} matrix. We use a pipe model in the same dimension shown in Figure \ref{fig:cylinder} meshed with 24,450 elements with $N_{nd}=5462$. We use one core to extract the {\it global} matrix, and four cores to examine {\it local} matrices. As shown in Figure \ref{fig:globalocal}, the eigenvalue distributions of the {\it global} and {\it local} matrices are similar. Although the eigenvalues from the {\it global} and {\it local} matrices are not exactly the same, the distribution of eigenvalues of {\it local} matrices is a good approximations to the distribution of eigenvalues in the {\it global} matrix. 

\begin{acknowledgements}
This work was supported by NIH grant (NIH R01-EB018302), NSF SSI grants  1663671 and 1339824, and  NSF CDSE CBET 1508794. This work used the Extreme Science and Engineering Discovery Environment (XSEDE)\cite{XSEDE}, which is supported by National Science Foundation grant number ACI-1548562.
We thank Mahidhar Tatineni for assisting on building Trilinos on Comet cluster, which was made possible through the XSEDE Extended Collaborative Support Service (ECSS) program \cite{EECS}. 
The authors also thank Michael Saunders, Michael Heroux, Mahdi Esmaily, Ju Liu, and Vijay Vedula, for fruitful discussions that helped in the preparation of this paper. The authors would like to thank the two anonymous reviewers whose comments greatly contributed to improve the completeness of the present study. We also acknowledge support from the open source SimVascular project at www.simvascular.org.
\end{acknowledgements}

\bibliography{./Linear_solver}
\bibliographystyle{abbrv} 
\end{document}